\documentclass[11pt,a4paper,british]{article}

\newcommand{\blind}{0}

\newcommand{\BW}{0}

\usepackage{scalerel,tikz,xcolor,hyperref} 

\definecolor{lime}{HTML}{A6CE39}
\DeclareRobustCommand{\orcidicon}{%
    \thinspace\scalerel*{\includegraphics{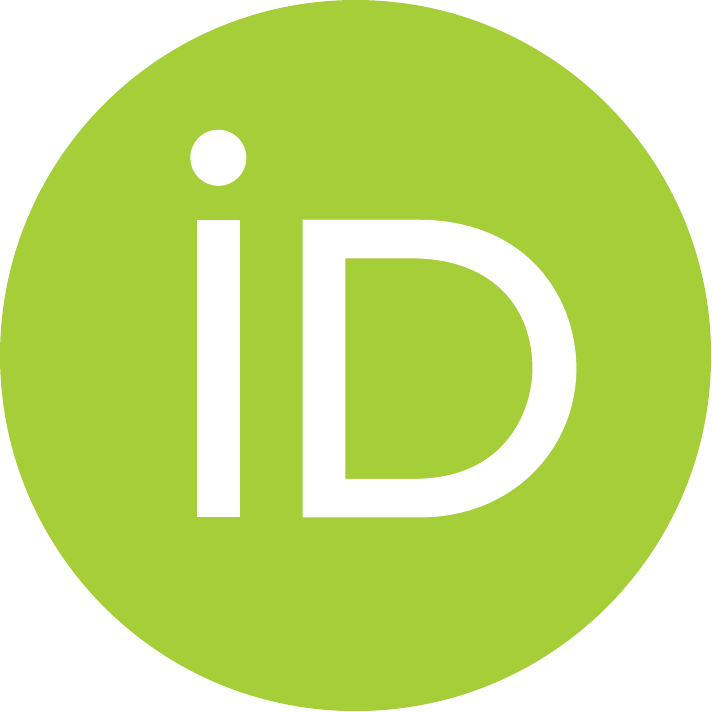}}{A}\!\! 
}

\foreach \x in {A, ..., Z}{%
	\expandafter\xdef\csname orcid\x\endcsname{\noexpand\href{https://orcid.org/\csname orcidauthor\x\endcsname}{\noexpand\orcidicon}}
}

\usepackage[T1]{fontenc}
\usepackage[latin9]{inputenc}
\usepackage{lmodern}
\usepackage{graphicx}
\usepackage{amsthm}
\usepackage{babel}
\usepackage{bm}

\usepackage{epstopdf}
\DeclareGraphicsExtensions{.eps,.pdf}
\if1\BW
	\graphicspath{{./Figures/Black_White/BW_}}
\else
	\graphicspath{{./Figures/Color/}}
\fi

\usepackage{authblk,bibentry,natbib}
\usepackage{mathrsfs}
\usepackage[titletoc,title]{appendix}
\usepackage{amsmath}
\usepackage{amssymb}
\usepackage{setspace}
\usepackage{float}
\usepackage[labelformat=simple]{subcaption}
\usepackage{hhline}
\usepackage{booktabs}
\usepackage{mathtools}
\usepackage{hyperref}
\usepackage{cleveref}
\usepackage{nameref}
\usepackage{thmtools}
\usepackage{pdflscape}
\usepackage[stable]{footmisc}
\usepackage{relsize}
\usepackage{enumitem}
\usepackage{tabularx}
\usepackage{bbm}
\usepackage{datetime}
\usepackage[ruled,vlined,linesnumbered]{algorithm2e}

\usepackage{chngcntr} 

\usepackage{geometry}
\PassOptionsToPackage{normalem}{ulem}
\usepackage{ulem}

\usepackage{color}
\definecolor{webbrown}{rgb}{0.65, 0.16, 0.16}
\definecolor{RoyalBlue}{rgb}{0.0, 0.14, 0.4}
\definecolor{webgreen}{rgb}{0.0, 0.5, 0.0}
\hypersetup{
colorlinks=true, linktocpage=true, pdfstartpage=1, pdfstartview=FitV,
breaklinks=true, pdfpagemode=UseNone, pageanchor=true, pdfpagemode=UseOutlines,
plainpages=false, bookmarksnumbered, bookmarksopen=true, bookmarksopenlevel=1,
hypertexnames=true, pdfhighlight=/O, urlcolor=webbrown, linkcolor=RoyalBlue, citecolor=webgreen}

\renewcommand{\theequation}{\arabic{equation}}

\newtheorem*{assumption*}{Assumption}
\crefname{equation}{equation}{equations}
\Crefname{algocf}{Algorithm}{Algorithms}

\newcommand{\nonl}{\renewcommand{\nl}{\let\nl\oldnl}}

\newcommand*{\doi}[1]{\url{https://doi.org/#1}}

\usepackage{soul}

\newcommand{\double}{0}

\newcommand{\finalv}{0}

\if1\double
    
\fi

\begin{document}

\title{Customer Price Sensitivities in Competitive Automobile Insurance Markets}
\if0\blind
    \author{\large Robert Matthijs Verschuren\orcidA{}\thanks{
    \if1\double
        \protect\linespread{1.5}\protect\selectfont 
    \fi
    Corresponding author: Amsterdam School of Economics, University of Amsterdam, Roetersstraat 11, 1018 WB, Amsterdam, The Netherlands. E-mail: \href{mailto:r.m.verschuren@uva.nl}
    {\tt r.m.verschuren@uva.nl}}}
    \affil{\it Amsterdam School of Economics, University of Amsterdam}
\fi
\if1\blind
    \author{}
\fi
\if0\finalv
    \date{This version is released on \usdate\today.}
\fi
\if1\finalv
    \date{}
\fi
\maketitle

\begin{abstract}

\noindent Insurers are increasingly adopting more demand-based strategies to incorporate the indirect effect of premium changes on their policyholders' willingness to stay. However, since in practice both insurers' renewal premia and customers' responses to these premia typically depend on the customer's level of risk, it remains challenging in these strategies to determine how to properly control for this confounding. We therefore consider a causal inference approach in this paper to account for customers' price sensitivity and to deduce optimal, multi-period profit maximizing premium renewal offers. More specifically, we extend the discrete treatment framework of \citet{guelman2014} by Extreme Gradient Boosting, or XGBoost, and by multiple imputation to better account for the uncertainty in the counterfactual responses. We additionally introduce the continuous treatment framework with XGBoost to the insurance literature to allow identification of the exact optimal renewal offers and account for any competition in the market by including competitor offers. The application of the two treatment frameworks to a Dutch automobile insurance portfolio suggests that a policy's competitiveness in the market is crucial for a customer's price sensitivity and that XGBoost is more appropriate to describe this than the traditional logistic regression. Moreover, an efficient frontier of both frameworks indicates that substantially more profit can be gained on the portfolio than realized, also already with less churn and in particular if we allow for continuous rate changes. A multi-period renewal optimization confirms these findings and demonstrates that the competitiveness enables temporal feedback of previous rate changes on future demand.

\end{abstract}

\textbf{Keywords:} Causal inference, renewal optimization, price sensitivity, customer churn prediction, Extreme Gradient Boosting, automobile insurance. 
\section{Introduction} \label{Section1}

In the non-life insurance industry dynamic pricing has traditionally focused purely on risk classification of individual policyholders. Insurance companies have therefore often employed a cost-based pricing strategy to determine the necessary risk premium for their new and existing customers using, for instance, Generalized Linear Models (GLMs) (see, e.g., \citet{englund2008}; \citet{antonio2012}; \citet{verschuren2021}). This, in turn, may lead to an updated premium that policyholders must pay at renewal of their policy and that can directly increase the profits of the insurer.

However, a change in these premia can affect the insurer's profitability indirectly as well. Policyholders that receive a large premium increase, for instance, may decide to lapse their policy, whereas if they were offered a lower premium increase or a premium decrease they might have prolonged it. Attracting a new customer can be up to twelve or twenty times as costly as retaining an existing one \citep{torkzadeh2006,vafeiadis2015}. A change, or even no change, in the premium can thus influence the profits of the insurance company indirectly through the demand of policyholders. Insurers have become increasingly aware of, and interested in, this indirect effect and have therefore been shifting their pricing strategies towards a more demand-based method.

The main objective of these demand-based pricing strategies is to account for the price sensitivity of individual customers. More specifically, insurers aim to deduce optimal, multi-period profit maximizing premium renewal offers from these price sensitivities for their customers. Crucial to this price optimization is therefore the response of customers if they had been given an alternative renewal offer, which we do not observe in practice. Suppose that we consider the premium offered to a customer as the treatment and the customer's decision whether to churn or not as the response. To determine what premium, or treatment, the insurer can best offer a customer, we would need to have some expectation of how this customer would respond to unobserved, counterfactual premium offers. As such, the premium renewal offer optimization can be seen as a causal inference problem.

In a randomized controlled experiment, the counterfactual responses are easy to infer since the treatment groups are formed randomly. However, the premium renewal offers in non-life insurance result from a risk-based pricing exercise and therefore introduce confounding in both the treatment assignment mechanism and the response mechanism. The standard observational approach attempts to control for this confounding by modeling the customer churn rate in terms of the premium renewal offer and the risk characteristics of the customer \citep{smith2000,yeo2001}. In practice, though, an insurer tends to offer higher premia to high risk customers and lower premia to low risk customers, since we expect the latter group to be less costly and more sensitive to premium increases than the former group. As a result, the confounders will generally be insufficiently balanced across treatments and it will be problematic to extrapolate the causal effects from the standard observational approach \citep{rubin1979,rubin1997,morgan2007,guo2009}.

A solution to this problem is provided in the causal inference approach of \citet{guelman2014}. They discretize the percentage premium changes and employ propensity score matching in the context of Rubin's model for causal inference \citep{rosenbaum1983,gu1993}. The general principle behind this matching algorithm is to pair customers based only on their conditional treatment assignment probability (the propensity score), rather than on all their risk characteristics, to infer the potential response to any counterfactual rate change. When the premium changes are continuous and known in practice, we can adopt the Generalized Propensity Score (GPS) as a function of the treatment dose \citep{hirano2004}. This GPS allows us to specify a continuous conditional distribution for assignment to every possible treatment dose \citep{imai2004}.

Despite its advantages, applications of this continuous treatment framework are sparse in non-life insurance. This paper therefore adds to the literature by adopting it for a price sensitivity analysis in Dutch automobile insurance and by comparing it to a multiple imputation extension of the discrete approach of \citet{guelman2014}, both with Extreme Gradient Boosting, or XGBoost, developed by \citet{chen2016}. More specifically, multiple imputation is considered for the discrete framework to better account for the uncertainty in the counterfactual responses. We additionally introduce the competitiveness of each renewal offer into these frameworks by incorporating competitor offers. As such, this paper not only investigates how to optimally set renewal premia over multiple periods but also how sensitive these are in practice to premia offered by competitors.

The remainder of this paper is organized as follows. In \Cref{Section2}, we elaborate on the causal inference framework and discuss the concepts of (generalized) propensity scores and matching based on these scores. While we describe the Dutch automobile insurance portfolio and provide details on the exact optimization procedure in \Cref{Section3}, \Cref{Section4} applies this methodology and elaborates on the results. To conclude this paper, we discuss the most important findings and implications in the final section. 
\section{Causal inference framework} \label{Section2}

\subsection{Customer price sensitivity} \label{Section2.1}

Since premium changes can affect the demand of customers, insurers want to account for this indirect effect to maximize their profits. In other words, they are interested in measuring the causal effect of rate changes on a customer's response, or a customer's price sensitivity. It is thus of key importance to have some expectation of how customers will respond to unobserved, counterfactual rate changes. In a randomized controlled experiment, this is easily deduced from the responses of customers that did receive the rate change of interest. However, in automobile insurance this is less straightforward due to the confounding in the treatment assignment mechanism and the response mechanism.

Let the random variable $Y_i(t) \in \{0, 1\}$ represent the potential response of policy $i$ to any treatment $t \in \mathcal{T}$. Suppose that $T_i$ and $\bm{X}_i$ denote the actual treatment and risk characteristics observed for policy $i$, respectively, where the probability of receiving a certain treatment given these risk characteristics (the propensity score) is defined as $\pi(t, \bm{X}_i) \coloneqq \mathbb{P}\left[ T_i = t | \bm{X}_i \right]$. Causal inference is still feasible in this context under two assumptions. The first assumption of weak unconfoundedness requires the potential response to be independent of the actually assigned treatment conditional on the risk characteristics, or
\begin{equation*}
    Y_i(t) \perp T_i | \bm{X}_i \quad \forall t \in \mathcal{T}, \quad \forall i
\end{equation*}
\citep{imbens2000,hirano2004,guardabascio2014}. This requirement implies that the treatment assignment mechanism may depend on observed risk characteristics and other potential responses, but not on any unobserved covariates. While we cannot test for this unconfoundedness, it most likely holds in practice since the rate changes $T_i$ in automobile insurance result from a risk-based pricing exercise using observed and available risk characteristics. The second assumption of common support concerns the distribution of observed covariates over the assigned treatments. More specifically, it requires each policy to have a possibility of receiving every treatment, or
\begin{equation*}
    0 < \pi(t, \bm{X}_i) < 1 \quad \forall t \in \mathcal{T}, \quad \forall i
\end{equation*}
\citep{rosenbaum1983,imbens2004,wager2018}. While common support may not hold globally for the risk characteristics, causal inference is still possible for the subset in which it does hold. Together, weak unconfoundedness and common support lead to the property of strong ignorability, which is necessary for the identification of average treatment effects \citep{rosenbaum1983,imai2004,rosenbaum2010}.

Assuming strong ignorability holds, a standard approach in automobile insurance is to estimate the price sensitivity for $C$ treatment categories in a portfolio of $N$ policies through the logistic regression
\begin{equation} \label{Equation_1}
    \mathrm{logit}\left(\mathbb{E}\left[Y_i(T_i)|\bm{X}_i \right] \right) = \bm{X}_i^{\prime} \bm{\beta} + \sum_{c = 1}^{C - 1} \gamma_{c} \mathbbm{1}\left[ T_i = t_{c} \right] \quad \textrm{for} \quad i = 1, \dots, N,
\end{equation}
where $Y_i(T_i) = 1$ when rate change $T_i$ yields a lapse on policy $i$ (see, e.g., \citet{smith2000}; \citet{yeo2001}). The response variable in this approach is assumed to be independently Bernoulli distributed conditional on the risk characteristics $\bm{X}_i$. The parameter $\gamma_{c}$ denotes the causal effect of categorical rate change $t_{c}$ that we are interested in and it can be estimated straightforwardly by Maximum Likelihood.

The risk characteristics are included in this approach to adjust for the confounding that is generally present in automobile insurance. However, this adjustment is often still insufficient due to balancedness and extrapolation problems, and additional steps need to be taken to control for the confounding \citep{rubin1979,rubin1997,morgan2007,guo2009}. We therefore consider two alternative approaches for inferring a customer's potential response to counterfactual rate changes, where one is based on discrete treatment effects and the other on continuous treatment effects.

\subsection{Discrete treatment categories} \label{Section2.2}

The discrete treatment approach relies on discretization of the exact treatment dose, or in our case rate change, into a finite number of categories. More specifically, it discretizes the observed treatment doses $T_1, \dots, T_N \in \mathbb{R}$ into the $C \in \mathbb{N}_{+}$ ordered categories $\{ t_{1}, \dots, t_{C} \}$. For each of these categories, customers that did not receive the treatment of interest are paired with similar customers that did receive the treatment based on their risk characteristics. A typical measure for this multivariate similarity is the (rank-based) Mahalanobis distance \citep{rubin1979,rosenbaum2010}. Ideally these pairs are matched perfectly when the risk characteristics are identical for two customers, but this can be challenging or even impossible in practice when we have many risk characteristics to match.

Another method to express the similarity between customers uses the propensity score. This score indicates how likely customers are to receive a certain treatment given their risk characteristics. Usually, the propensity score is unknown in observational studies and is therefore estimated by a logistic or multinomial regression, but Gradient Boosting Models (GBMs) and causal forests have been adopted for this as well (see, e.g., \citet{guelman2014}; \citet{wager2018}). Interestingly, this score has a balancing property that states that if customers have the same propensity score, then the difference in their risk characteristics is independent of the assigned treatment, or
\begin{equation*}
    T_i \perp \bm{X}_i | \pi(t_{c}, \bm{X}_i) \quad \forall c \in \{1, \dots, C\}
\end{equation*}
\citep{imbens2004,guo2009,rosenbaum2010}. This balancing property, in turn, implies that it is sufficient to match customers only on their propensity score, rather than on all their risk characteristics, in terms of, for instance, the absolute distance
\begin{equation} \label{Equation_2}
    \left| \hat{\pi}(t_{c}, \bm{X}_i) - \hat{\pi}(t_{c}, \bm{X}_j) \right| \quad \text{for} \quad i \neq j.
\end{equation}
This property can also be used to check for any misspecification of the propensity score by testing whether the covariates, on average, differ significantly between treatment categories, but this significance test can be imprecise in practice \citep{mccaffrey2013,guelman2014}. More importantly, this balancing property holds independently of the strong ignorability property, and if strong ignorability holds conditional on the risk characteristics, then it also holds conditional on the propensity score. Identification of the average treatment effects is thus still possible with the propensity score, while reducing the matching algorithm to a one-dimensional problem.

Given the customer matches, we can infer the counterfactual responses and develop a global response model to measure the price sensitivity \citep{guelman2014}. For the global response model, we can, for instance, apply the logistic regression in \Cref{Equation_1} or a technique from the field of machine learning (see, e.g., \citet{verbeke2011}; \citet{vafeiadis2015}; \citet{bolance2016}). However, this approach only imputes a counterfactual response once and does not incorporate any uncertainty of its true value into the price sensitivity estimates. This paper therefore adopts multiple imputation to at least partially account for this uncertainty and to provide more representative estimates. 

While multiple imputation has mainly been applied to impute missing covariates, it can also be used to infer counterfactual responses \citep{westreich2015}. In this multiple imputation extension, we randomly sample with replacement from the $I$ closest responses in terms of the propensity score distance in \Cref{Equation_2}. We perform this imputation $M$ times and estimate a global response model for each imputation. Using Rubin's rule \citep{rubin1987}, we can combine the $M$ estimators by
\begin{equation*}
    \bar{\bm{\delta}} = \frac{1}{M} \sum_{m = 1}^{M} \hat{\bm{\delta}}_m \quad \textrm{and} \quad \mathrm{Var}(\bar{\bm{\delta}}) = \bar{W} + \left(1 + \frac{1}{M} \right) B,
\end{equation*}
where the row vector $\hat{\bm{\delta}}_m = (\hat{\bm{\beta}}_m, \hat{\bm{\gamma}}_m)$ with corresponding variance estimates $\hat{W}_m$ for the $m$-th imputation and
\begin{equation*}
    \bar{W} = \frac{1}{M} \sum_{m = 1}^{M} \hat{W}_m \quad \textrm{and} \quad B = \frac{1}{M - 1} \sum_{m = 1}^{M} \left(\hat{\bm{\delta}}_m - \bar{\bm{\delta}} \right)^{\prime} \left(\hat{\bm{\delta}}_m - \bar{\bm{\delta}} \right)
\end{equation*}
denote the within- and between-imputation variance, respectively. While the within-imputation variance reflects the uncertainty of parameter estimation, the between-imputation variance accounts for the imputation uncertainty in the counterfactual responses. The discrete treatment approach in this paper thus largely follows the causal inference framework in \citet{guelman2014}, but extends it by incorporating, at least partially, the imputation uncertainty in the counterfactual responses.

\subsection{Continuous treatment doses} \label{Section2.3}

While the literature traditionally focuses on discrete, categorical treatments, a continuous treatment dose can be considered as well. The advantage of this continuous approach is that it avoids any compression of information from discretization. More specifically, suppose the observed treatment doses $T_1, \dots, T_N$ are defined in the continuum of potential treatment doses $[\mspace{2mu} \underline{T}, \overline{T} \mspace{2mu}]$. Following the same reasoning as in the discrete approach, the continuous treatment assignment mechanism can be modeled by a propensity score function $\pi(t, \bm{X}_i)$ for every observed and unobserved treatment dose $t$. In addition, it has the balancing property, which we can use for misspecification tests between treatment dose intervals, and allows for strong ignorability if it already holds conditional on the risk characteristics \citep{hirano2004}. A variety of models can again be used to estimate this function, but typically a Gaussian linear model or a GLM is used (see, e.g., \citet{fryges2008}; \citet{guardabascio2014}; \citet{zhu2015}). The continuous counterpart of the propensity score, the GPS, follows from this function and the observed treatment dose as $\pi(T_i, \bm{X}_i)$.

\citet{hirano2004} argue that a global response model can be formed with this propensity score function, but now conditional on only the GPS instead of all the risk characteristics. More specifically, they suggest to approximate the conditional response quadratically by
\begin{equation*}
    \mathbb{E}\left[ Y_i(T_i) | \pi(T_i, \bm{X}_i) \right] = \beta_0 + \beta_1 \pi(T_i, \bm{X}_i) + \beta_2 \pi(T_i, \bm{X}_i)^2 + \beta_3 T_i + \beta_4 T_i^2 + \beta_5 \pi(T_i, \bm{X}_i) T_i
\end{equation*}
and to estimate it by Ordinary Least Squares (OLS) using the estimated GPS $\hat{\pi}(T_i, \bm{X}_i)$, but more advanced approaches have been applied as well (see, e.g., \citet{kreif2015}; \citet{zhao2018}). \citet{hirano2004} stress that this global response model has no causal interpretation due to the fixed GPS estimates $\hat{\pi}(T_i, \bm{X}_i)$, but that it is possible to estimate an average dose-response function by using the estimated propensity score function $\hat{\pi}(t, \bm{X}_i)$. This average dose-response function represents the average potential response to a certain treatment dose $t$ and is estimated as
\begin{equation*}
   \widehat{\mathbb{E}\left[Y(t)\right]} = \frac{1}{N} \sum_{i = 1}^{N} \left( \hat{\beta}_0 + \hat{\beta}_1 \hat{\pi}(t, \bm{X}_i) + \hat{\beta}_2 \hat{\pi}(t, \bm{X}_i)^2 + \hat{\beta}_3 t + \hat{\beta}_4 t^2 + \hat{\beta}_5 \hat{\pi}(t, \bm{X}_i) t \right),
\end{equation*}
where $\hat{\pi}(t, \bm{X}_i)$ is evaluated at the, potentially counterfactual, treatment dose of interest $t$. Its corresponding confidence bounds can additionally be determined through a bootstrap procedure.

Despite the remark made by \citet{hirano2004}, the global response model can still be used to predict potential changes in the propensity scores. These predicted changes, in turn, enable us to determine customers' expected potential responses. In other words, the continuous approach allows us to identify the rate change in the premium offer optimization, whereas the discrete approach can only indicate the best rate change category but not the exact rate change itself. We therefore adopt the continuous treatment framework next to the discrete treatment framework in this paper to benefit from this advantage for premium renewal offer optimization in practice. 
\section{Data and empirical considerations} \label{Section3}

\subsection{Automobile insurance} \label{Section3.1}

To compare the discrete treatment approach to its continuous analogue, we apply both frameworks to policy renewals in non-life insurance. More specifically, we analyze an automobile insurance portfolio from a large Dutch insurer with data on individual policy renewals. We consider the risk factors in \Cref{Table_1} since these are commonly used in the design of renewal premia in the Netherlands, which include, among other covariates, the type of policy and the policyholder's current level of risk. This level of risk represents the insurer's estimate of a customer's riskiness before his or her policy is renewed and it is related to the Bonus-Malus level currently occupied by the policyholder. 

Besides policyholders' level of risk, insurers also incorporate the premia that are offered by competitors in the market into their renewal pricing strategies. The simulation-based data mining approaches of \citet{owadally2019}, for instance, show that insurers adjust their premia based on the offers of competitors in the previous policy year and that, hence, a game ensues between the insurers on the market. Contrary to previous studies in the literature, this portfolio therefore contains the premia offered by six of the largest competitors in the market. Using these premia, we deduce the competitiveness of each renewal offer before any rate changes $(A)$ from its relative difference with the current cheapest competing offer $(B)$, or $(B - A) / A$. We consider the renewal offers before any competitiveness adjustments as well as any autonomous corrections, such as those for the age and Bonus-Malus level of the insured driver, to avoid confounding with the total rate changes. More specifically, the total rate changes are defined as the sum of the autonomous corrections and the competitiveness adjustments, and it may therefore lead to simultaneity issues if the renewal offers in this measure contain these alterations, in particular when there are no competitiveness adjustments. It is more intuitive to consider the renewal offers before any autonomous and competitiveness alterations as well since policyholders also only observe the premium before and after renewal of their policy. In addition to this measure, we adopt the underpricedness of the renewal offers before competitiveness adjustments but after autonomous corrections $(C)$ compared to the cheapest and second-cheapest competitor in the market $(D)$, or $(D - C)^{+}$. We now do include the autonomous corrections in the renewal offers since these are used in practice in the design of the renewal premia and are thus needed to explain the treatment assignment mechanism.

\begin{table}[t!]
    \caption{Description of the key variables (top) and risk factors (bottom) used for automobile insurance.}
    \label{Table_1}\vspace{-6pt}
	\centerline{\scalebox{0.80}{\begin{tabularx}{1.25\textwidth}{l l l }
		\toprule \addlinespace[1ex] \vspace{1pt}
		\textbf{Variable}\hspace{46pt}\hspace{10pt} & \textbf{Values}\hspace{25pt}\hspace{10pt} & \textbf{Description} \\ \hline \addlinespace[0.4ex]
		\texttt{Churn} & $2$ categories & Whether the policyholder churned or not. \\
		\texttt{Rate\_Change} & Continuous & Percentage rate change offered at renewal of the policy. \\
		\texttt{Expenses} & Continuous & Expected policy expenses to the insurer in case of renewal in euros. \\
		\texttt{Competitiveness} & Continuous & Competitiveness of the renewal offer before any rate changes ($A$) relative to the \\
		& & cheapest competitor ($B$), i.e., $(B - A) / A$. \\
		\texttt{Premium\_Old} & Continuous & Premium offered at previous renewal of the policy in euros. \\
		\hline \addlinespace[0.4ex]
		\texttt{Premium\_New\_Base} & Continuous & Renewal offer for the policy after risk factor corrections, but before any \\
		& & competitiveness adjustments in euros. \\
		\texttt{Undershooting\_1} & Continuous & Degree of underpricedness of renewal offer \texttt{Premium\_New\_Base} ($C$) compared to the \\
		& & cheapest competitor ($D_{1}$) in euros, i.e., $(D_{1} - C)^{+}$. \\
		\texttt{Undershooting\_2} & Continuous & Degree of underpricedness of renewal offer \texttt{Premium\_New\_Base} ($C$) compared to the \\
		& & second-cheapest competitor ($D_{2}$) in euros, i.e., $(D_{2} - C)^{+}$. \\
		\texttt{Risk\_Level} & $4$ categories & Current level of risk of the policyholder, i.e., very low, low, medium, or high. \\
		\texttt{Policy\_Type} & $3$ categories & Type of the policy, i.e., a policy for employees or a second car, or a regular policy. \\
		\bottomrule
	\end{tabularx}}}\vspace{-3pt}
\end{table}

\begin{figure}[t!]
    \centering
    \begin{subfigure}{\textwidth}
        \centering
        \begin{tabular}{c c}
            \centering
            \includegraphics[width=0.475\textwidth]{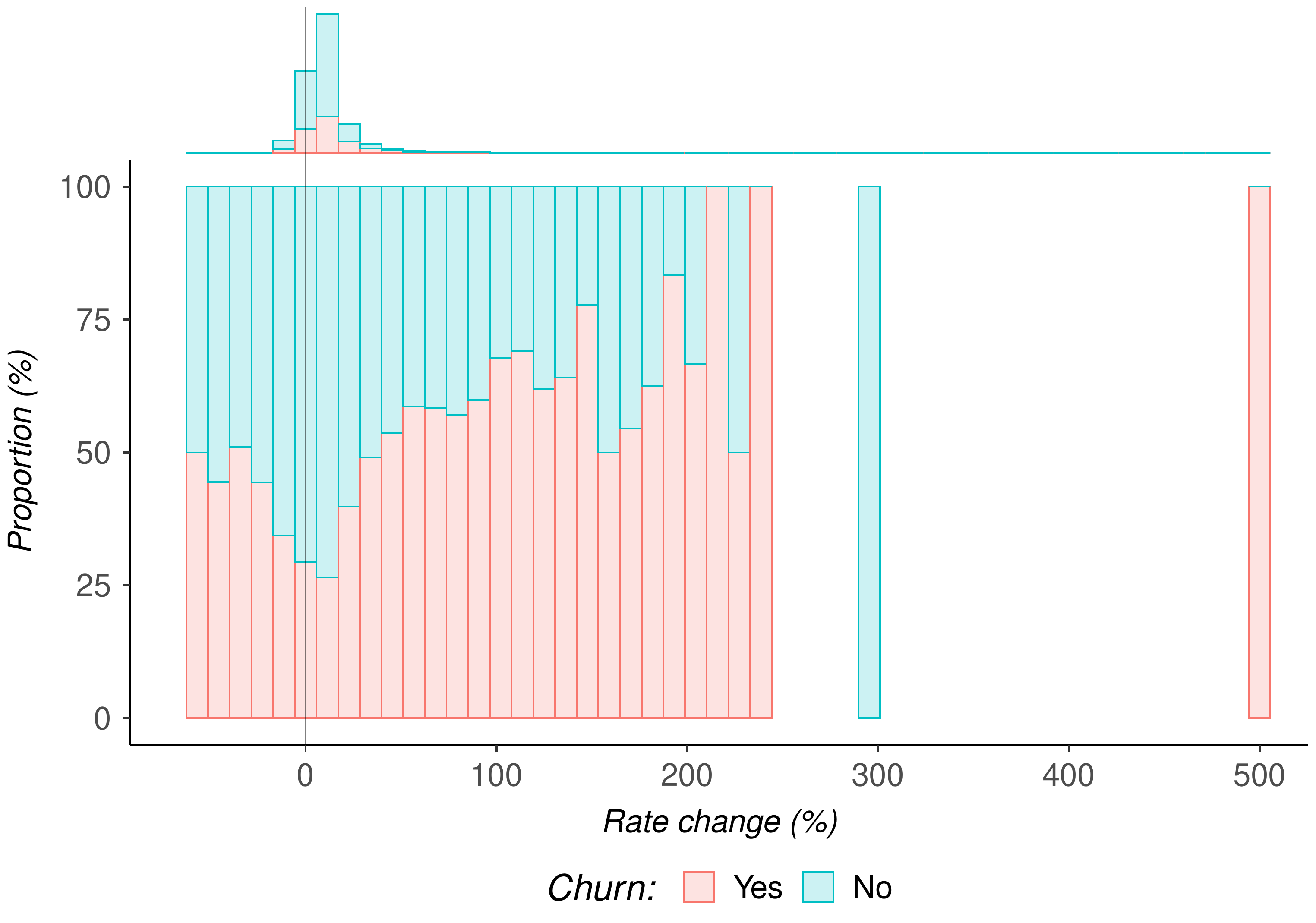}&
            \includegraphics[width=0.475\textwidth]{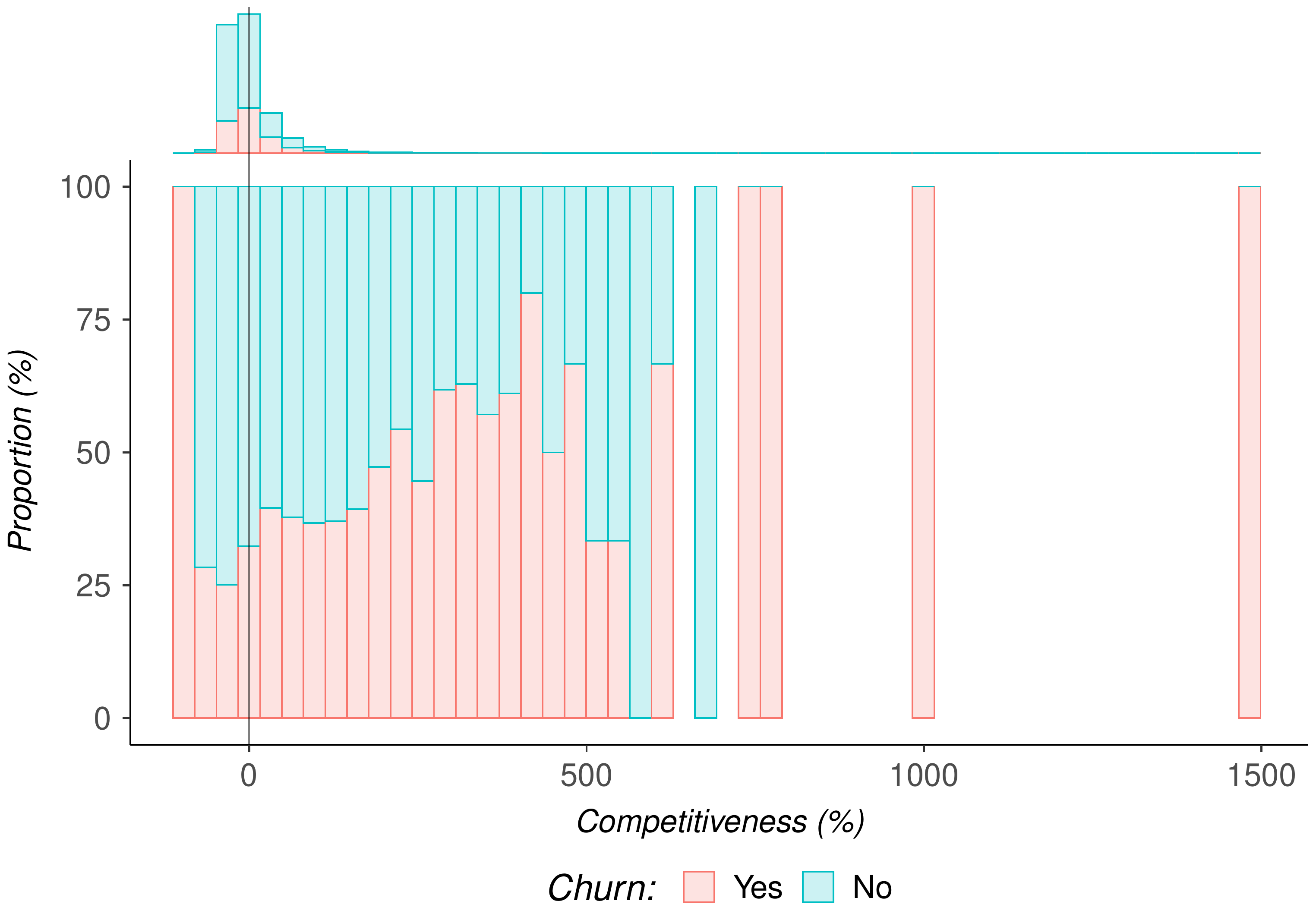}
        \end{tabular}\vspace{-8pt}
        \caption{Churn proportions and marginal distributions for rate changes (left) and competitiveness (right)}
        \label{Figure_1a}
    \end{subfigure}
    
    \vspace{-9pt}
    
    \begin{subfigure}{\textwidth}
        \centering
        \begin{tabular}{c c}
            \centering\hspace{-45pt}
            \includegraphics[width=0.675\textwidth]{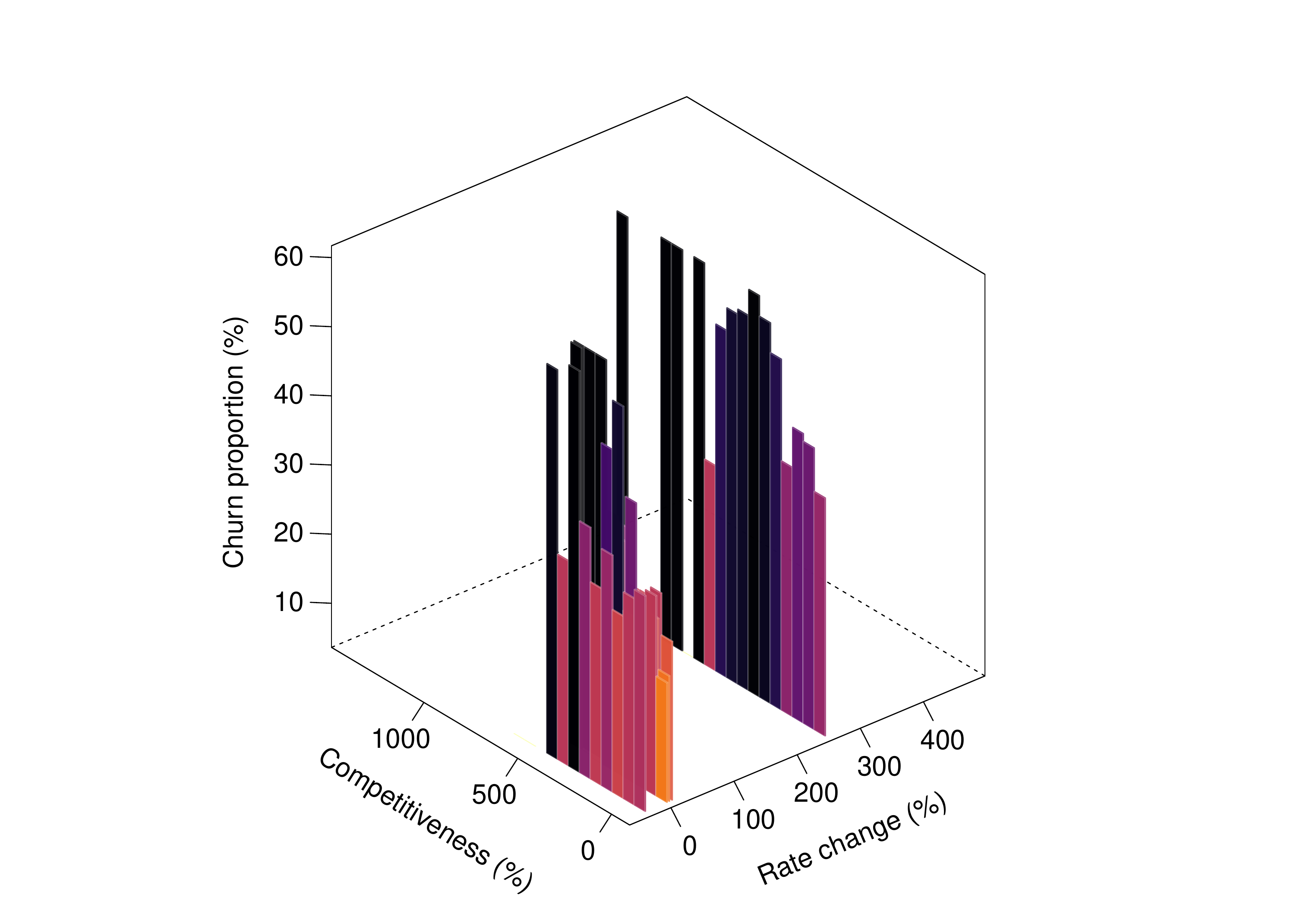}&\hspace{-85pt}
            \includegraphics[width=0.675\textwidth]{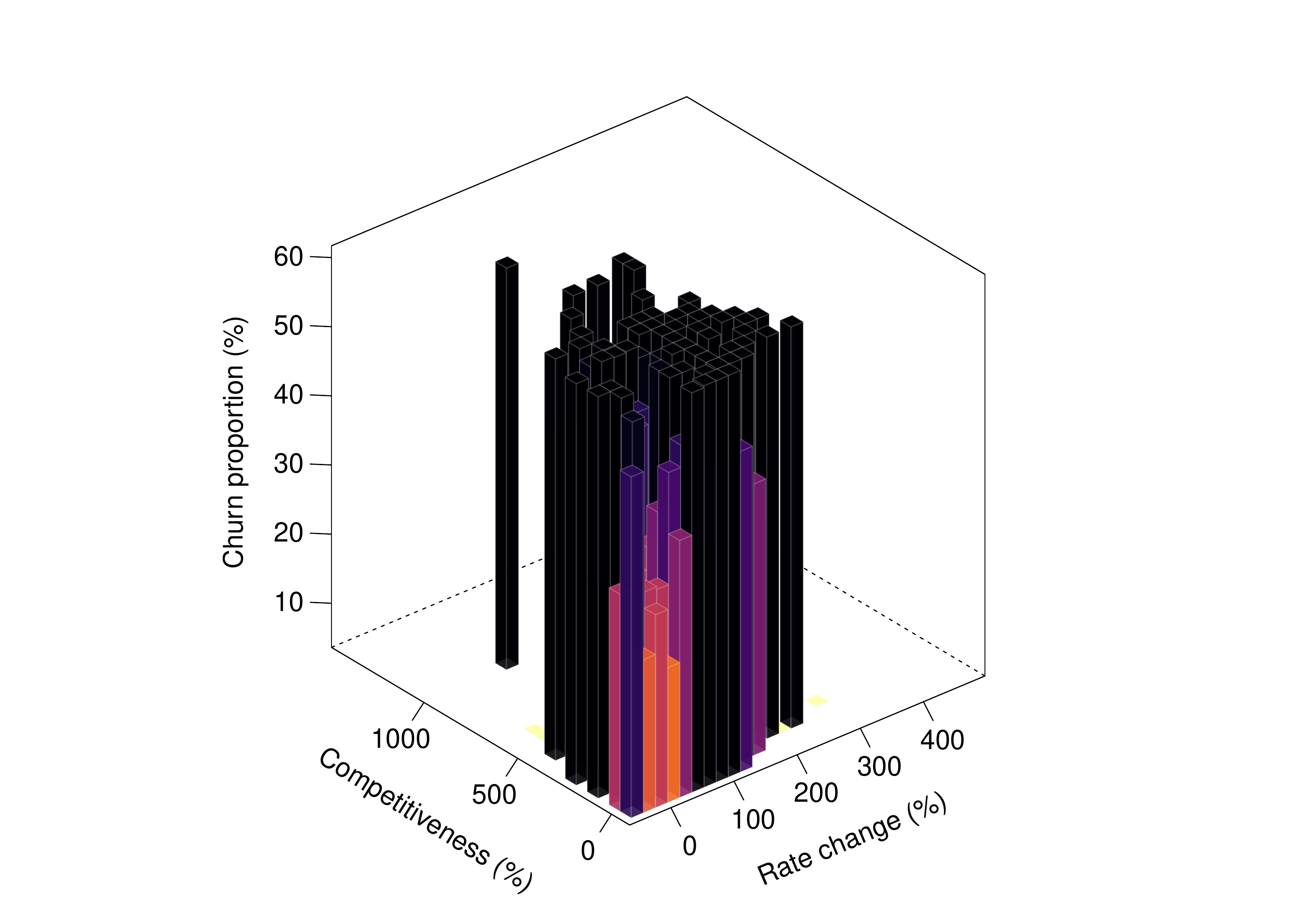}
        \end{tabular}\vspace{-8pt}
        \caption{Joint (truncated) churn proportions for quintile interval (left) and continuous (right) rate changes and competitiveness}
	    \label{Figure_1b}
    \end{subfigure}\vspace{-3pt}
    \caption{Churn proportions and marginal distributions for rate changes (left) and competitiveness (right) (panel (a)) with their joint (truncated) proportions for quintile interval (left) and continuous (right) rate changes (panel (b)) before outlier removal in automobile insurance.}
	\label{Figure_1}
\end{figure}

\begin{figure}[t!]
    \centering
    \begin{subfigure}{\textwidth}
        \centering
        \begin{tabular}{c c}
            \centering
            \includegraphics[width=0.475\textwidth]{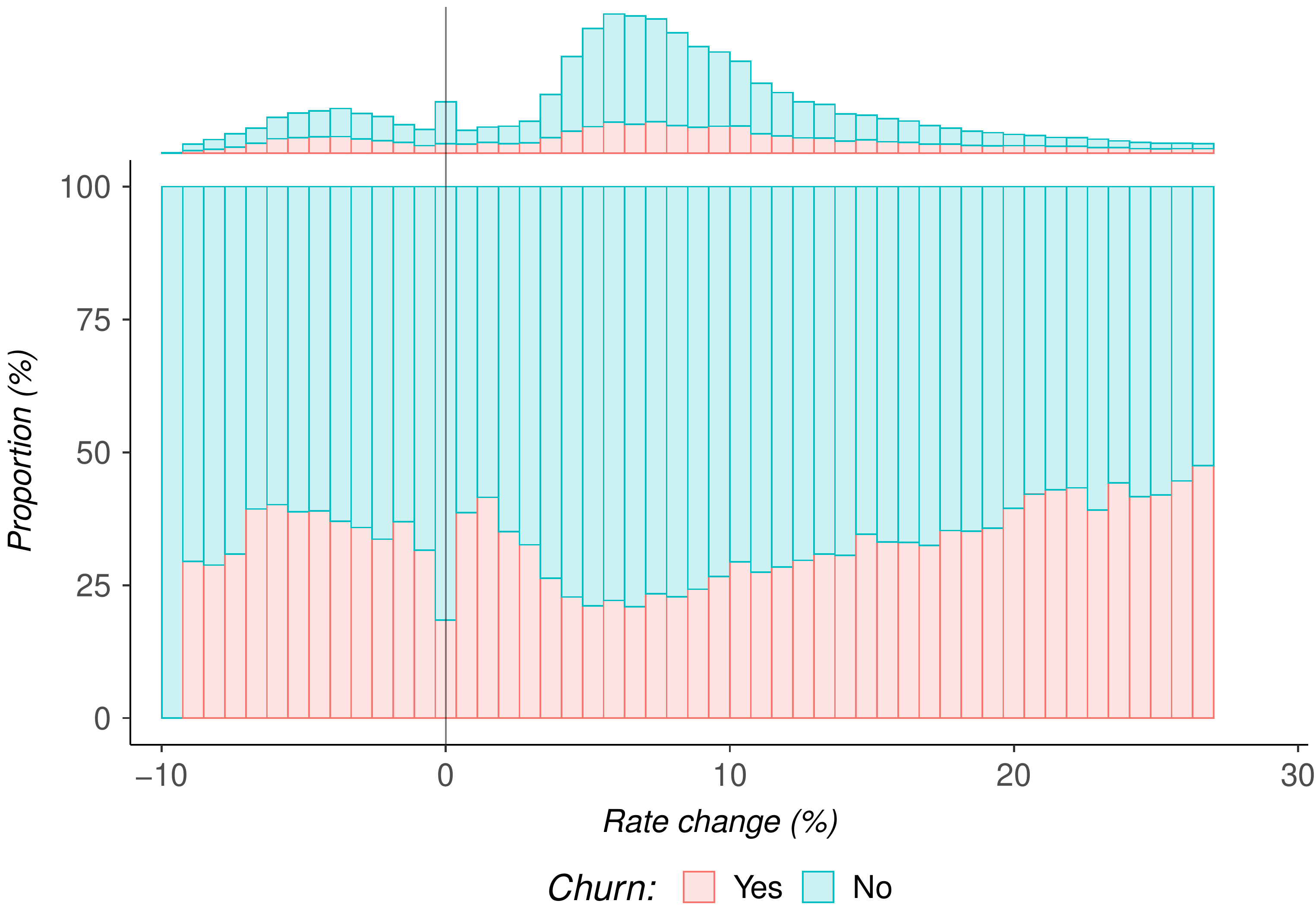}&
            \includegraphics[width=0.475\textwidth]{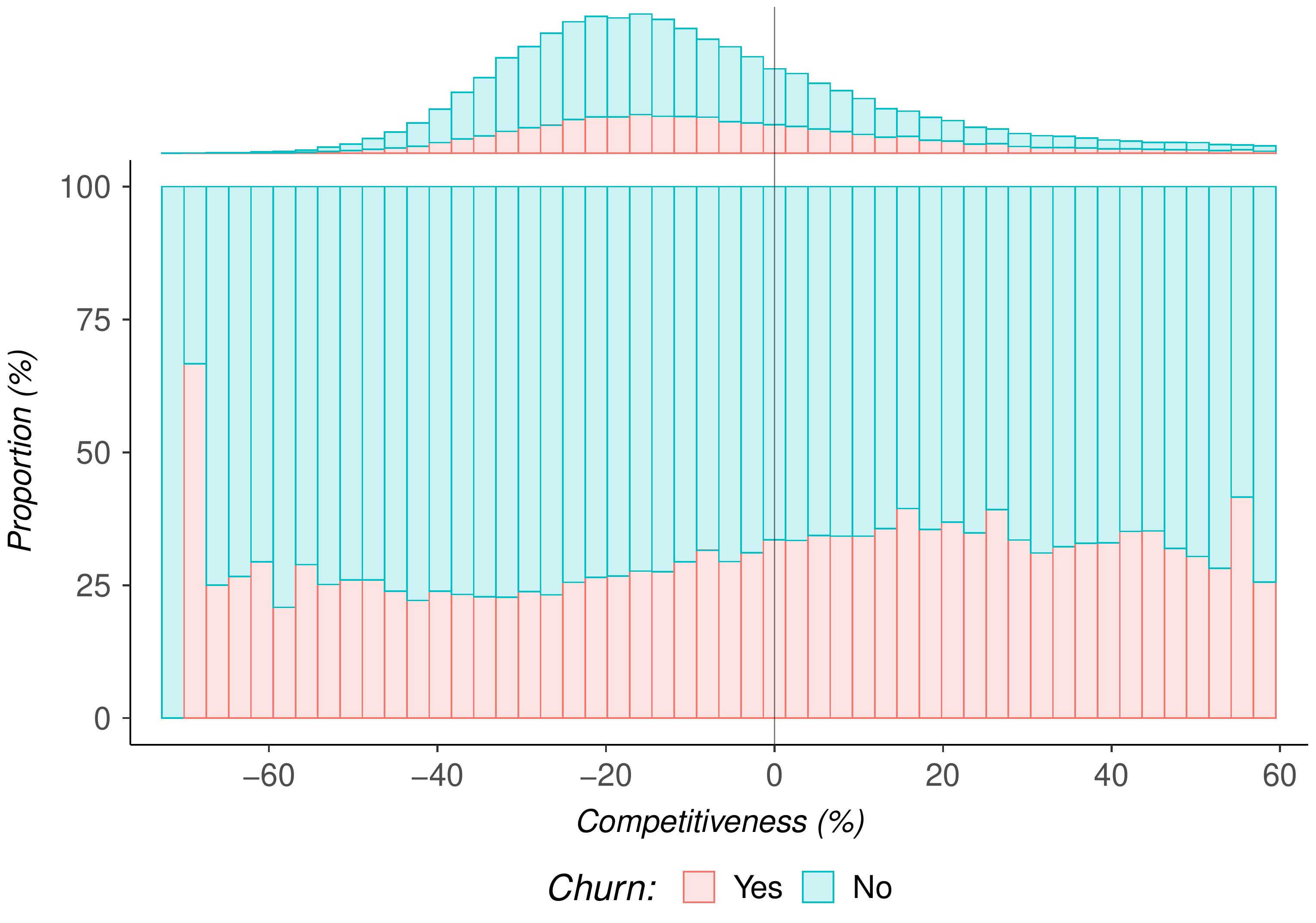}
        \end{tabular}\vspace{-8pt}
        \caption{Churn proportions and marginal distributions for rate changes (left) and competitiveness (right)}
        \label{Figure_2a}
    \end{subfigure}
    
    \vspace{-9pt}
    
    \begin{subfigure}{\textwidth}
        \centering
        \begin{tabular}{c c}
            \centering\hspace{-45pt}
            \includegraphics[width=0.675\textwidth]{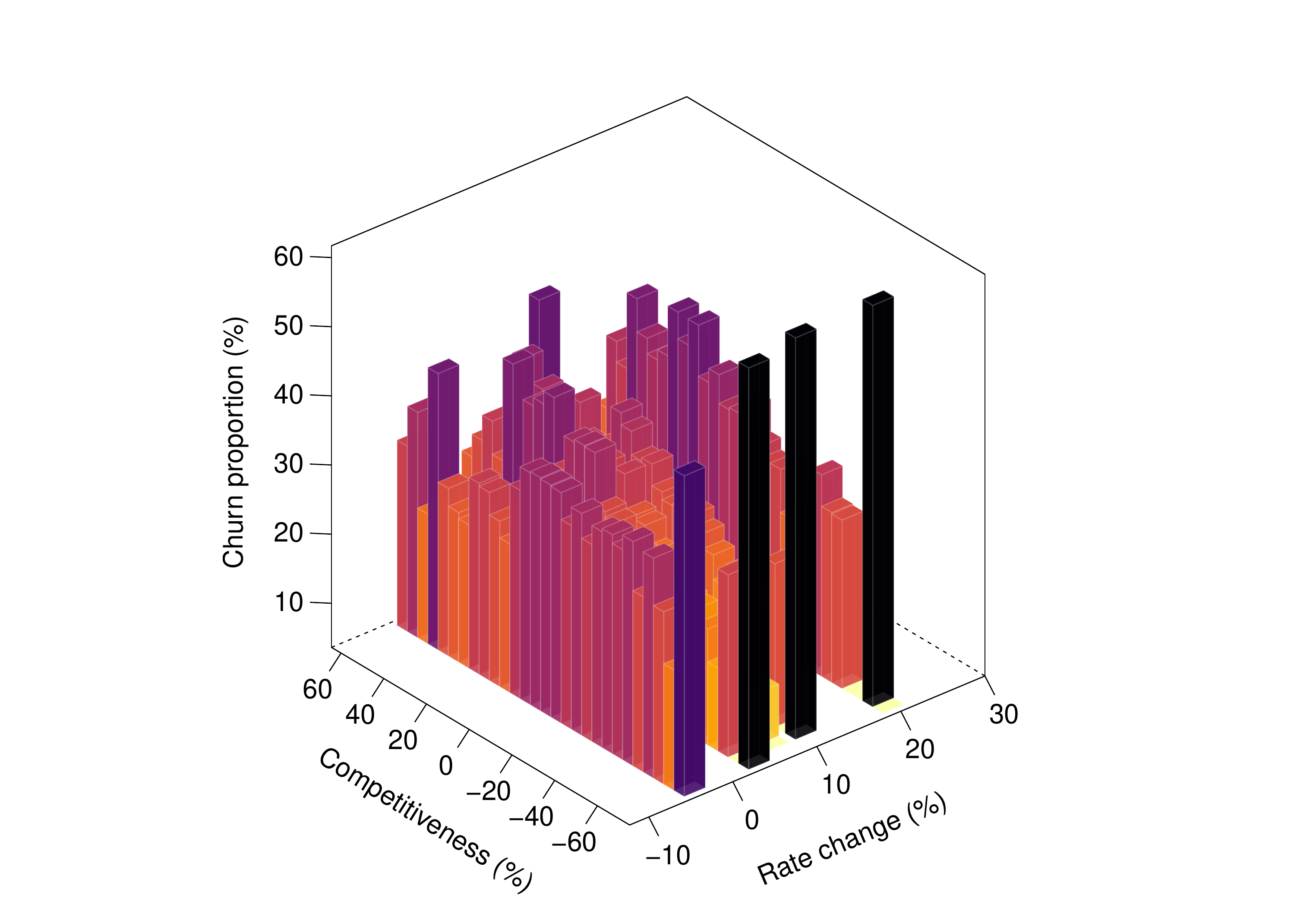}&\hspace{-85pt}
            \includegraphics[width=0.675\textwidth]{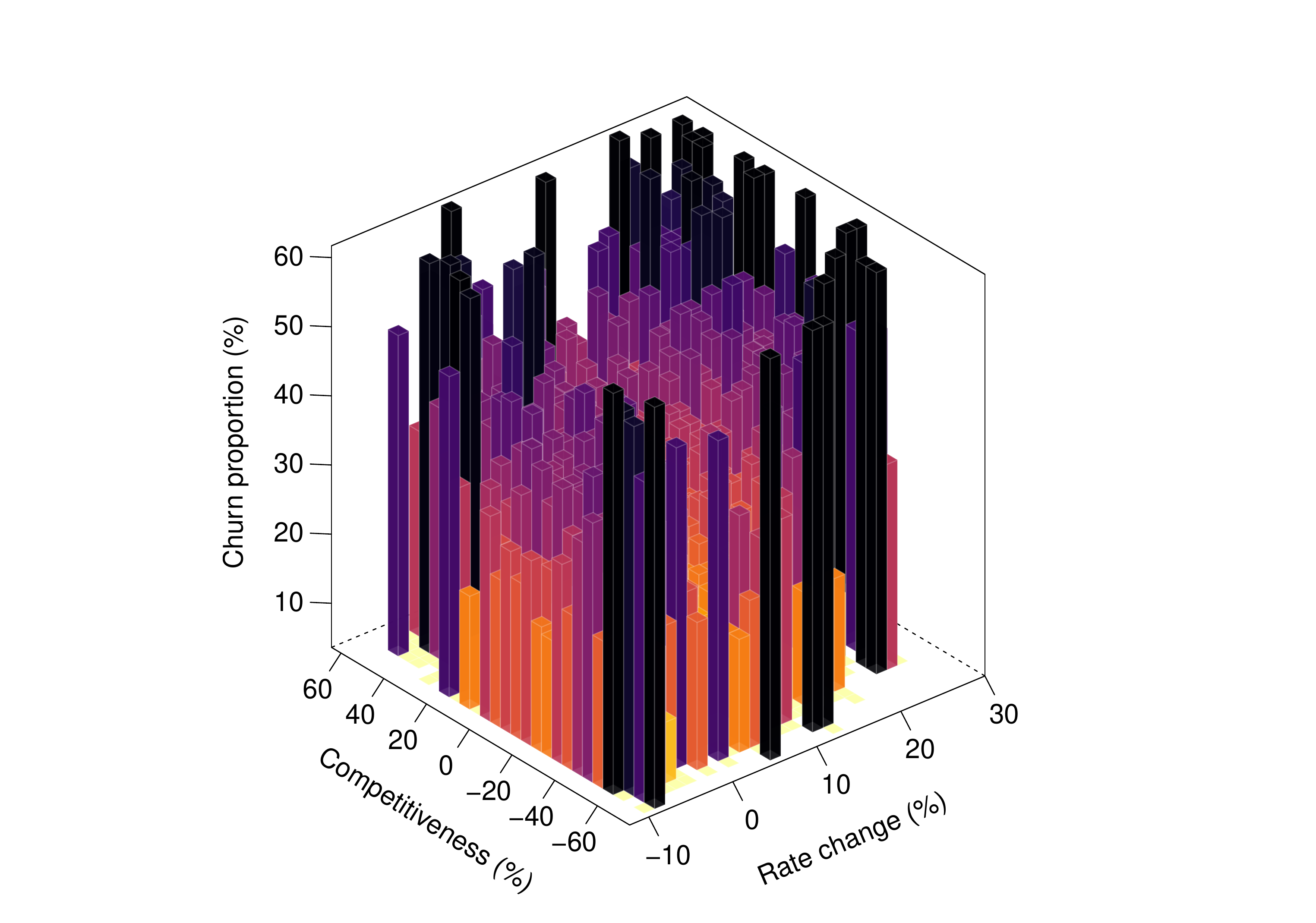}
        \end{tabular}\vspace{-8pt}
        \caption{Joint (truncated) churn proportions for quintile interval (left) and continuous (right) rate changes and competitiveness}
	    \label{Figure_2b}
    \end{subfigure}\vspace{-3pt}
    \caption{Churn proportions and marginal distributions for rate changes (left) and competitiveness (right) (panel (a)) with their joint (truncated) proportions for quintile interval (left) and continuous (right) rate changes (panel (b)) after outlier removal in automobile insurance.}
	\label{Figure_2}
\end{figure}

The Dutch automobile insurance portfolio used in this paper spans the period of 2017 up to and including 2019. Its policies generally have a duration of one year and need to be renewed annually. Customer churn is defined here as lapsing the policy before renewal but after receipt of the rate change. The rate changes and competitiveness of these policies as well as the ratio between policy lapses and renewals are given in \Cref{Figure_1}. From \Cref{Figure_1} we can clearly see that some extremely high or low rate changes and competitiveness levels are rarely observed in the portfolio. Since this lack of exposure may cause issues for the propensity score framework described earlier, this paper focuses on all rate changes and competitiveness levels within $1.5$ times the interquartile range, or $[-9.28\%, 27.01\%]$ and $[-73.08\%, 59.48\%]$, respectively, equivalent to removing all outliers in a standard boxplot procedure (see, e.g., \citet{wienand2014}). As such, our portfolio effectively retains a total of $71{,}522$ ($86.60\%$) policy renewals by $30{,}738$ ($94.62\%$) customers with $20{,}649$ ($28.87\%$) policy lapses, with rate changes and competitiveness as shown in \Cref{Figure_2}. This, in turn, yields a much more balanced and less skewed distribution of rate changes and competitiveness levels, with a more stable, slower increasing churn ratio and a large inflection point at small rate changes in \Cref{Figure_2a}. This is in line with the portfolio observed in \citet{guelman2014}, although they do not observe this inflection point and their churn ratio increases substantially slower due to an overall churn rate of only $5.92\%$. However, we observe the same pattern, including this inflection point, in \Cref{Figure_2b} when we consider the rate changes in quintile intervals (left), similar to \citet{guelman2014}, and continuously (right) jointly with the competitiveness levels. The inflection point at small rate changes thus seems to be a persistent novelty in this automobile insurance portfolio.

An explanation for this inflection point may be that insurance customers respond boundedly rational to rate changes, i.e., they react more to the fact that their premium changes than to the exact size of the rate change. It may additionally be a sign of information asymmetry since an insurer's rate change causes customers to focus their attention on the price, and hence competitiveness, of their policy and to become more aware of competing offers in the market \citep{brockett2008,robson2015}. This, in turn, may explain the irregular, non-monotonic behavior of insurance customers at small rate changes. We expect the continuous framework to be better able to describe this non-monotonicity since it does not require us to aggregate responses over an entire interval, like in the discrete framework. Instead, the continuous framework allows for a continuum of rate changes and it may therefore have more flexibility to capture non-monotonicities within rate change intervals.

\subsection{Optimization methodology} \label{Section3.2}

Using the causal inference framework, we estimate the customer price sensitivities in the Dutch automobile insurance portfolio. We consider both a discrete and continuous treatment (dose) approach to determine the optimal renewal offers, where XGBoost is adopted to model the (generalized) propensity score. Moreover, after a variable selection procedure we estimate the global response model in the discrete framework by a multiply imputed GLM, whereas XGBoost is used to model the conditional dose-response function in the continuous framework as well. This, in turn, allows us to flexibly model the customer price sensitivities and to estimate the causal effect as accurately as possible.

Both treatment approaches thus rely on two components, namely a (generalized) propensity score model and a global response model or conditional dose-response function. Traditionally, a logistic (Gaussian) regression is used to describe the (generalized) propensity score, but \citet{mccaffrey2004} and \citet{guelman2014} propose using Gradient Boosting Models (GBMs). They argue that GBMs are more suited for this purpose since they allow for flexible non-linear effects of the risk factors, identify complex interactions during the sequential tree-learning algorithm, and have a built-in variable selection procedure. A more flexible and much faster extension of GBMs, XGBoost, is therefore adopted in this paper to model the (generalized) propensity score, which can be performed efficiently in \texttt{R} with the package \texttt{xgboost} and it is described in detail in \ref{Appendix_A}.

While a (generalized) propensity score model aims to predict the true propensity score as accurately as possible, its primary goal is to maximize the balance among the risk factors. We therefore adopt early stopping once the Average Standardized Absolute Mean (ASAM) difference in the risk factors in \Cref{Equation_A.4} has not improved for $250$ consecutive decision trees, analogous to \citet{mccaffrey2004,mccaffrey2013} and \citet{guelman2014} and with at most $10{,}000$ decision trees in total. Moreover, we follow the suggestion of \citet{boehmke2019} by first tuning the learning rate and the tree-specific parameters, secondly the stochastic features, and lastly the regularization penalties while considering the parameter grids and default values in \Cref{Table_2} sequentially with $10$-fold cross-validation. Finally, we minimize the negative generalized Bernoulli, or multinoulli, and truncated Gaussian log-likelihood in \Cref{Equation_A.2,Equation_A.3} in the discrete and continuous framework, respectively. The continuous framework thus follows the principle behind the homoskedastic approach of \citet{hirano2004}, but we additionally allow for heteroskedasticity between the treatment intervals after truncation and renormalization of the Gaussian distribution.

\begin{table}[t!]
    \caption{Sequential parameter grids and default values for XGBoost in discrete (continuous) treatment optimization with $10$-fold cross-validation, using the \texttt{xgboost} package in \texttt{R} with $\texttt{early\_stopping\_rounds} = 250$.}
    \label{Table_2}\vspace{-6pt}
	\centerline{\scalebox{0.80}{\begin{tabularx}{1.25\textwidth}{l X r X r X r X r}
	    \toprule \addlinespace[1ex] \vspace{1pt}
	    && && && \multicolumn{3}{c}{\textbf{Optimal}} \\ \cline{7-9} \addlinespace[1ex]
	    \textbf{Parameter} && \multicolumn{1}{c}{\textbf{Search grid}} && \multicolumn{1}{c}{\textbf{Default}} && \multicolumn{1}{c}{\textbf{(G)PS}} && \multicolumn{1}{c}{\textbf{Churn}} \\ \hline \addlinespace[0.4ex]
		\textit{Learning rate} && && && & \\ 
		\hspace{10pt}\textit{-\texttt{eta}} ($\mathit{\eta}$) && $\{0.01, 0.02, \dots,  0.05, 0.1, \dots, 0.25, 0.5\}$ && && $0.01$ ($0.5$) && ($0.03$) \\ \hline \addlinespace[0.4ex]
		\textit{Tree-specific parameters} && && && & \\ 
		\hspace{10pt}\textit{-\texttt{max\_depth}} ($\mathit{d}$) && $\{0, 1, 2, 4, \dots, 10, 25, 50\}$ && $6$ && $4$ ($50$) && ($10$) \\
		\hspace{10pt}\textit{-\texttt{min\_child\_weight}} ($\mathit{\nu}$) && $\{0, 1, \dots, 5, 10, 25, 50\}$ && $1$ && $1$ ($4$) && ($0$) \\ \hline \addlinespace[0.4ex]
		\textit{Stochastic features} && && && & \\ 
		\hspace{10pt}\textit{-\texttt{subsample}} ($\mathit{\delta}$) && $\{0.1, 0.2, \dots, 1\}$ && $1$ && $0.1$ ($0.9$) && ($0.9$) \\
		\hspace{10pt}\textit{-\texttt{colsample\_bytree}} ($\mathit{\kappa}$) && $\{0.1, 0.2, \dots, 1\}$ && $1$ && $1$ ($0.6$) && ($0.6$) \\ \hline \addlinespace[0.4ex]
		\textit{Regularization penalties} && && && & \\ 
		\hspace{10pt}\textit{-\texttt{gamma}} ($\mathit{\gamma}$) && $\{0, 0.1, 1, 10, 100\}$ && $0$ && $0.1$ ($0$) && ($0$) \\
		\hspace{10pt}\textit{-\texttt{lambda}} ($\mathit{\lambda}$) && $\{0, 0.1, 1, 10, 100\}$ && $1$ && $100$ ($1$) && ($0$) \\
		\hspace{10pt}\textit{-\texttt{alpha}} ($\mathit{a}$) && $\{0, 0.1, 1, 10, 100\}$ && $0$ && $10$ ($0$) && ($0$) \\ \hline \addlinespace[0.4ex]
		\textit{Number of trees} && && && & \\
		\hspace{10pt}\textit{-\texttt{nrounds}} ($\mathit{L}$) && $\{1, 2, \dots, 10{,}000\}$ && $10{,}000$ && $2{,}118$ ($38$) && ($959$) \\
		\bottomrule
	\end{tabularx}}}
\end{table}

Based on this model, we employ propensity score matching to infer the potential responses to the counterfactual treatment categories. More specifically, we find the $10$ closest matches based on the absolute distance in \Cref{Equation_2} and randomly draw $M = 10$ potential responses with replacement from these matches. Next, we estimate a logistic GLM for the average potential responses with Least Absolute Shrinkage and Selection Operator (LASSO) developed by \citet{tibshirani1996}, using, for instance, the \texttt{glmnet} package in \texttt{R}. We include up to second order and level interactions of the risk factors in \Cref{Table_1} with the treatment categories and competitiveness to identify any complex, non-linear relations and determine the optimal penalty from the grid $\{e^{x}: x \in \{-20, -19.99, \dots, 5\}\}$ by the one-standard-error rule in \citet{hastie2009}. In other words, we select the largest penalty within one standard error of the negative Binomial log-likelihood minimizer in $10$-fold cross-validation to obtain the most parsimonious model closest to the optimal model. Given the selected hyperparameters, we estimate a logistic GLM for each imputation and combine all estimators using Rubin's rule. From this, we can find the, potentially non-linear, average treatment effects as a function of the competitiveness and determine the optimal renewal offer from the median rate change in each treatment category.

Contrary to the global response model, the conditional dose-response function has no causal interpretation, making XGBoost actually highly suited to model this as well. In other words, our aim is again to merely predict the potential response as accurately as possible and XGBoost allows us to pursue this. We therefore adopt XGBoost for the conditional dose-response function in the continuous framework as well, using the setup described earlier but with the negative Bernoulli log-likelihood in \Cref{Equation_A.1} as loss function and without the ASAM stopping rule. By averaging over all the, potentially counterfactual, dose-responses, we are able to estimate the customer price sensitivity for every potential treatment dose in the continuous framework and to determine the optimal renewal offer as a function of the competitiveness more accurately. 
\section{Applications in automobile insurance} \label{Section4}

\subsection{Propensity score matching} \label{Section4.1}

\begin{figure}[t!]
    \centering
    \begin{subfigure}{\textwidth}
        \centering
	    \includegraphics[width=0.96\textwidth]{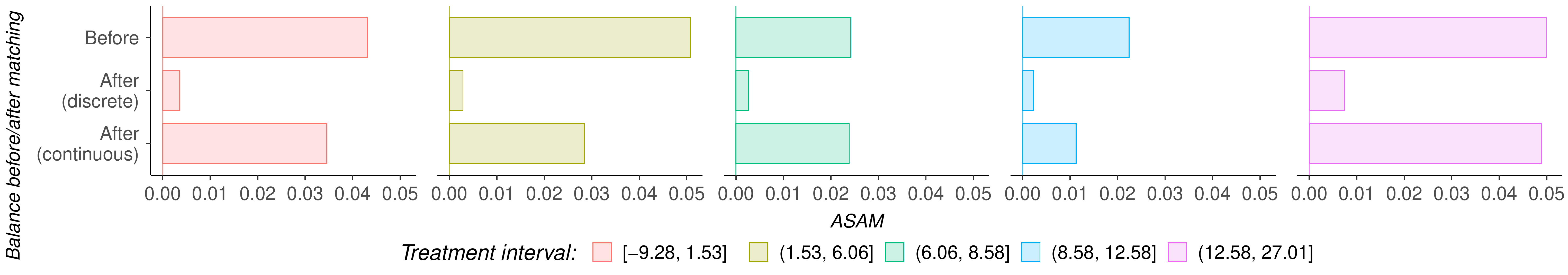}\vspace{-3pt}
        \caption{Mean balance before and after matching}
        \label{Figure_3a}
    \end{subfigure}
    
    \vspace{3pt}
    
    \begin{subfigure}{\textwidth}
        \centering\hspace{24pt}
        \includegraphics[width=0.905\textwidth]{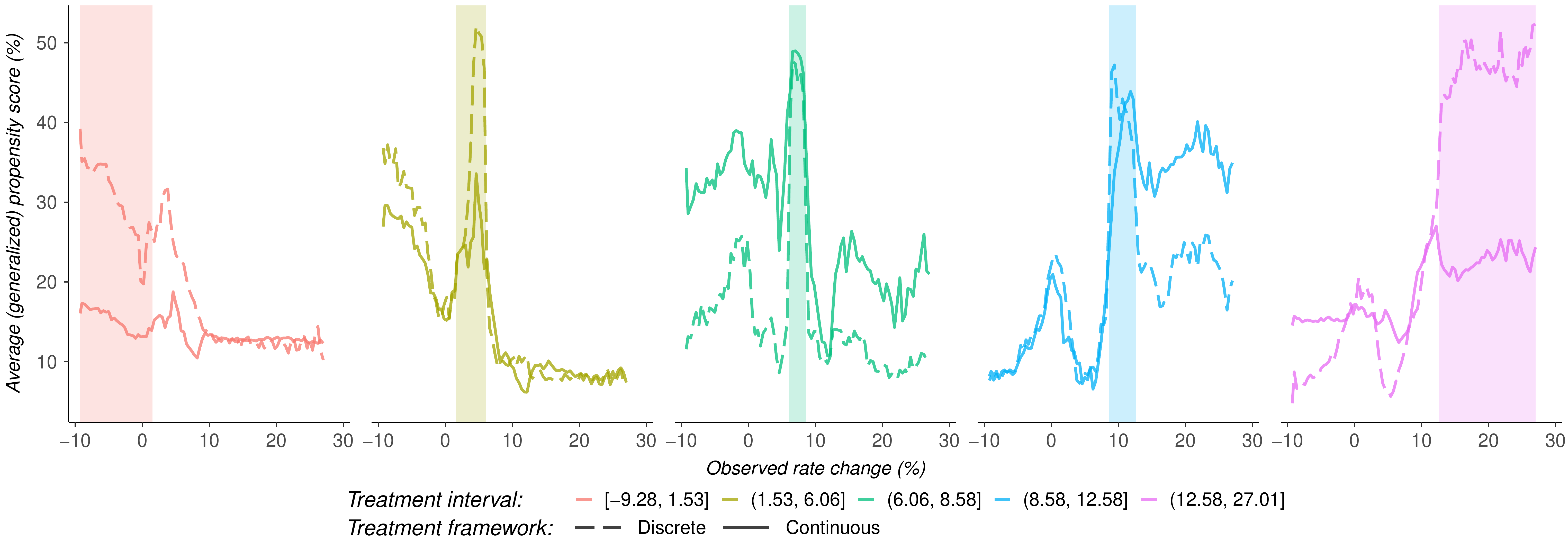}\vspace{-3pt}
        \caption{Average (generalized) propensity scores}
        \label{Figure_3b}
    \end{subfigure}\vspace{-3pt}
    \caption{Mean balance before and after matching (panel (a)) with average (generalized) propensity scores for observed treatments and the true intervals marked in color (panel (b)) for each rate change interval with discrete and continuous rate changes in automobile insurance.}
	\label{Figure_3}
\end{figure}

Following the methodology proposed earlier, we explore the properties of the (generalized) propensity score models for the Dutch automobile insurance portfolio. While we adopt the quintiles of the selected rate changes to construct five discrete treatment categories, namely $[-9.28\%, 1.53\%]$, $(1.53\%, 6.06\%]$, $(6.06\%, 8.58\%]$, $(8.58\%, 12.58\%]$, and $(12.58\%, 27.01\%]$, we consider the corresponding treatment intervals for the continuous approach. Using these intervals, we compare the balance of the risk factors in \Cref{Table_1} used in the Dutch automobile insurance market before and after (generalized) propensity score matching. As such, we report both the marginal ASAM for each risk factor and the overall ASAM for every rate change interval in \Cref{Table_3} and \Cref{Figure_3a}, respectively, for both the discrete and continuous approach. We additionally display the average (generalized) propensity score distributions in \Cref{Figure_3b} to explore their ability to distinguish between the five treatment intervals.

\begin{table}[t!]
    \caption{Balance of the risk factors before and after (generalized) propensity score matching, with means shown for discrete treatment categories (continuous treatment dose intervals).}
    \label{Table_3}\vspace{-6pt}
	\centerline{\scalebox{0.628}{\begin{tabularx}{1.5973\textwidth}{l r@{}l r@{}l c r@{}l r@{}l c r@{}l r@{}l c r@{}l r@{}l c r@{}l r@{}l }
	    \toprule \addlinespace[1ex] \vspace{1pt}
	    & \multicolumn{4}{c}{$\mathbf{t \in [-9.28\%, 1.53\%]}$} &  & \multicolumn{4}{c}{$\mathbf{t \in (1.53\%, 6.06\%]}$}  &  & \multicolumn{4}{c}{$\mathbf{t \in (6.06\%, 8.58\%]}$}  &  & \multicolumn{4}{c}{$\mathbf{t \in (8.58\%, 12.58\%]}$}  &  & \multicolumn{4}{c}{$\mathbf{t \in (12.58\%, 27.01\%]}$} \\
	    \cline{2-5} \cline{7-10} \cline{12-15} \cline{17-20} \cline{22-25} \addlinespace[0.4ex]
		\textbf{Risk factor} & \multicolumn{2}{c}{\textbf{Before}} & \multicolumn{2}{c}{\textbf{After}} &  & \multicolumn{2}{c}{\textbf{Before}} & \multicolumn{2}{c}{\textbf{After}}  &  & \multicolumn{2}{c}{\textbf{Before}} & \multicolumn{2}{c}{\textbf{After}}  &  & \multicolumn{2}{c}{\textbf{Before}} & \multicolumn{2}{c}{\textbf{After}}  &  & \multicolumn{2}{c}{\textbf{Before}} & \multicolumn{2}{c}{\textbf{After}} \\ \hline \addlinespace[0.4ex]
		\textit{\texttt{Premium\_New\_Base}} & 606.7311& & 490.4429& &  & 636.9227& & 481.1366& &  & 447.9680& & 470.5754& &  & 363.9404& & 467.7113& &  & 327.7379& & 457.4721& \\
		& (606.7311&) & (567.4517&) &  & (636.9227&) & (557.6668&) &  & (447.9680&) & (449.2777&) &  & (363.9404&) & (457.5248&) &  & (327.7379&) & (359.2658&) \\
		\textit{\texttt{Undershooting\_1}} & 47.3295& & 36.4911& &  & 21.9470& & 37.9439& &  & 24.6958& & 35.3999& &  & 36.7176& & 37.2466& &  & 54.5401& & 34.7362& \\
		& (47.3295&) & (46.3789&) &  & (21.9470&) & (27.2446&) &  & (24.6958&) & (24.7761&) &  & (36.7176&) & (32.0867&) &  & (54.5401&) & (57.3157&) \\
		\textit{\texttt{Undershooting\_2}} & 85.8819& & 64.8561& &  & 43.3591& & 66.2022& &  & 44.0948& & 64.4642& &  & 61.4869& & 64.6085& &  & 83.5309& & 59.4180& \\
		& (85.8819&) & (84.5597&) &  & (43.3591&) & (50.9704&) &  & (44.0948&) & (44.2415&) &  & (61.4869&) & (55.6155&) &  & (83.5309&) & (89.0896&) \\
		\textit{\texttt{Risk\_Level}} & && & && &  & && & && \\
		\hspace{10pt}\textit{- Very low} & 0.7279& & 0.7333& &  & 0.6863& & 0.6995& &  & 0.7410& & 0.7184& &  & 0.7468& & 0.7242& &  & 0.6978& & 0.7198& \\
		& (0.7279&) & (0.7338&) &  & (0.6863&) & (0.7126&) &  & (0.7410&) & (0.7393&) &  & (0.7468&) & (0.7365&) &  & (0.6978&) & (0.6940&) \\
		\hspace{10pt}\textit{- Low} & 0.2066& & 0.1881& &  & 0.2738& & 0.1870& &  & 0.2059& & 0.1716& &  & 0.1435& & 0.1719& &  & 0.1204& & 0.1681& \\
		& (0.2066&) & (0.1918&) &  & (0.2738&) & (0.2221&) &  & (0.2059&) & (0.2071&) &  & (0.1435&) & (0.1585&) &  & (0.1204&) & (0.1200&) \\
		\hspace{10pt}\textit{- Medium} & 0.0646& & 0.0779& &  & 0.0387& & 0.1125& &  & 0.0522& & 0.1075& &  & 0.1083& & 0.1009& &  & 0.1779& & 0.1009& \\
		& (0.0646&) & (0.0733&) &  & (0.0387&) & (0.0637&) &  & (0.0522&) & (0.0528&) &  & (0.1083&) & (0.1026&) &  & (0.1779&) & (0.1811&) \\
		\hspace{10pt}\textit{- High} & 0.0010& & 0.0007& &  & 0.0012& & 0.0011& &  & 0.0009& & 0.0025& &  & 0.0014& & 0.0030& &  & 0.0038& & 0.0112& \\
		& (0.0010&) & (0.0011&) &  & (0.0012&) & (0.0016&) &  & (0.0009&) & (0.0009&) &  & (0.0014&) & (0.0024&) &  & (0.0038&) & (0.0049&) \\
		\textit{\texttt{Policy\_Type}} & && & && &  & && & && \\
		\hspace{10pt}\textit{- Regular} & 0.9093& & 0.8759& &  & 0.9223& & 0.8703& &  & 0.9042& & 0.8624& &  & 0.8345& & 0.8680& &  & 0.7766& & 0.8887& \\
		& (0.9093&) & (0.9087&) &  & (0.9223&) & (0.9069&) &  & (0.9042&) & (0.9034&) &  & (0.8345&) & (0.8857&) &  & (0.7766&) & (0.8216&) \\
		\hspace{10pt}\textit{- Employee} & 0.0447& & 0.0518& &  & 0.0434& & 0.0417& &  & 0.0512& & 0.0369& &  & 0.0684& & 0.0456& &  & 0.0289& & 0.0255& \\
		& (0.0447&) & (0.0447&) &  & (0.0434&) & (0.0410&) &  & (0.0512&) & (0.0517&) &  & (0.0684&) & (0.0595&) &  & (0.0289&) & (0.0235&) \\
		\hspace{10pt}\textit{- Second car} & 0.0460& & 0.0723& &  & 0.0343& & 0.0880& &  & 0.0446& & 0.1007& &  & 0.0970& & 0.0864& &  & 0.1946& & 0.0858& \\
		& (0.0460&) & (0.0466&) &  & (0.0343&) & (0.0521&) &  & (0.0446&) & (0.0450&) &  & (0.0970&) & (0.0548&) &  & (0.1946&) & (0.1549&) \\ \hline \addlinespace[0.4ex]
		\textit{ASAM} & 0.0432& & 0.0036& &  & 0.0507& & 0.0029& &  & 0.0242& & 0.0026& &  & 0.0224& & 0.0023& &  & 0.0500& & 0.0075& \\
		& (0.0432&) & (0.0345&) &  & (0.0507&) & (0.0284&) &  & (0.0242&) & (0.0238&) &  & (0.0224&) & (0.0113&) &  & (0.0500&) & (0.0490&) \\ \hline \addlinespace[0.4ex]
		\textit{Observations} & \multicolumn{4}{c}{14{,}305} & & \multicolumn{4}{c}{14{,}305} & & \multicolumn{4}{c}{14{,}306} & & \multicolumn{4}{c}{14{,}303} & & \multicolumn{4}{c}{14{,}303} \\
		\bottomrule
	\end{tabularx}}}
\end{table}

From the marginal and overall ASAMs in \Cref{Table_3} and \Cref{Figure_3a}, we observe that (generalized) propensity score matching can substantially improve the balance. More specifically, accounting for these scores increases the balance in almost every rate change interval and for each risk factor, and it can reduce the ASAM by up to $95\%$. The discrete approach additionally improves the balance substantially more than its continuous analogue, since it optimizes the rate change interval assignments directly in its estimation procedure and only has to distinguish between five categories. However, the continuous (discrete) approach still improves the balance (considerably) on average and the (generalized) propensity score therefore seems to be appropriately specified and estimated with XGBoost.

While the (generalized) propensity score is used to balance the risk factors, it is estimated based on the treatment assignment mechanism. \Cref{Figure_3b} therefore compares its estimates to the observed rate change assignments for each rate change interval in both the discrete and continuous framework. The leftmost, red curve, for instance, denotes the average probability in the portfolio of a rate change assignment from the lowest quintile, where the red area marks all rate changes that actually belong to this quintile. Based on these curves, we find that the discrete approach generally has much more discriminatory power and that rate change assignments can often be identified only reasonably well in the continuous approach. Similar to \citet{guelman2014}, the inner quintiles appear to be harder to distinguish from each other than from the outer quintiles, since their centers are closer together with smaller intervals and the average (generalized) propensity score can be relatively high outside of these quintiles, especially in the continuous approach. Nonetheless, both approaches clearly exhibit common support in \Cref{Figure_3b} and strong ignorability thus holds. This, in turn, implies that identification of the average treatment effects is possible in this automobile insurance application.

\begin{figure}[t!]
    \centering
    \begin{subfigure}{\textwidth}
        \centering
	    \includegraphics[width=0.95\textwidth]{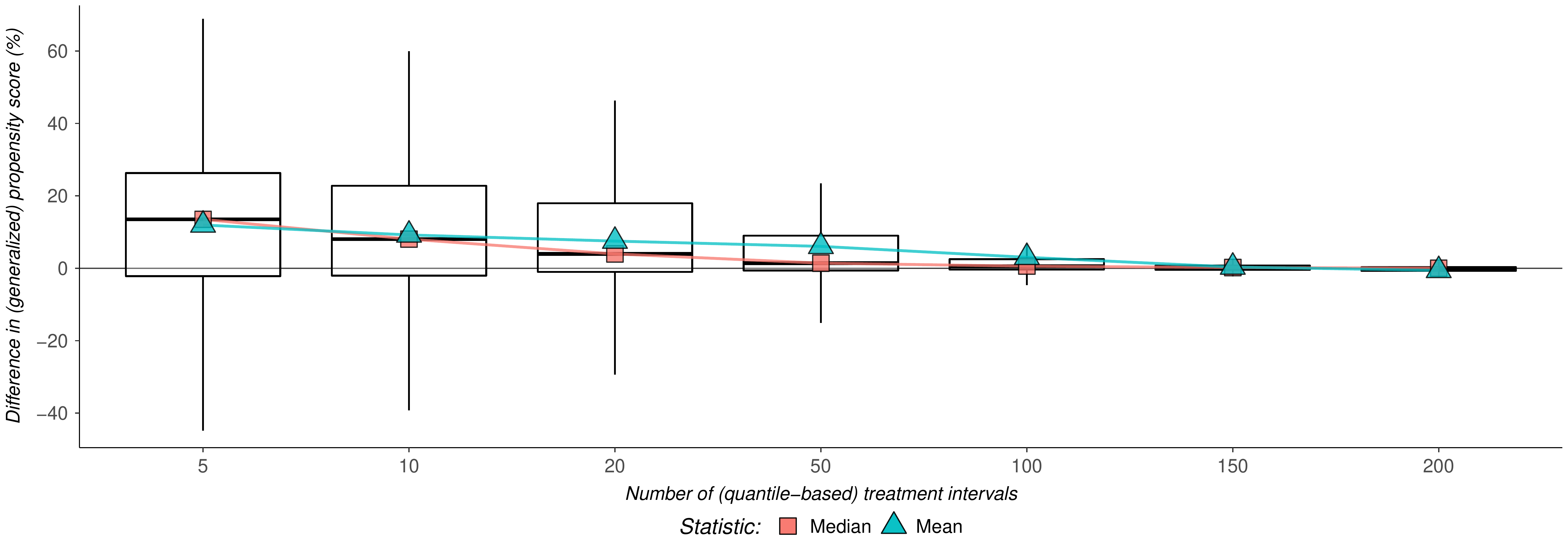}\vspace{-3pt}
        \caption{Distribution of differences in (generalized) propensity score}
        \label{Figure_4a}
    \end{subfigure}
    
    \vspace{3pt}
    
    \begin{subfigure}{\textwidth}
        \centering
        \includegraphics[width=0.95\textwidth]{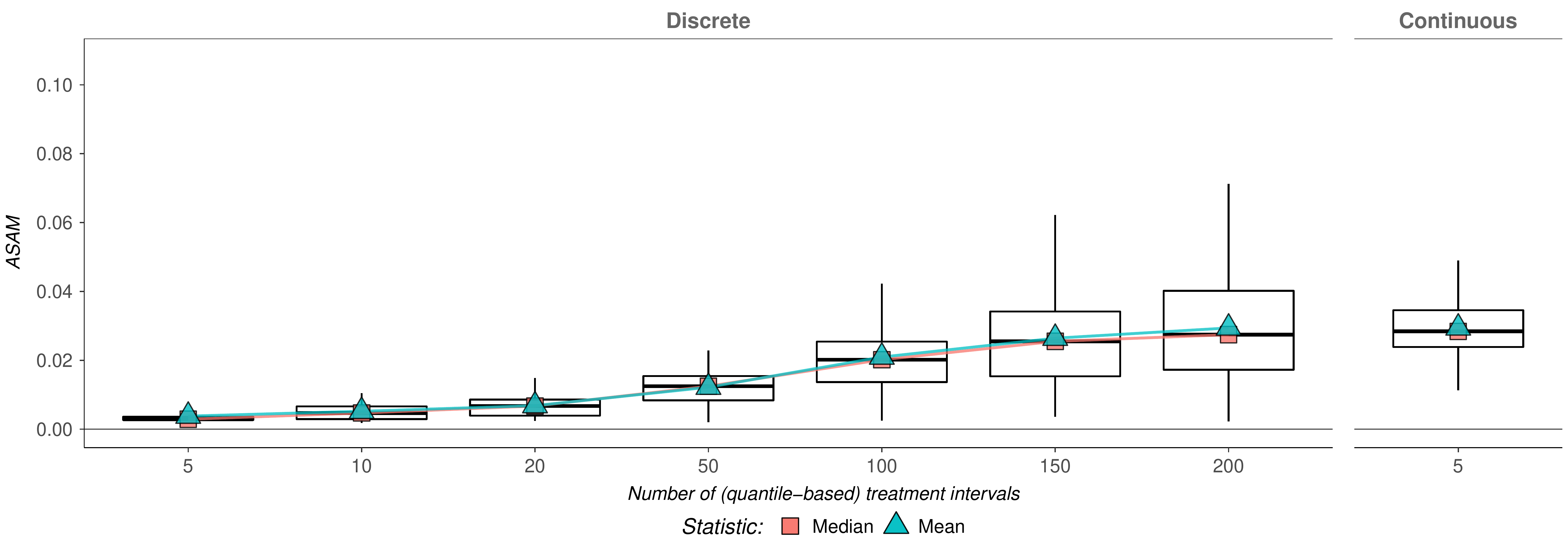}\vspace{-3pt}
        \caption{Distribution of ASAMs}
        \label{Figure_4b}
    \end{subfigure}\vspace{-3pt}
    \caption{Distribution of differences in (generalized) propensity score (panel (a)) and ASAMs (panel (b)) for a varying number of (quantile-based) treatment intervals with discrete and continuous rate changes in automobile insurance.}
	\label{Figure_4}
\end{figure}

Although the two approaches can lead to considerably different (generalized) propensity scores, we expect the discrete approach to approximate the continuous approach when the number of treatment categories increases. The intuition behind this is that the continuous approach can be seen as a limiting case of the discrete approach since it effectively adopts an infinite number of treatment intervals by assuming continuous rate changes. To investigate this hypothesis, we adjust the (generalized) propensity score models to consider $C \in \{5, 10, 20, 50, 100, 150, 200\}$ treatment intervals based on the quantiles of the selected rate changes. More specifically, we adopt the hyperparameters optimized previously for the two models but now consider $C$ intervals in the performance measures. While \Cref{Figure_4a} shows how the differences in the (generalized) propensity score between these discrete and continuous models behave with respect to the number of treatment intervals, \Cref{Figure_4b} depicts how the ASAM, or the balance in each interval, is affected by this number. In line with our hypothesis, we find in \Cref{Figure_4a} that the (generalized) propensity scores from the discrete approach converge to those from the continuous approach when the number of treatment categories increases. From \Cref{Figure_4b} we also observe that the ASAM (balance) increases (decreases) on average for the discrete approach as the number of categories increases, simply because it becomes more difficult to distinguish between each category when we have more categories to consider. The balance in the discrete approach in fact converges to that in the continuous approach with more categories, although we observe noticeably less variance in the ASAM in the continuous approach. More treatment categories have been considered for the continuous approach as well, but this only slightly increases the ASAM. The increasing (decreasing) trend in the ASAM (balance) thus seems to persist in the discrete approach and we therefore expect the continuous approach to further outperform the discrete approach with even more treatment categories, since this setting will increasingly resemble the design of the continuous approach.

\begin{figure}[t!]
    \centering
    \includegraphics[width=0.95\textwidth]{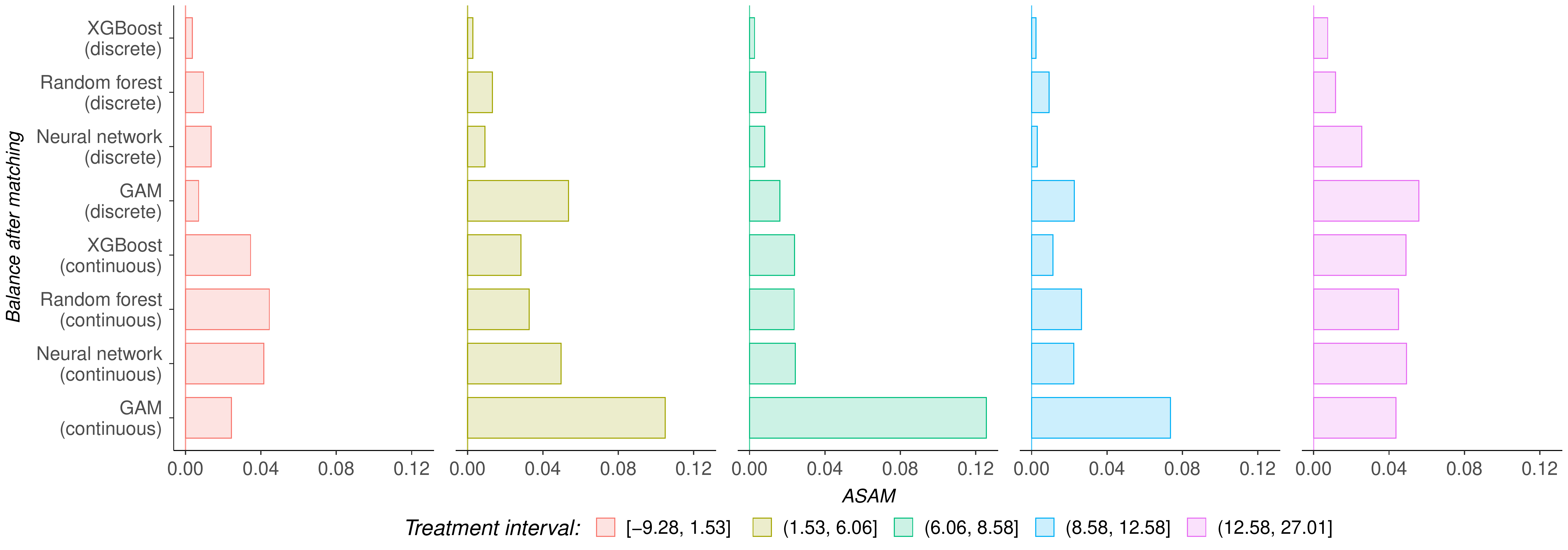}\vspace{-3pt}
    \caption{Mean balance after matching using XGBoost, random forest, neural network, and GAM for each rate change interval with discrete and continuous rate changes in automobile insurance.}
	\label{Figure_5}
\end{figure}

Besides the number of treatment intervals, we may also explore how robust the proposed (generalized) propensity score model is with respect to other methods. To investigate this, we examine how well the XGBoost model can distinguish between the five treatment intervals in comparison to an approach based on random forests, neural networks, or Generalized Additive Models (GAMs). We use a setup similar to the XGBoost method to estimate these models, where we adopt $10$-fold cross-validation with early stopping to optimize over the parameter grids in \Cref{Table_B.1} in supplementary \ref{Appendix_B} for the random forest and neural network approach and consider natural cubic splines for all second order and level interactions of the risk factors in \Cref{Table_1} for the GAM. The resulting ASAM, or balance, in each rate change interval is shown in \Cref{Figure_5} for each of these methods and for both the discrete and continuous framework. From this comparison we find that the GAM leads to a substantially higher ASAM (lower balance) than the other methods in almost every rate change interval and in both frameworks, and that it is unable to improve the balance observed before the matching procedure. The other methods do improve the balance after matching, indicating that these more complex, non-linear techniques are in fact necessary to adequately capture the treatment assignment mechanism. However, even though both the random forest and neural network approach lead to a noticeably lower ASAM (higher balance) in almost every rate change interval after matching, the proposed XGBoost method improves the balance considerably more. The (generalized) propensity score model thus seems to be more capable of describing the treatment assignment mechanism, and hence more suited to implement, with XGBoost. We therefore adopt the (generalized) propensity score model with XGBoost in the remainder of this analysis.

\subsection{Customer price sensitivities} \label{Section4.2}

\begin{figure}
    \vspace{-24pt}
    \centering
    \begin{subfigure}{\textwidth}
        \centering
        \begin{tabular}{c c}
            \centering
            {\hspace{-45pt}\footnotesize\underline{\hspace{18.5mm}\textbf{Discrete$_{_{}}$}\hspace{18.5mm}}} & {\hspace{-85pt}\footnotesize\underline{\hspace{16.0mm}\textbf{Continuous$_{_{}}$}\hspace{16.0mm}}} \vspace{-13pt}\\
            \centering\hspace{-45pt}
            \includegraphics[width=0.675\textwidth]{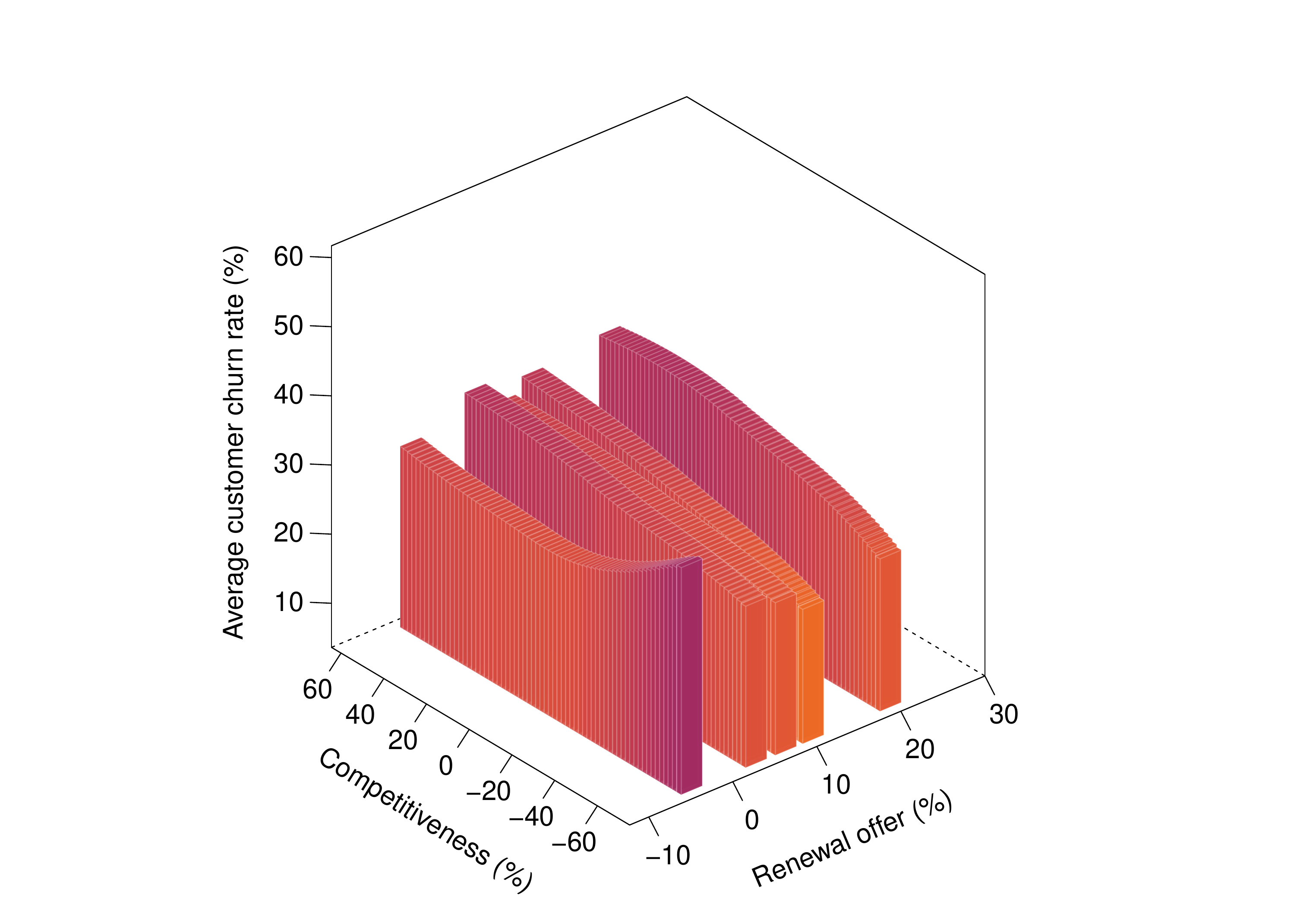}&\hspace{-85pt}
            \includegraphics[width=0.675\textwidth]{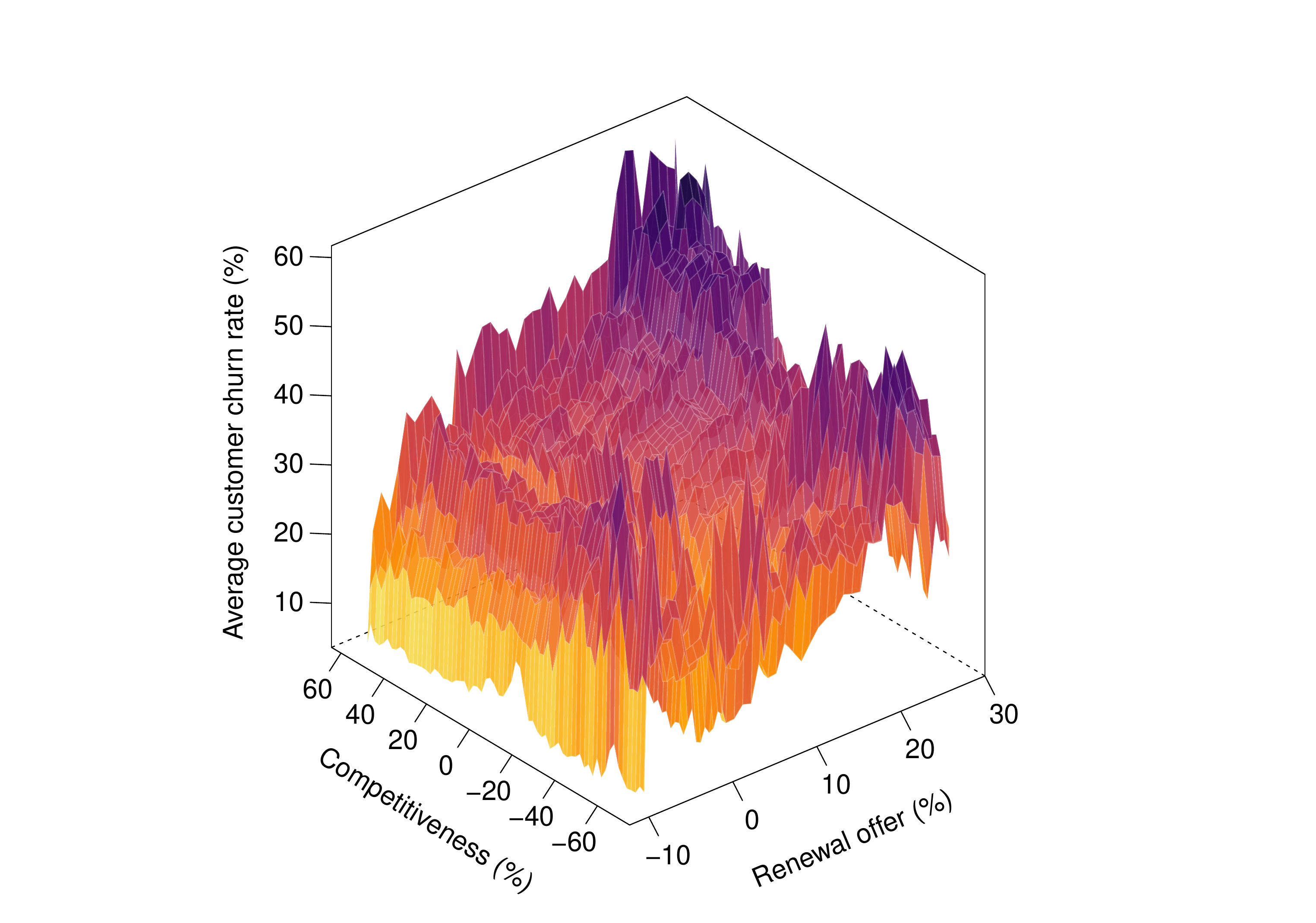}
        \end{tabular}\vspace{-8pt}
        \caption{Average customer churn for renewal offers and competitiveness}
	    \label{Figure_6a}
    \end{subfigure}

    \vspace{-9pt}

    \begin{subfigure}{\textwidth}
        \centering
        \begin{tabular}{c c}
            \centering\hspace{-45pt}
            \includegraphics[width=0.675\textwidth]{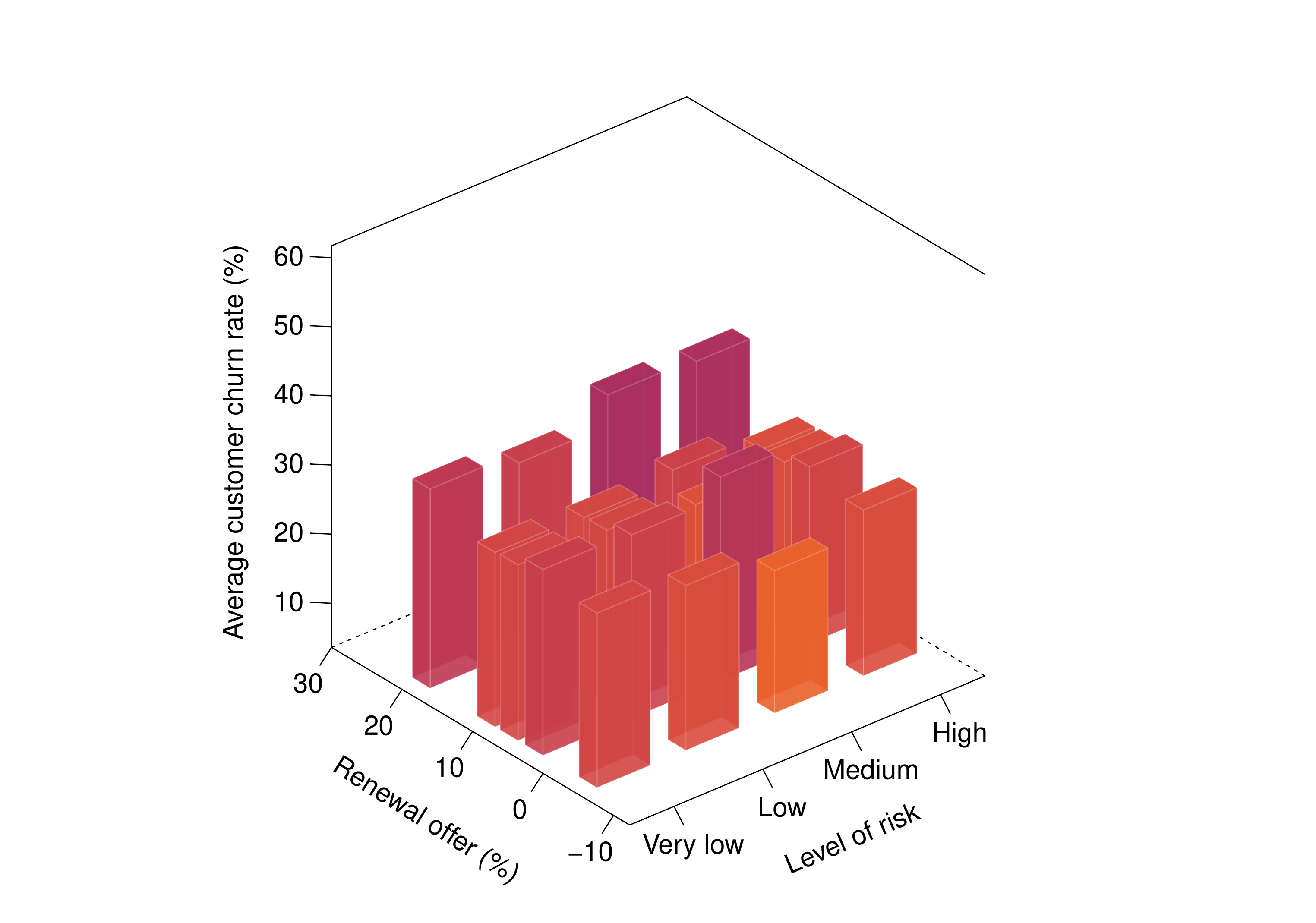}&\hspace{-85pt}
            \includegraphics[width=0.675\textwidth]{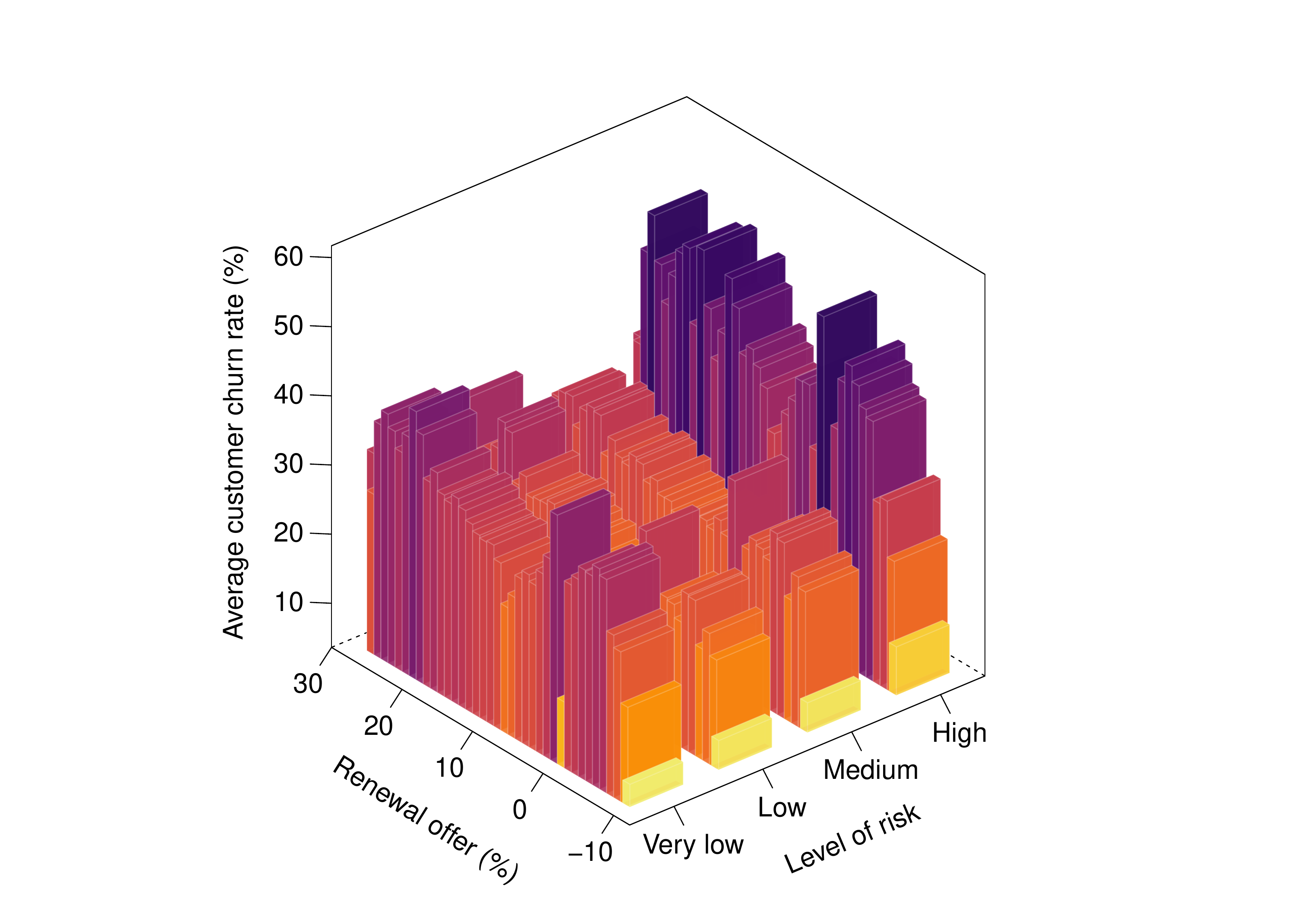}
        \end{tabular}\vspace{-8pt}
        \caption{Average customer churn for renewal offers and level of risk}
        \label{Figure_6b}
    \end{subfigure}
    
    \vspace{-9pt}
    
    \begin{subfigure}{\textwidth}
        \centering
        \begin{tabular}{c c}
            \centering\hspace{-45pt}
            \includegraphics[width=0.675\textwidth]{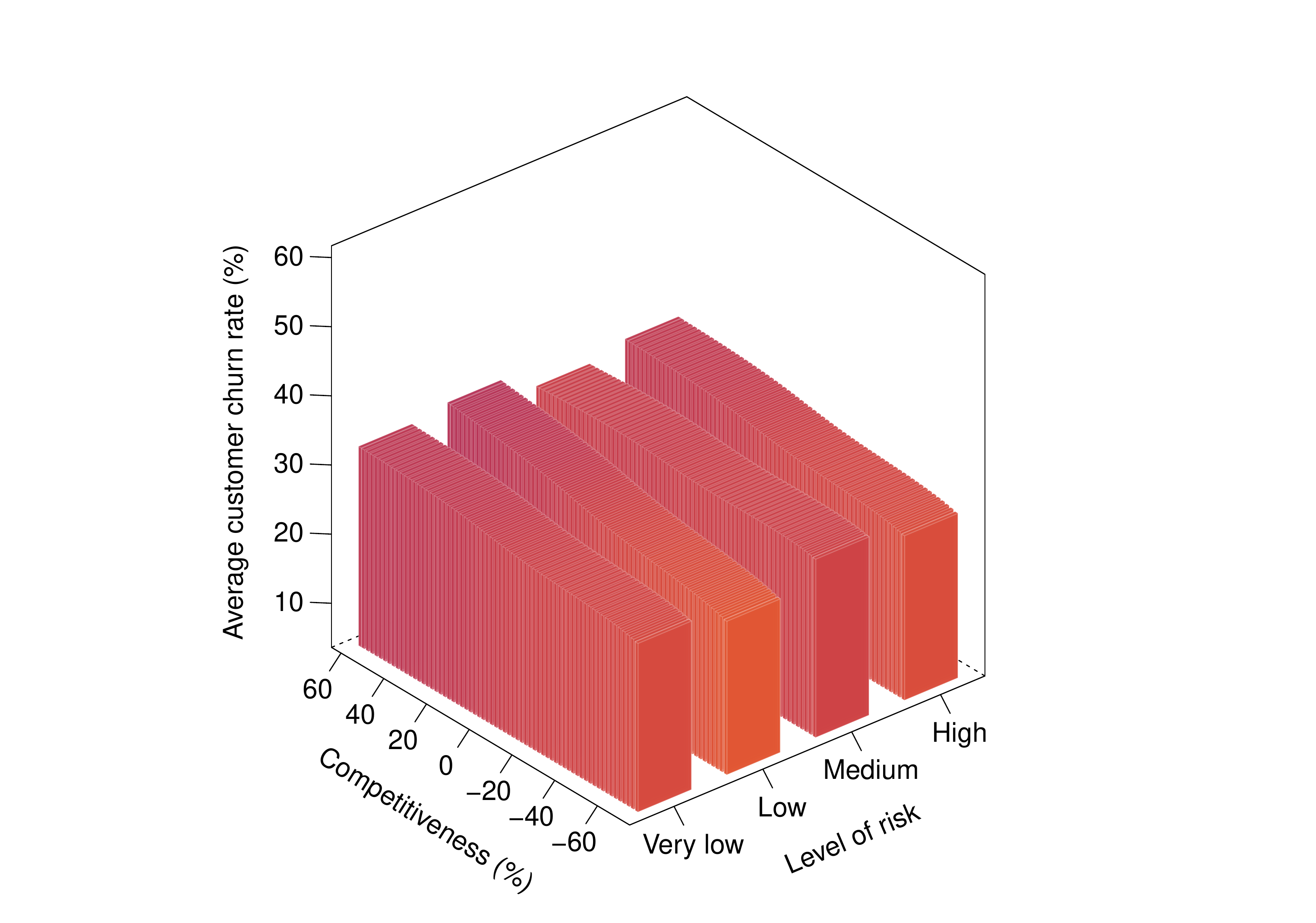}&\hspace{-85pt}
            \includegraphics[width=0.675\textwidth]{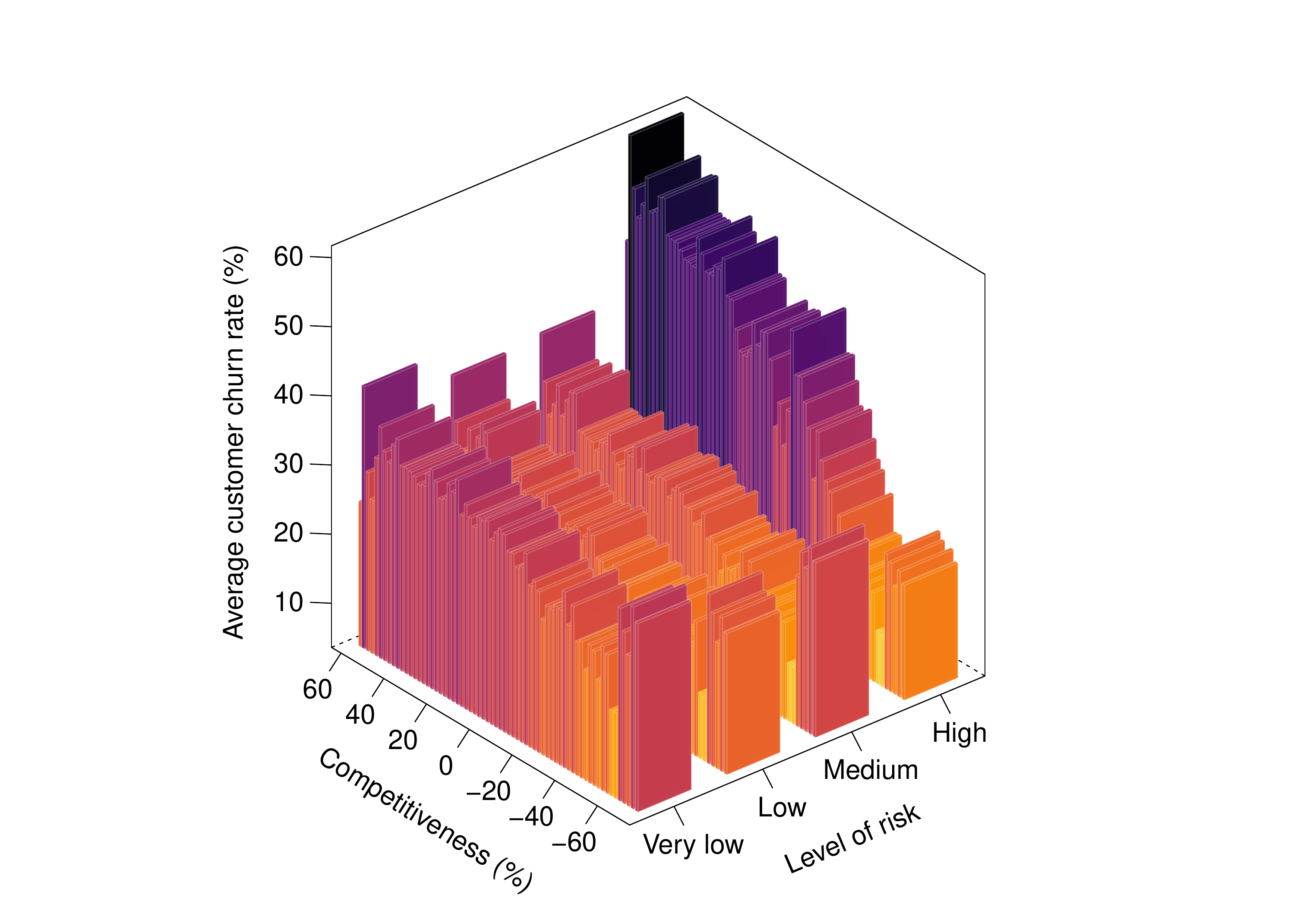}
        \end{tabular}\vspace{-8pt}
        \caption{Average customer churn for level of risk and competitiveness}
        \label{Figure_6c}
    \end{subfigure}\vspace{-3pt}
    \caption{Average customer churn estimate for each renewal offer and every competitiveness (panel (a)) and level of risk (panel (b)) as well as aggregated over all renewal offers (panel (c)) with discrete (left) and continuous (right) rate changes in automobile insurance.}
	\label{Figure_6}
\end{figure}

\begin{figure}[t!]
    \vspace{-22pt}
    \centering
    \begin{tabular}{c c}
            \centering\hspace{-45pt}
            \includegraphics[width=0.675\textwidth]{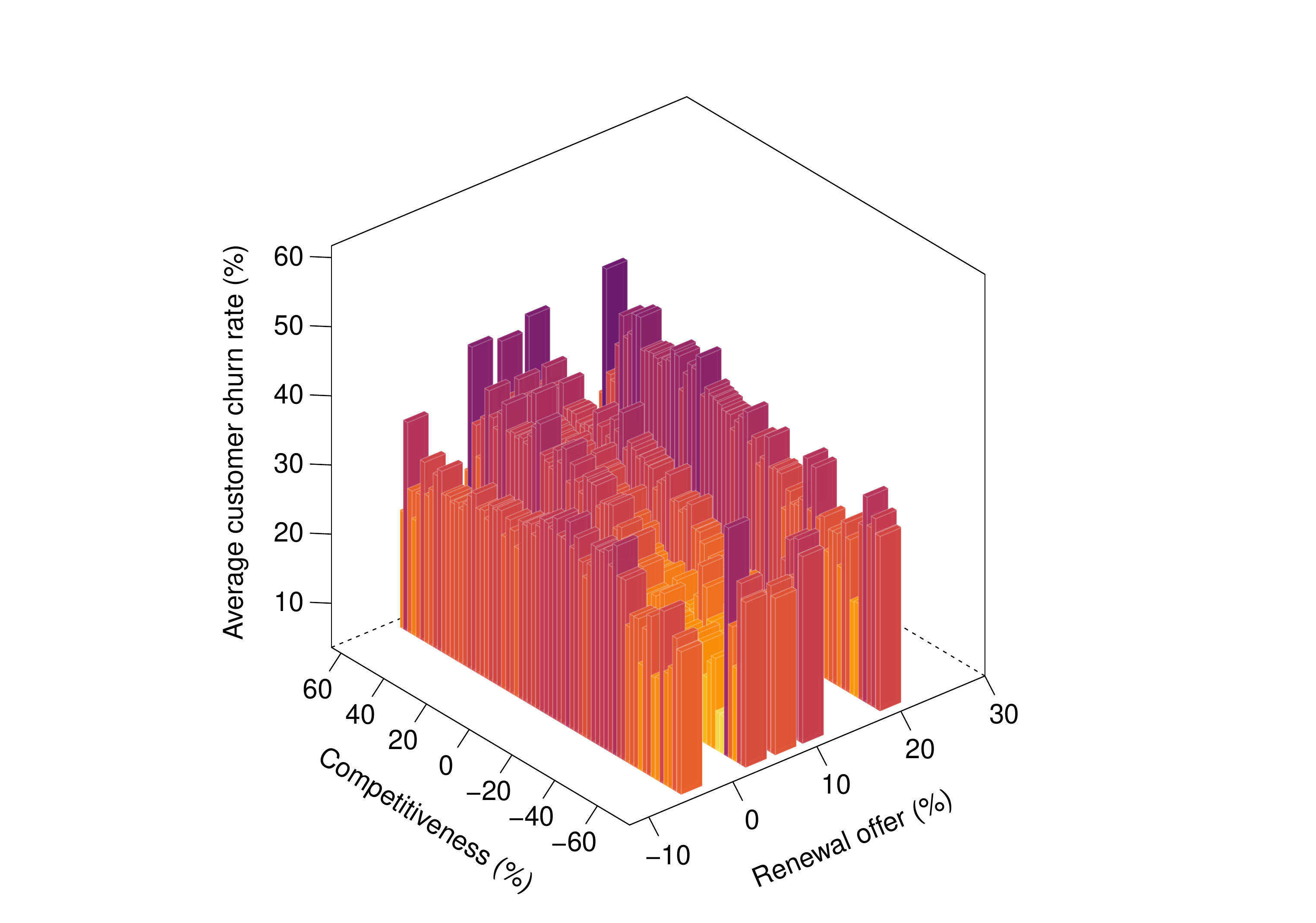}&\hspace{-85pt}
            \includegraphics[width=0.675\textwidth]{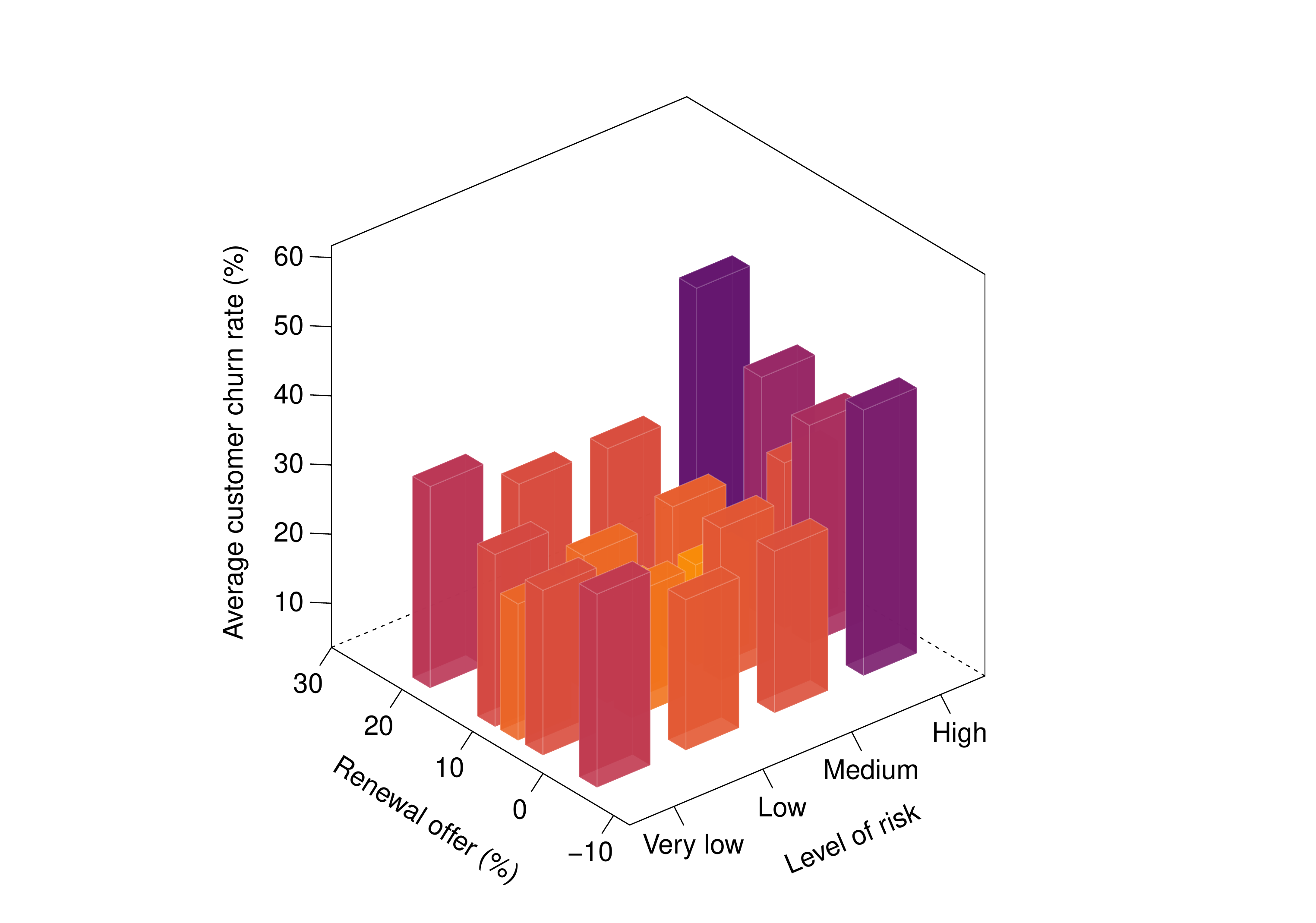}
        \end{tabular}\vspace{-8pt}
    \caption{Average customer churn estimate from continuous approach in \Cref{Figure_6} for each rate change interval and every competitiveness (left) and level of risk (right) in automobile insurance.}
    \label{Figure_7}
\end{figure}

Given the (generalized) propensity scores, we can model customers' churning behavior based on their potential responses. We perform LASSO and multiple imputation with these responses to identify key interactions and to include any imputation uncertainty into the discrete global response model. For the continuous approach, we simply apply XGBoost to the observed responses and GPSs to estimate the (conditional) dose-response function. The resulting average treatment effects, or customer churn rates, from these two approaches are shown for all renewal offers and for both a policy's competitiveness and a customer's level of risk in \Cref{Figure_6}. A weighted average of the continuous response surface over the five rate change intervals is additionally displayed in \Cref{Figure_7}. For more details on the LASSO procedure and the discrete global response model, see \Cref{Figure_B.1} and \Cref{Table_B.2}, respectively.

From the customer churn rates in \Cref{Figure_6a}, we observe that the competitiveness can heavily influence customers' willingness to stay. More specifically, we find that customers are generally much more likely to churn when their policy is already very competitive and (one of) the cheapest in the market, especially in the continuous framework. With respect to the renewal offers, we recover a slightly varying but moderately increasing churn surface in the discrete approach, similar to \citet{guelman2014}, whereas the continuous surface has a clear and much steeper positive slope. However, both surfaces are in line with the churn ratios observed earlier in \Cref{Figure_2} and the discrete surface actually appears to approximate the weighted average of the continuous surface over the five rate change intervals in \Cref{Figure_7}. Moreover, we display the churn surface for the discrete approach with XGBoost and without multiple imputation in \Cref{Figure_B.3}, from which we observe that this approximation becomes much closer when we adopt XGBoost for the discrete global response model as well and that multiple imputation only has a marginal impact. Note that this discrete XGBoost model has been applied to the average potential responses given the optimized hyperparameters for the continuous approach with $504$ decision trees only for illustrative purposes and for more details on the global response model without multiple imputation, see \Cref{Figure_B.2} and \Cref{Table_B.3}. All surfaces additionally indicate a relatively large inflection point at small rate changes, similar to the inflection point in \Cref{Figure_2}. This non-monotonicity is particularly well captured in the continuous approach since it avoids any discretization or aggregation, and thereby exploits more information, and can distinguish between the rate changes in each rate change interval. Furthermore, the continuous (discrete) surface in \Cref{Figure_6b} and the weighted average of the continuous surface in \Cref{Figure_7} show that customers are (typically) more likely to lapse their policy at a higher renewal rate and that this likelihood can be substantially (somewhat) larger for very low and high risks. Finally, \Cref{Figure_6c} indicates that in both frameworks the churn rate is predicted to increase with the competitiveness for every level of risk and that customers associated with the highest, and to a lesser extent with the lowest, amount of risk are particularly sensitive to premium changes, even at lower competitiveness levels.

The intuition behind these results is threefold. First, it becomes more worthwhile for customers to switch insurers at higher rate changes, even though this incentive decreases as the rate change increases in absolute sense \citep{guelmanetal2014,delallave2019,leiria2021}. However, contrary to \citet{guelman2014}, this effect can be substantially larger at small rate changes since customers must be notified of any price changes, and they will be more inclined to lapse their policy once they realize that it will become more expensive or barely cheaper. Insurers are therefore usually reluctant to change their premia (by too much) in practice and prefer to let sleeping dogs lie, to avoid raising (more than necessary) this price awareness in their customers \citep{brockett2008,robson2015}. Secondly, policies are priced very competitively at inception to attract new customers, who typically compare the premia of different insurers before entering into a contract \citep{hinterhuber2004,gamble2009,guelmanetal2014,laas2016}. Similarly, competitive policies are not priced too expensively at other insurers, so it is relatively attractive to switch insurers \citep{spedicato2018,delallave2019,leiria2021}. Customers with very competitively priced policies thus seem to be more aware of, and sensitive to, competing offers in the market. Thirdly, the new policies generally become more expensive during their lifetimes since they are initially underpriced by design \citep{thomas2012,leiria2021}. Even though very low risk customers will probably never claim, their premium will therefore become higher than necessary. In addition, high risk customers are typically charged a substantial premium to reflect their previous claims, but this is usually to compensate for past losses on the policy and will be more than needed to cover the future risk on the policy. This amplifies the let sleeping dogs lie effect and it therefore tends to be relatively attractive for very low and high risk customers to lapse their (expensive) policy and to switch insurers, and perhaps even more attractive when considering more insurers than the six largest competitors in the market \citep{guillen2008,guillen2012,paredes2018,jeong2018,delallave2019}. This, in turn, seems to suggest that the level of competitiveness and risk are two key drivers of a customer's price sensitivity.

\begin{figure}[t!]
    \centering
    \begin{subfigure}{\textwidth}
        \centering
        \begin{tabular}{c c}
            \centering
            {\footnotesize\underline{\hspace{18.5mm}\textbf{Discrete$_{_{}}$}\hspace{18.5mm}}} & {\footnotesize\underline{\hspace{16.0mm}\textbf{Continuous$_{_{}}$}\hspace{16.0mm}}} \vspace{5pt}\\
            \centering
            \includegraphics[width=0.45\textwidth]{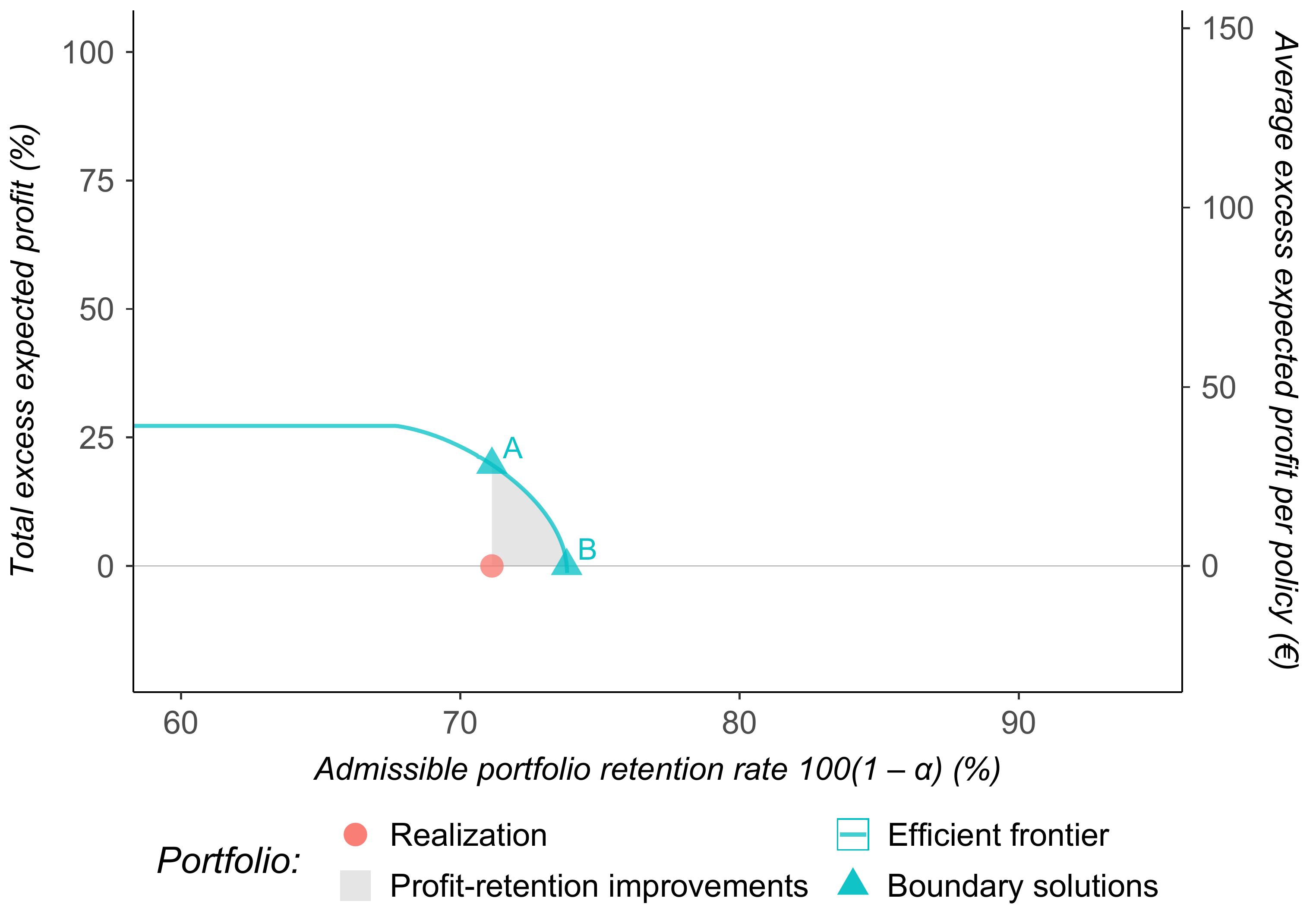}&
            \includegraphics[width=0.45\textwidth]{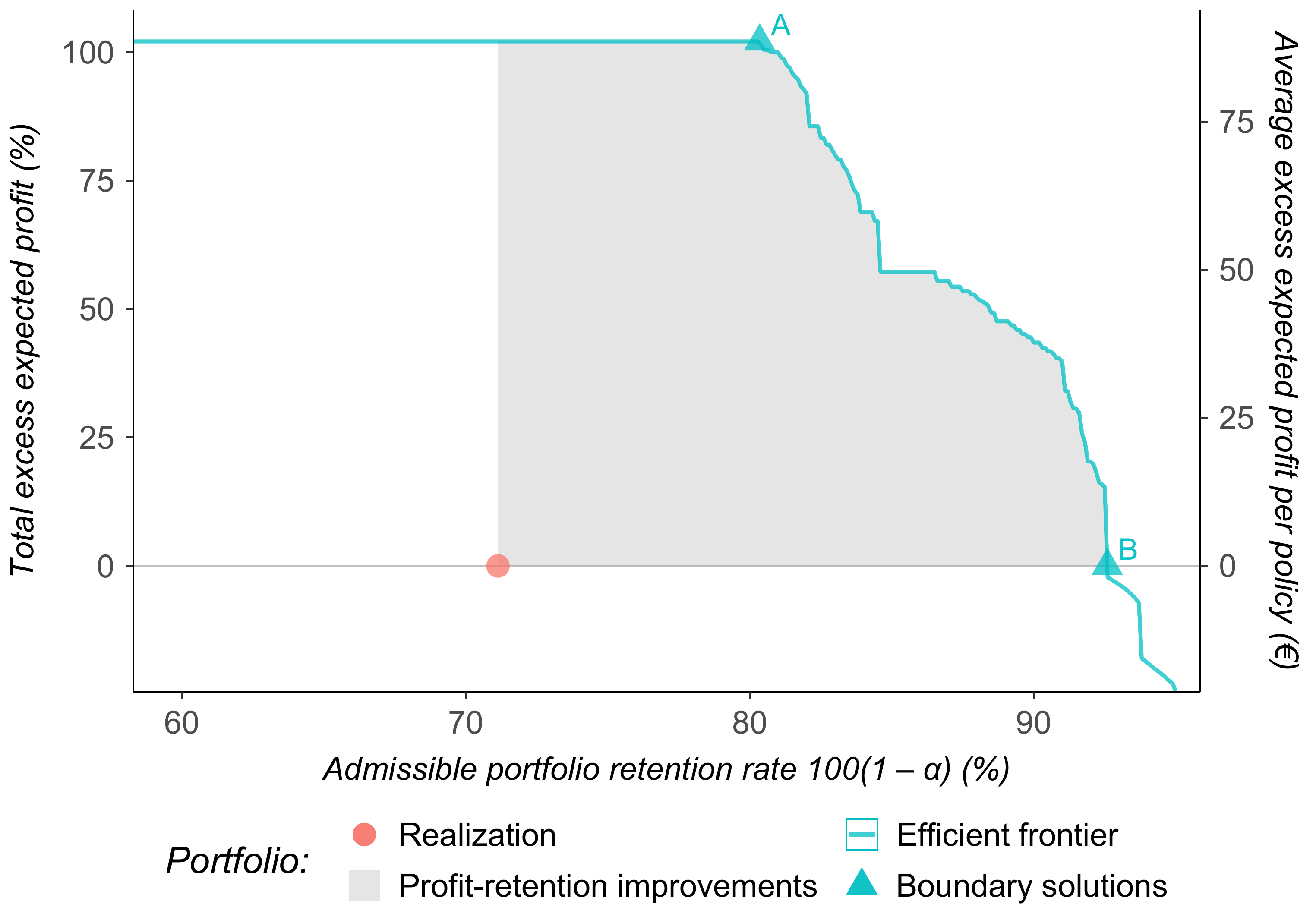}
        \end{tabular}\vspace{-6pt}
        \caption{Expected profit and portfolio retention efficient frontier}
        \label{Figure_8a}
    \end{subfigure}
    
    \vspace{3pt}
    
    \begin{subfigure}{\textwidth}
        \centering
        \begin{tabular}{c c}
            \centering
            \includegraphics[width=0.45\textwidth]{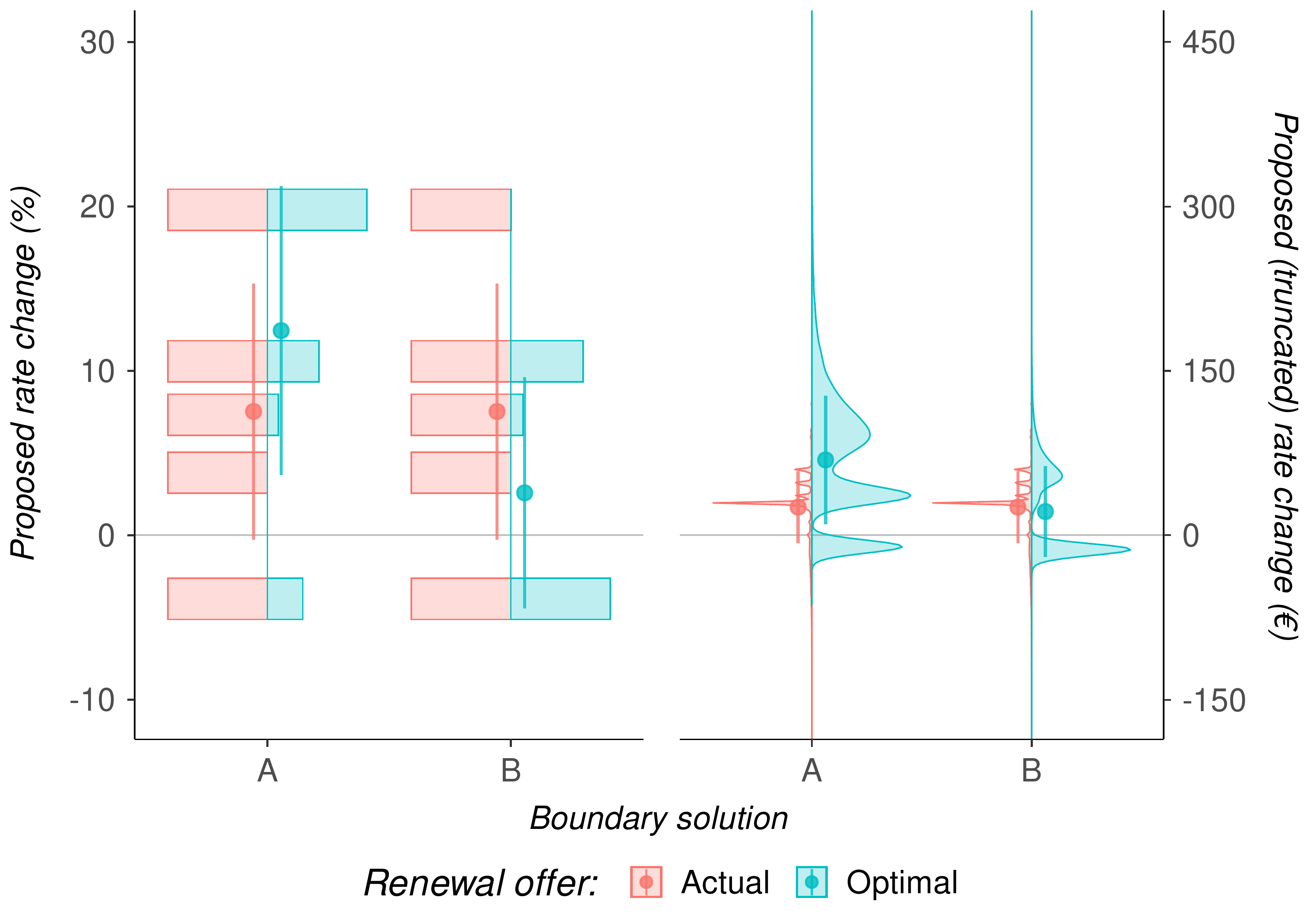}&
            \includegraphics[width=0.45\textwidth]{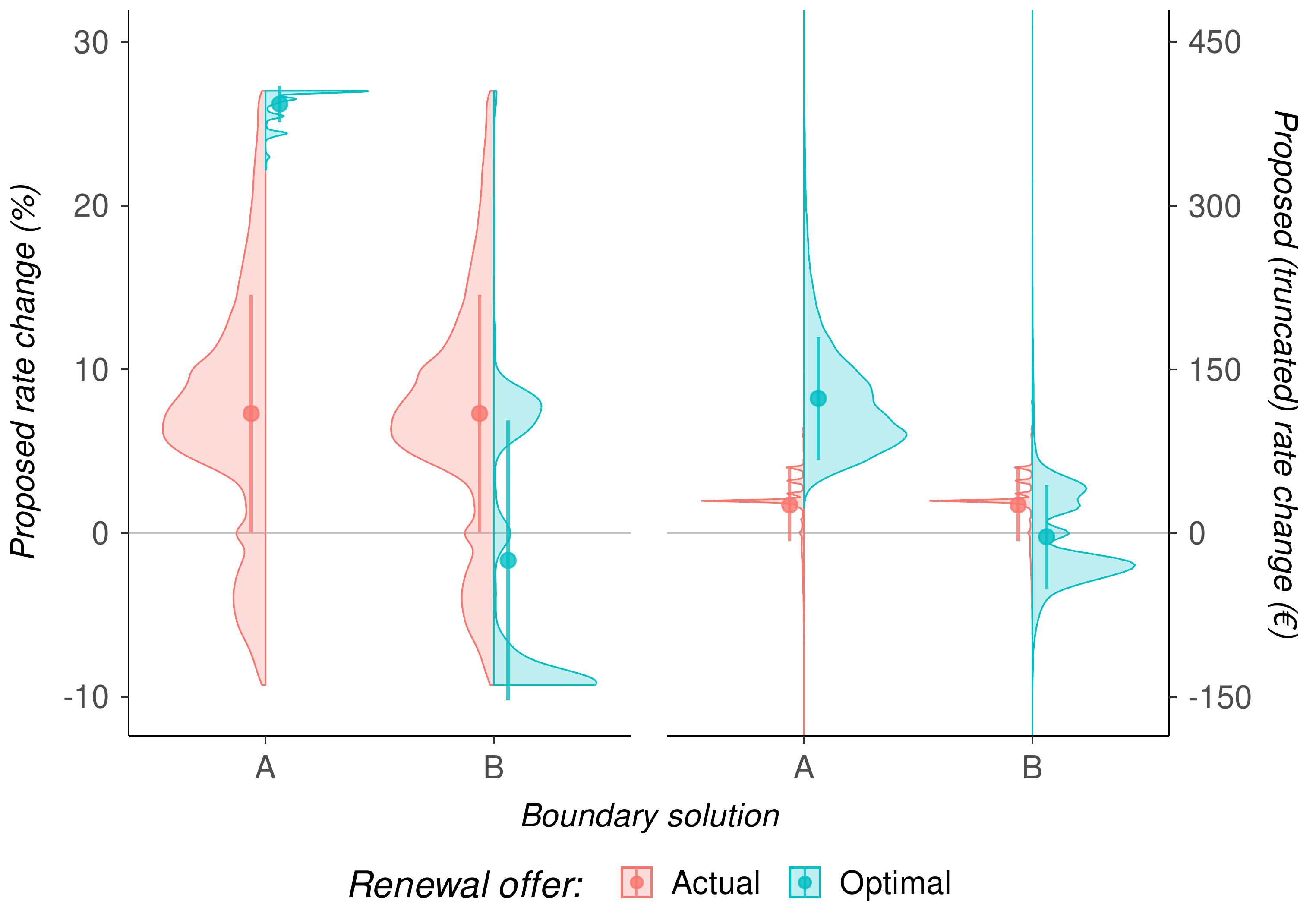}
        \end{tabular}\vspace{-6pt}
        \caption{Proposed rate changes in boundary solutions of efficient frontier}
        \label{Figure_8b}
    \end{subfigure}\vspace{-3pt}
    \caption{Expected profit and portfolio retention efficient frontier (panel (a)) and proposed rate changes in its boundary solutions (panel (b)) with discrete (left) and continuous (right) rate changes in automobile insurance. We assume here that insurers can always increase customer churn without losing profit in a horizontal frontier.}
	\label{Figure_8}
\end{figure}

\begin{figure}[t!]
    \centering
    \begin{subfigure}{\textwidth}
        \centering
        \begin{tabular}{c c}
            \centering
            {\footnotesize\underline{\hspace{11.875mm}\textbf{Efficient frontier$_{_{}}$}\hspace{11.875mm}}} & {\footnotesize\underline{\hspace{9.375mm}\textbf{Boundary solutions$_{_{}}$}\hspace{9.375mm}}} \vspace{5pt}\\
            \centering
            \includegraphics[width=0.45\textwidth]{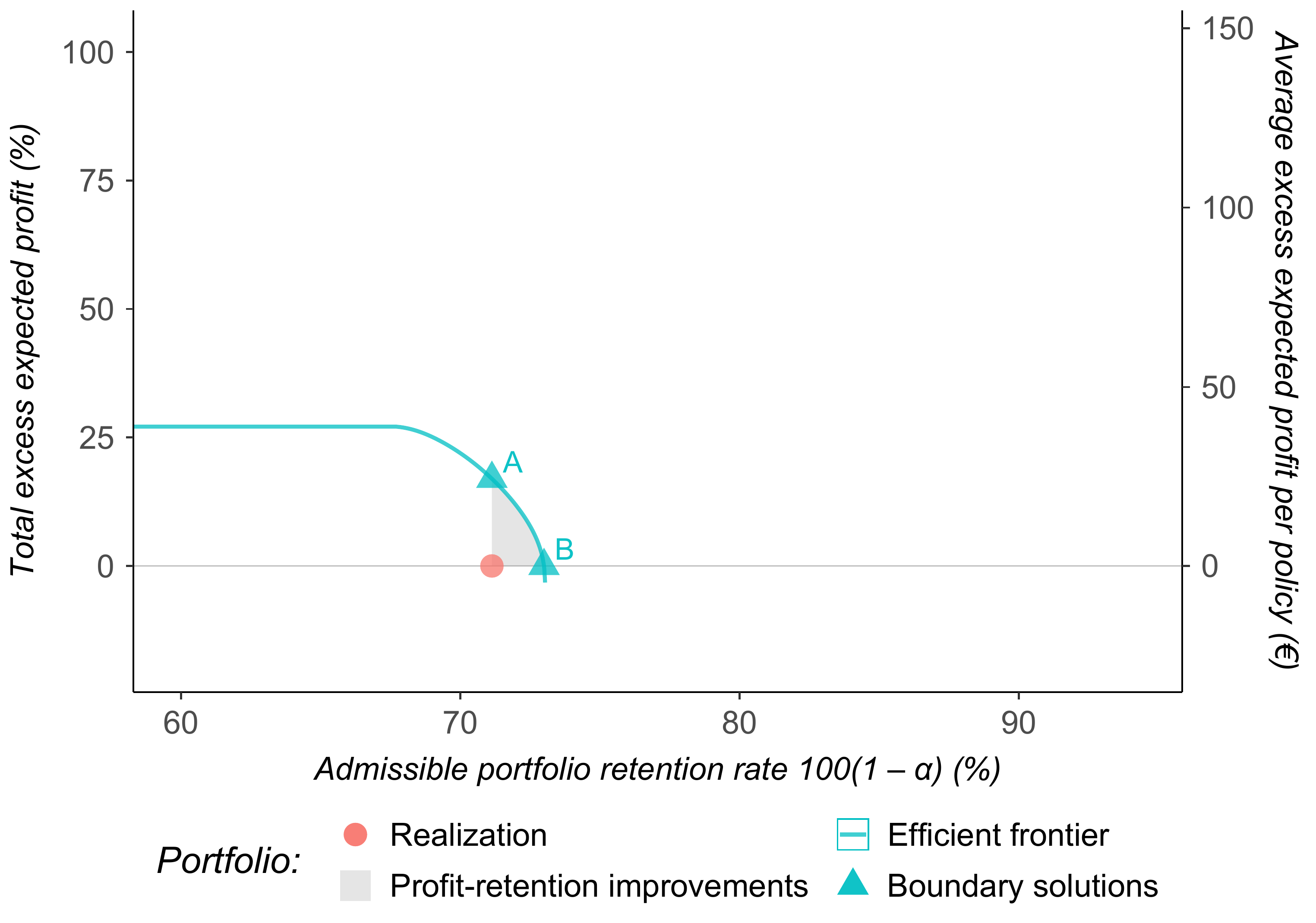}&
            \includegraphics[width=0.45\textwidth]{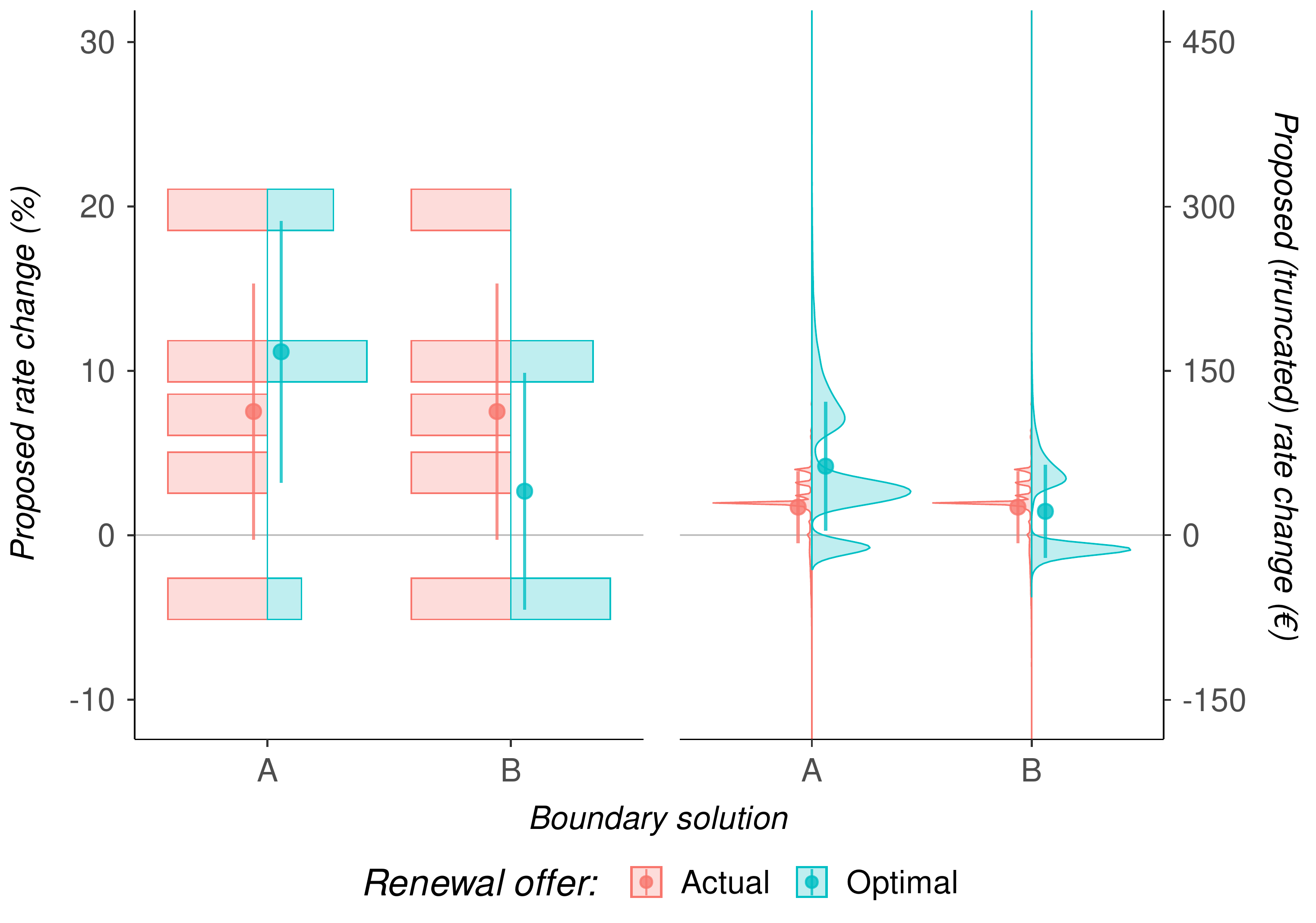}
        \end{tabular}\vspace{-6pt}
        \caption{Expected profit efficient frontier solutions for discrete global response model without multiple imputation}
        \label{Figure_9a}
    \end{subfigure}
    
    \vspace{3pt}
    
    \begin{subfigure}{\textwidth}
        \centering
        \begin{tabular}{c c}
            \centering
            \includegraphics[width=0.45\textwidth]{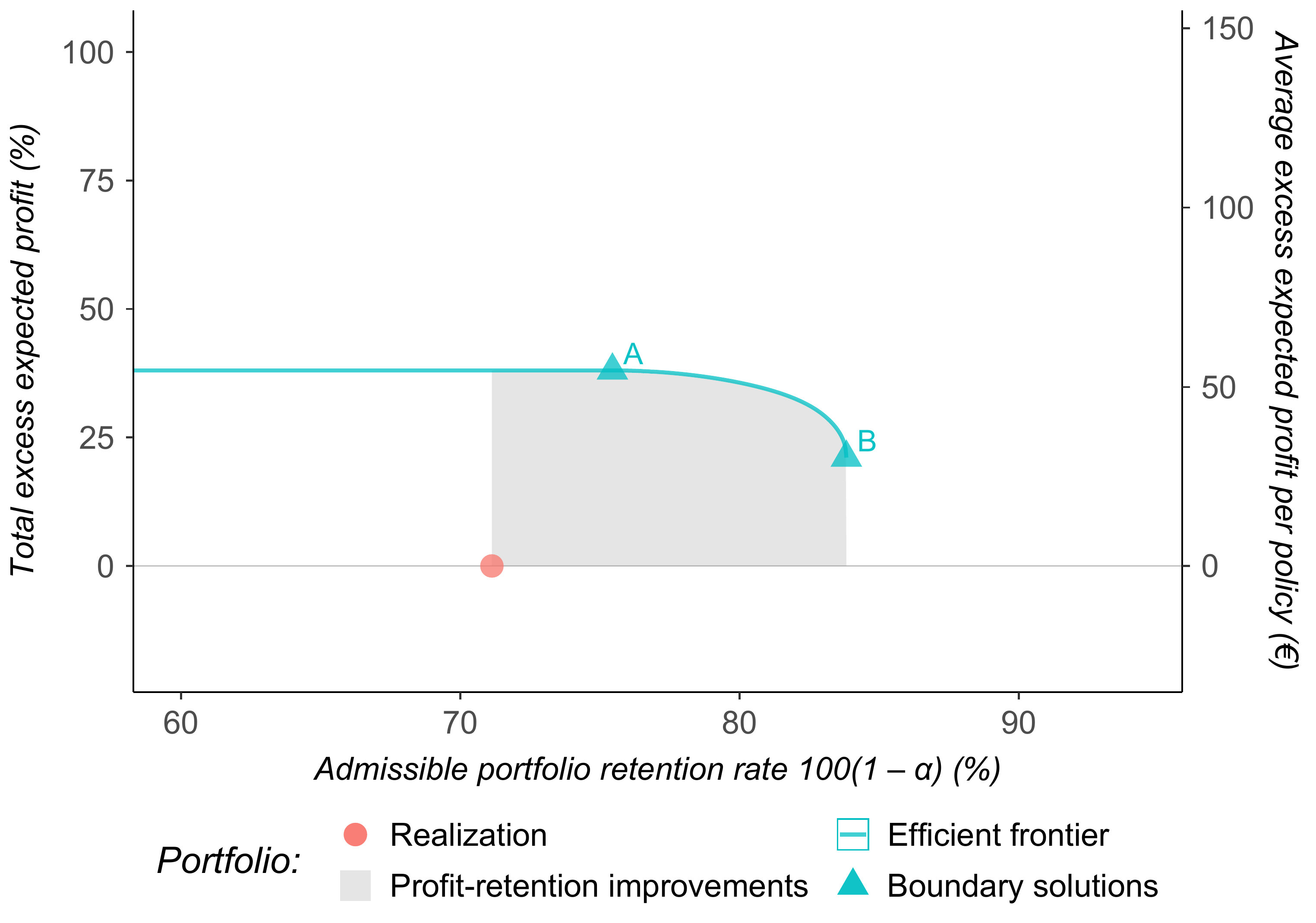}&
            \includegraphics[width=0.45\textwidth]{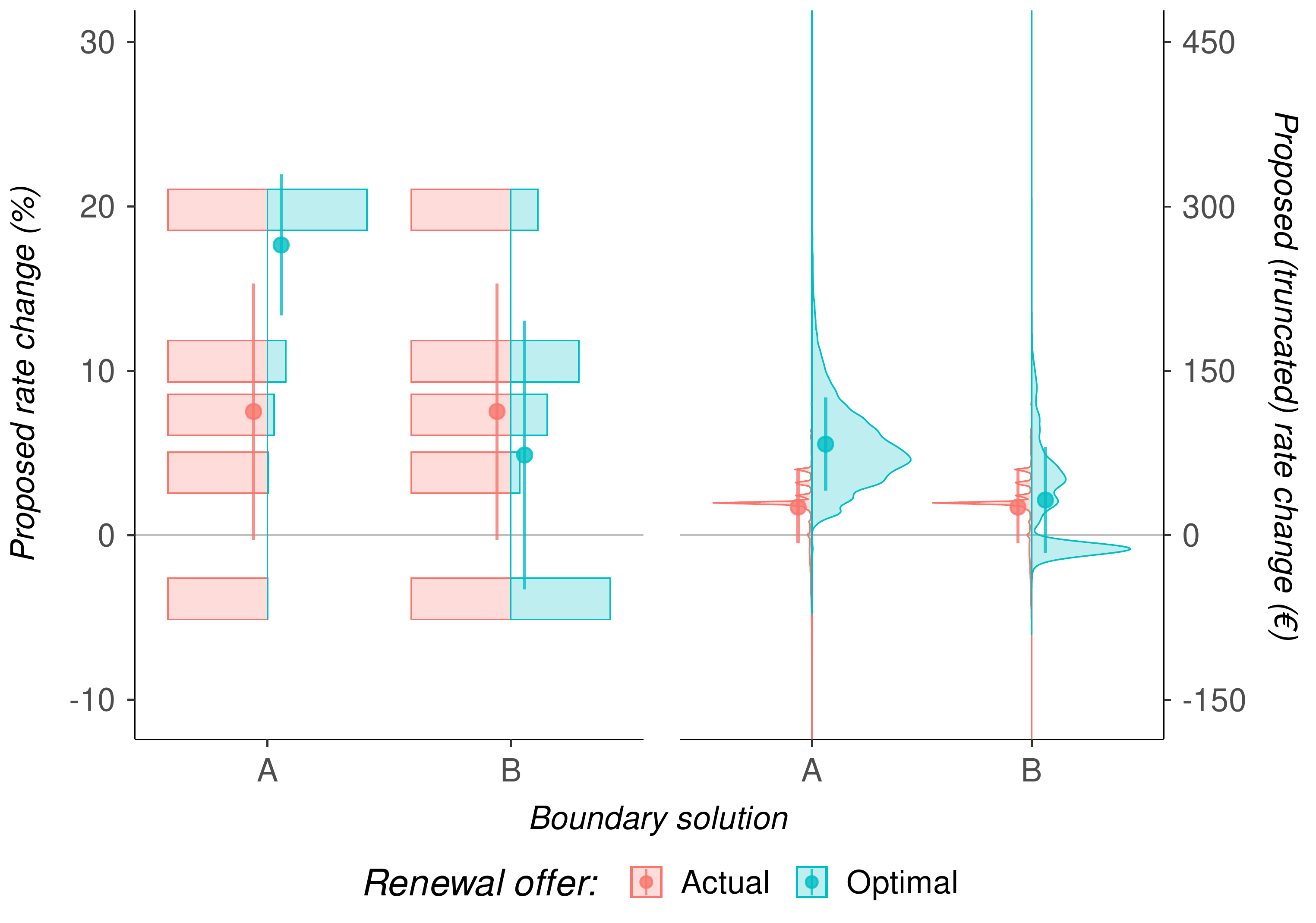}
        \end{tabular}\vspace{-6pt}
        \caption{Expected profit efficient frontier solutions for discrete global response model with XGBoost}
        \label{Figure_9b}
    \end{subfigure}
    
    \vspace{3pt}
    
    \begin{subfigure}{\textwidth}
        \centering
        \begin{tabular}{c c}
            \centering
            \includegraphics[width=0.45\textwidth]{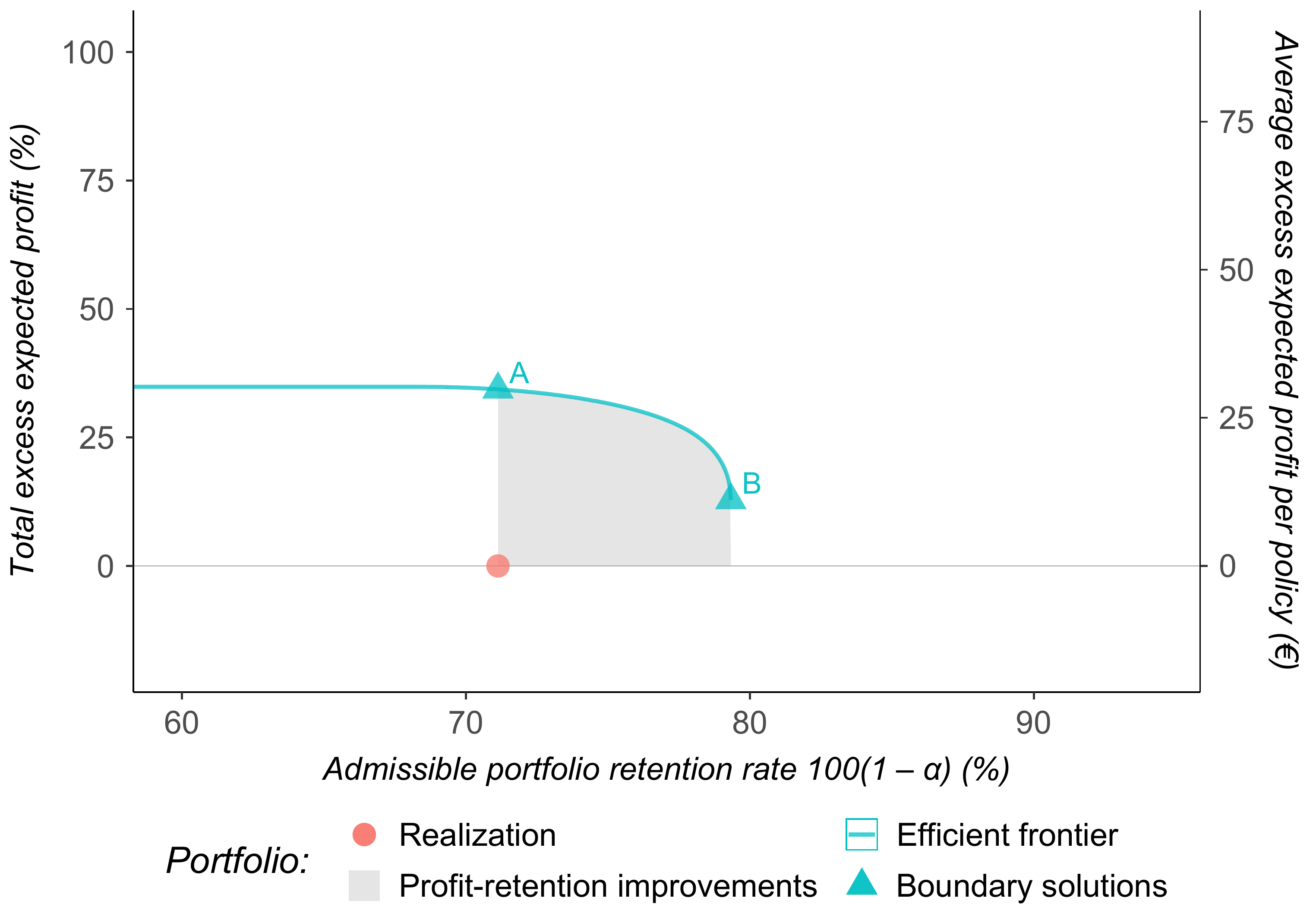}&
            \includegraphics[width=0.45\textwidth]{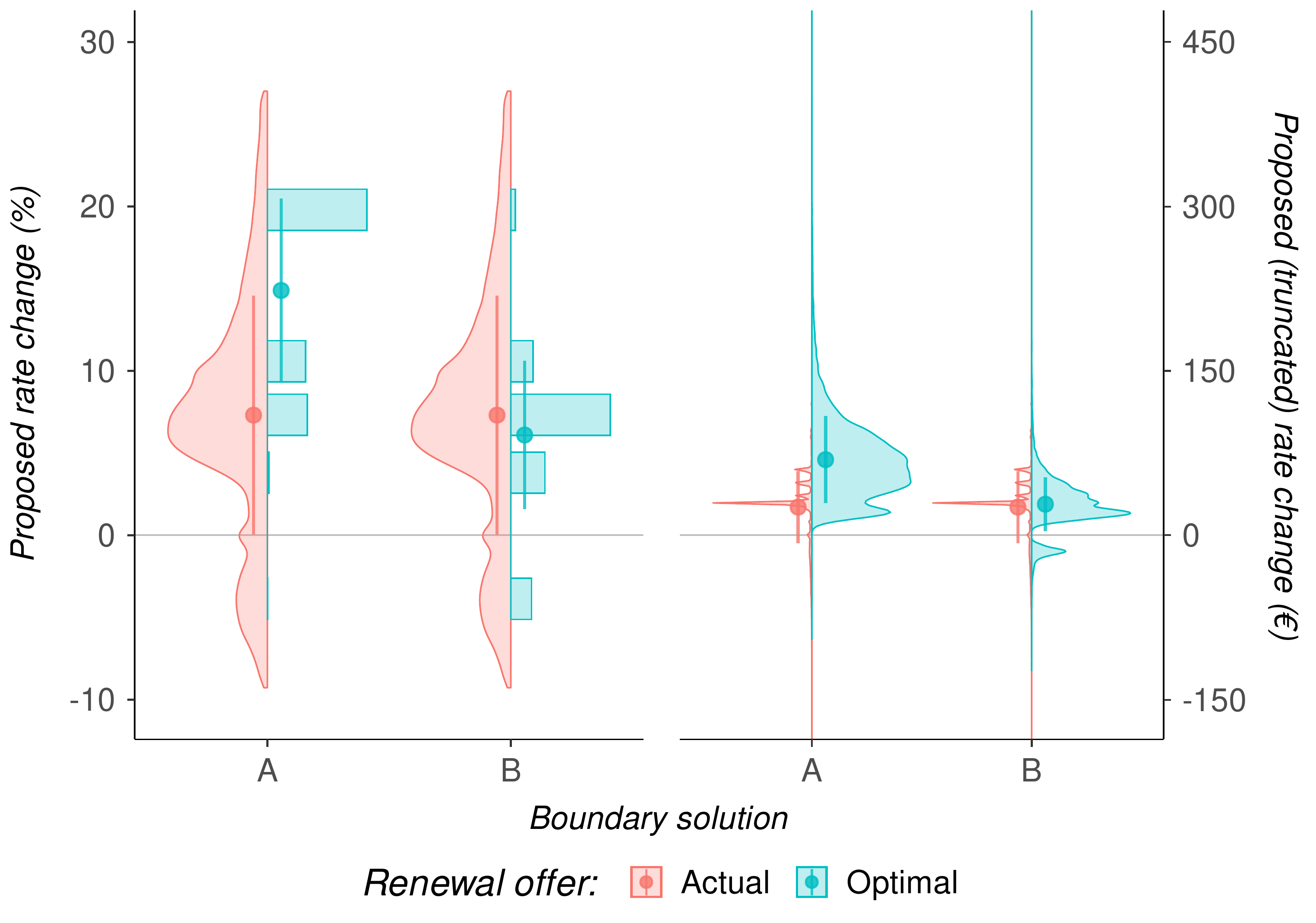}
        \end{tabular}\vspace{-6pt}
        \caption{Expected profit efficient frontier solutions for restricted continuous dose-response function}
        \label{Figure_9c}
    \end{subfigure}\vspace{-3pt}
    \caption{Expected profit and portfolio retention efficient frontier (left) and proposed rate changes in its boundary solutions (right) with discrete rate changes without multiple imputation (panel (a)) and with XGBoost (panel (b)), and continuous rate changes restricted to the five categorical rate change medians (panel (c)) in automobile insurance. We assume here that insurers can always increase customer churn without losing profit in a horizontal frontier.}
	\label{Figure_9}
\end{figure}

Based on these customer price sensitivities and their underlying drivers, we can identify what premia will optimize insurers' expected profit on a portfolio. A low premium may, for instance, lead to less customer churn but also to insufficient funds to cover the risk on the policies, whereas a high premium may yield more profit but also more customer churn. To show this trade-off between a portfolio's customer churn and profit potential, we therefore deduce the efficient frontier by numerically optimizing
\begin{equation*}
    \max_{\{t_{i}\}_{i = 1}^{N} \in \mathcal{T}^{N}} \left\{ \sum_{i = 1}^{N} \left(1 - \hat{Y_i}(t_{i}) \right) \left( P_{i} - E_{i} \right) \right\} \quad \textrm{subject to} \quad \frac{1}{N} \sum_{i = 1}^{N} \hat{Y_i}(t_{i}) \leq \alpha
\end{equation*}
for every policy $i$ in the portfolio in terms of the potential rate change $t_{i}$ and various levels of $\alpha$. In other words, we select the renewal premium $P_{i} = P^{\textrm{Old}}_{i} (1 + t_{i})$ that maximizes expected earnings after one year and after expenses $E_{i}$ associated with the policy and its level of risk, and limit the overall churn rate to at most $\alpha$, similar to \citet{guelman2014}. Note that $t_{i}$ is restricted to the medians $\{-3.87\%, 3.80\%, 7.32\%, 10.58\%, 19.79\% \}$ of the five rate change categories in the discrete framework, while $t_{i}$ can take on any value in the range of observed treatment doses $[-9.28\%, 27.01\%]$ in the continuous framework. Solving this optimization problem using Linear and Non-Linear Programming techniques, we obtain an approximation of the efficient frontier for the two frameworks in \Cref{Figure_8a}. In \Cref{Figure_8b} we display their boundary solutions, or the rate changes that maximize expected profit with exceeding the realized portfolio churn rate (A) and minimize expected churn without earning less than the realized portfolio profit (B). We additionally show the efficient frontiers and boundary solutions for the discrete approach without multiple imputation and with XGBoost, and the continuous approach when restricting the potential rate changes to the five categorical medians in \Cref{Figure_9}.

While the two treatment approaches propose similar rate changes in their boundary solutions, they do lead to somewhat different efficient frontiers. According to the discrete approach in \Cref{Figure_8}, for instance, we cannot achieve a lower expected churn rate on our portfolio than $26.20\%$ (B) without earning less than the realized portfolio profit, but can expect to earn from $0.73\%$ (B) up to $19.63\%$ (A) additional profit without exceeding the realized churn rate of $29.88\%$ (A). Although qualitatively similar to \citet{guelman2014}, the efficient frontier turns out to improve considerably with XGBoost and it decreases only slightly without multiple imputation in \Cref{Figure_9}. Not surprisingly, the continuous approach in \Cref{Figure_8} therefore indicates that the total excess expected profit can increase up to $102.07\%$ (A) at a portfolio churn rate of only $19.65\%$ (A) and that the realized total profit on our portfolio can also be attained at an expected churn rate of merely $7.40\%$ (B). The total profit and portfolio retention rate are expected to increase substantially more in the continuous approach not only due to XGBoost but also since it can distinguish between the rate changes within each rate change interval and is better able to describe the let sleeping dogs lie effect. When we restrict the potential rate changes to the five categorical rate change medians, we, for instance, indeed find a substantially lower efficient frontier in \Cref{Figure_9}, one that is even slightly lower than in the discrete XGBoost approach due to its lower balance. The (unrestricted) continuous approach thus appears to be able to compensate for this lower balance and can further increase the total profit expected in the discrete approach, while simultaneously decreasing the expected churn rate. These portfolio improvements are also denoted by the grey area, or the set of all potential portfolios that are expected to increase the total profit and/or decrease the customer churn relative to the realized portfolio, in \Cref{Figure_8a}. Nonetheless, the boundary solutions from the two approaches agree that larger rate changes maximize total expected profit and thus offset any increases in expected customer churn (A) and that considerably lower rate changes are needed to limit expected customer churn (B). More importantly, both approaches identify a vast profit potential for our portfolio, without having to increase churn and while even allowing for substantially less churn in the continuous approach.

\subsection{Multi-period renewal optimization} \label{Section4.3}

\begin{figure}[t!]
    \centering
    \begin{subfigure}{\textwidth}
        \centering
        \begin{tabular}{c c}
            \centering
            {\footnotesize\underline{\hspace{18.5mm}\textbf{Discrete$_{_{}}$}\hspace{18.5mm}}} & {\footnotesize\underline{\hspace{16.0mm}\textbf{Continuous$_{_{}}$}\hspace{16.0mm}}} \vspace{5pt}\\
            \centering
            \includegraphics[width=0.45\textwidth]{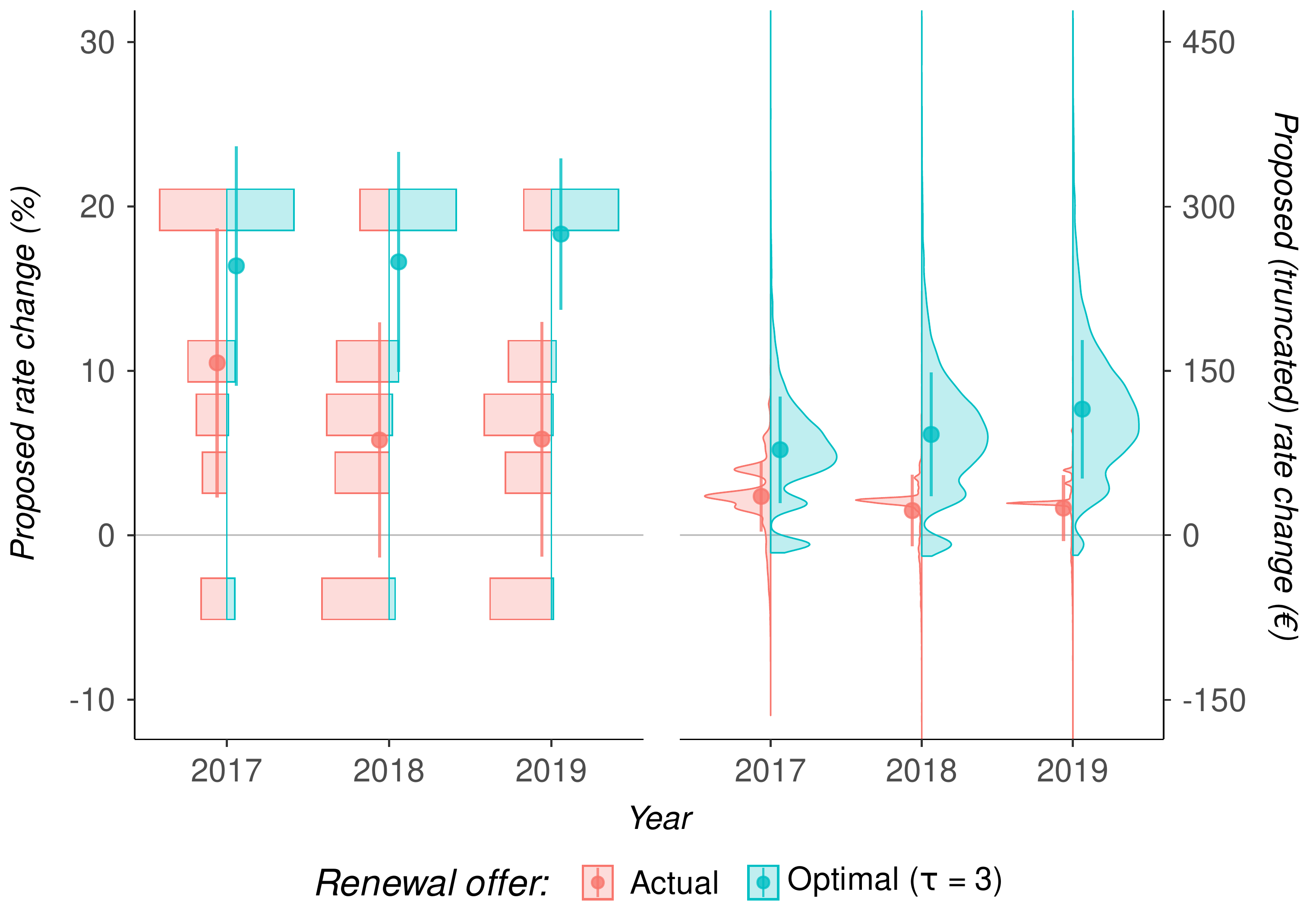}&
            \includegraphics[width=0.45\textwidth]{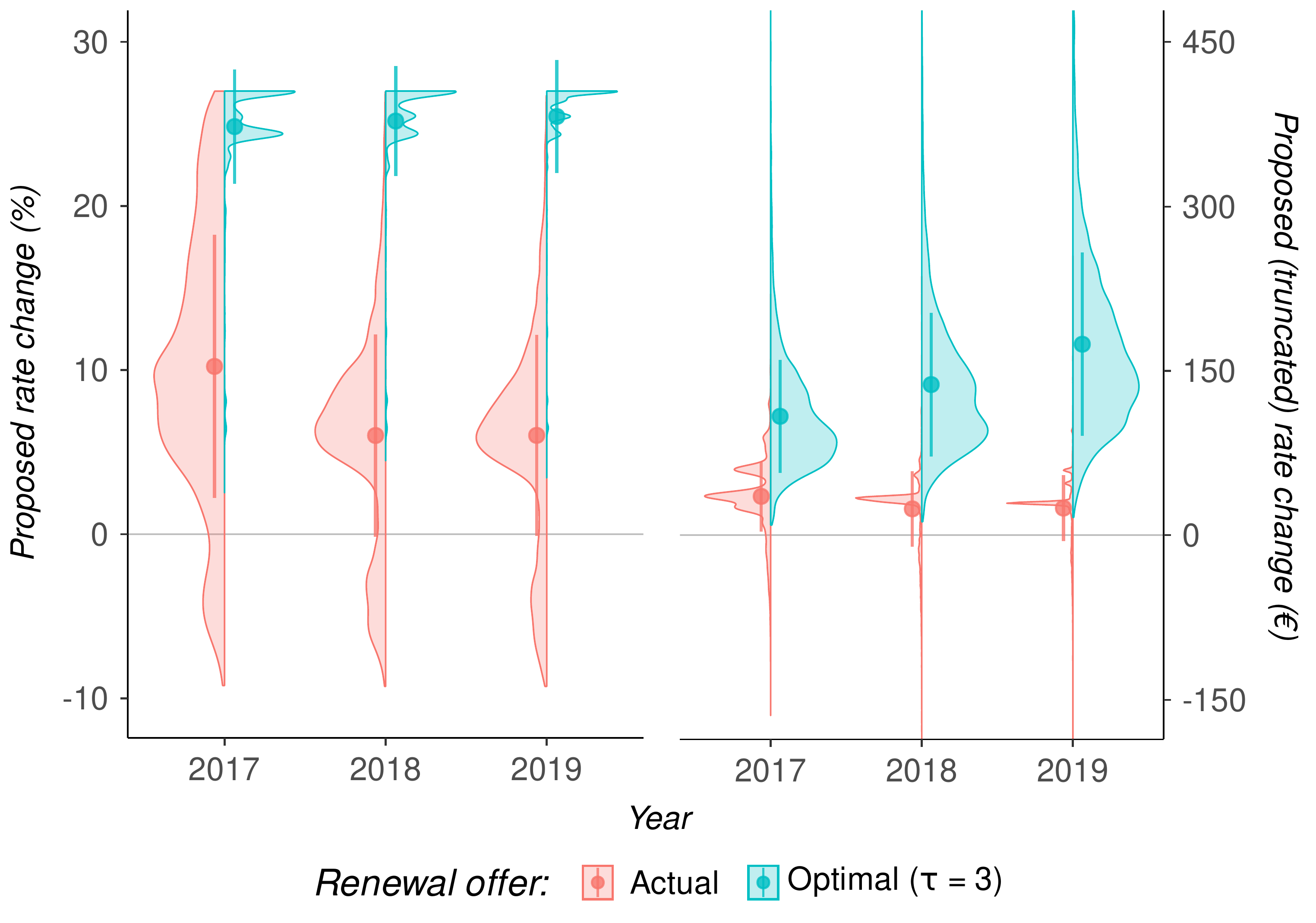}
        \end{tabular}\vspace{-6pt}
        \caption{Proposed rate changes for a forecasting horizon of $\tau = 3$ years}
        \label{Figure_10a}
    \end{subfigure}
    
    \vspace{3pt}
    
    \begin{subfigure}{\textwidth}
        \centering
        \begin{tabular}{c c}
            \centering
            \includegraphics[width=0.45\textwidth]{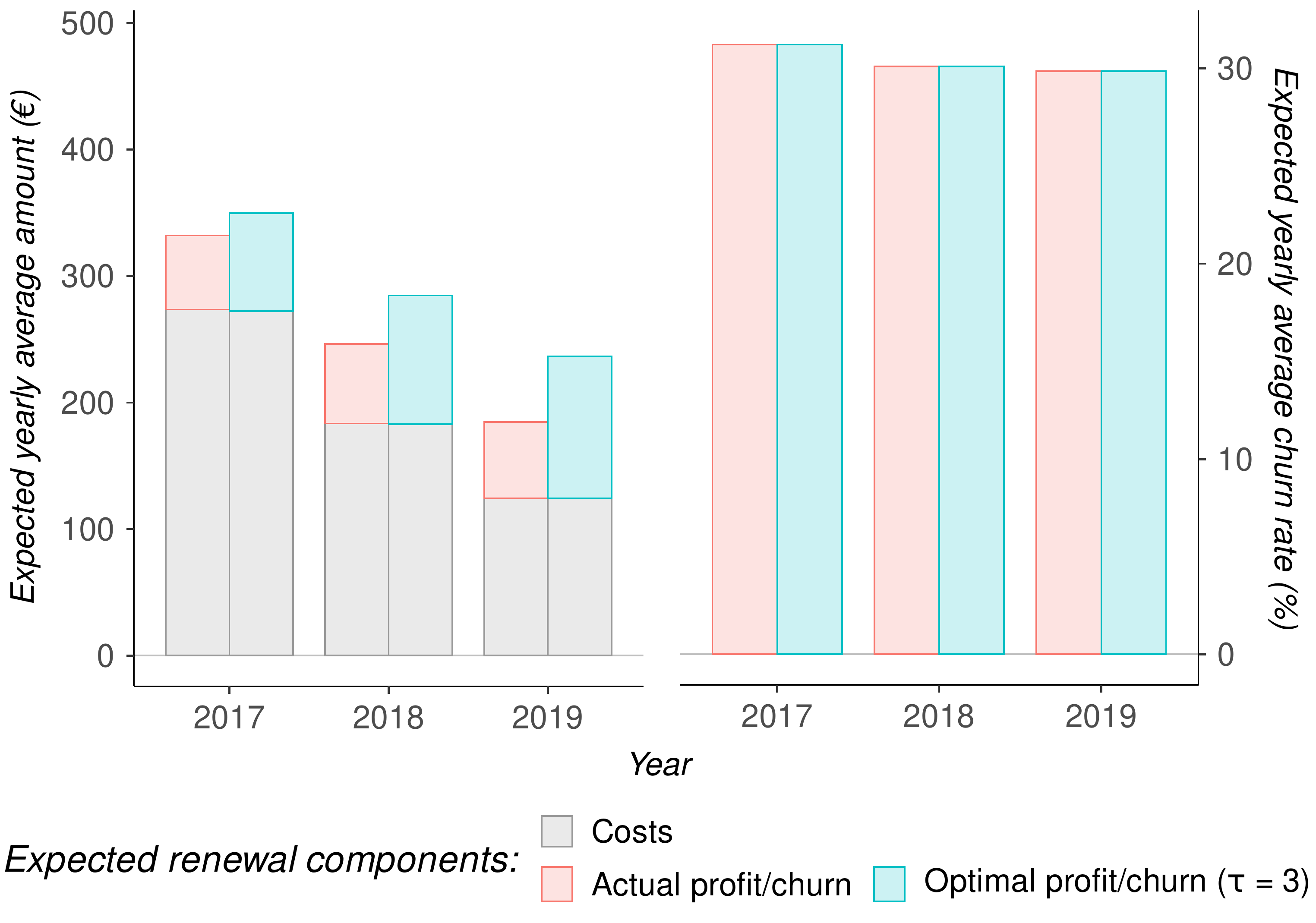}&
            \includegraphics[width=0.45\textwidth]{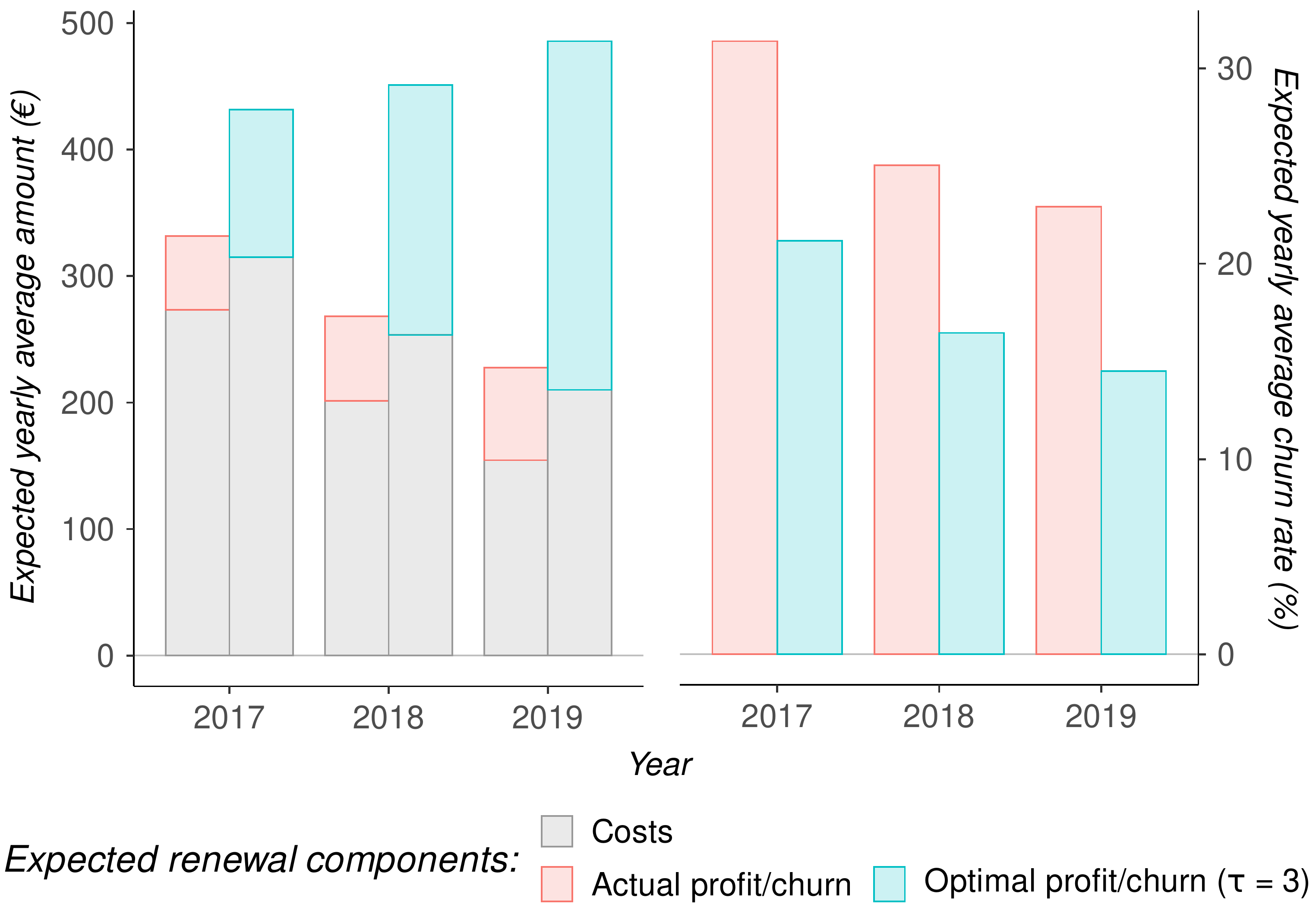}
        \end{tabular}\vspace{-6pt}
        \caption{Expected yearly average premia and churn rates for a forecasting horizon of $\tau = 3$ years}
        \label{Figure_10b}
    \end{subfigure}\vspace{-3pt}
    \caption{Proposed rate changes (panel (a)) and expected yearly average premia and churn rates (panel (b)) for $\tau = 3$ consecutive renewals with discrete (left) and continuous (right) rate changes in automobile insurance.}
	\label{Figure_10}
\end{figure}

\begin{figure}[t!]
    \centering
    \begin{subfigure}{\textwidth}
        \centering
        \begin{tabular}{c c}
            \centering
            {\footnotesize\underline{\hspace{6.875mm}\textbf{Proposed rate changes$_{_{}}$}\hspace{6.875mm}}} & {\footnotesize\underline{\hspace{5.75mm}\textbf{Expected yearly premia$_{_{}}$}\hspace{5.75mm}}} \vspace{5pt}\\
            \centering
            \includegraphics[width=0.45\textwidth]{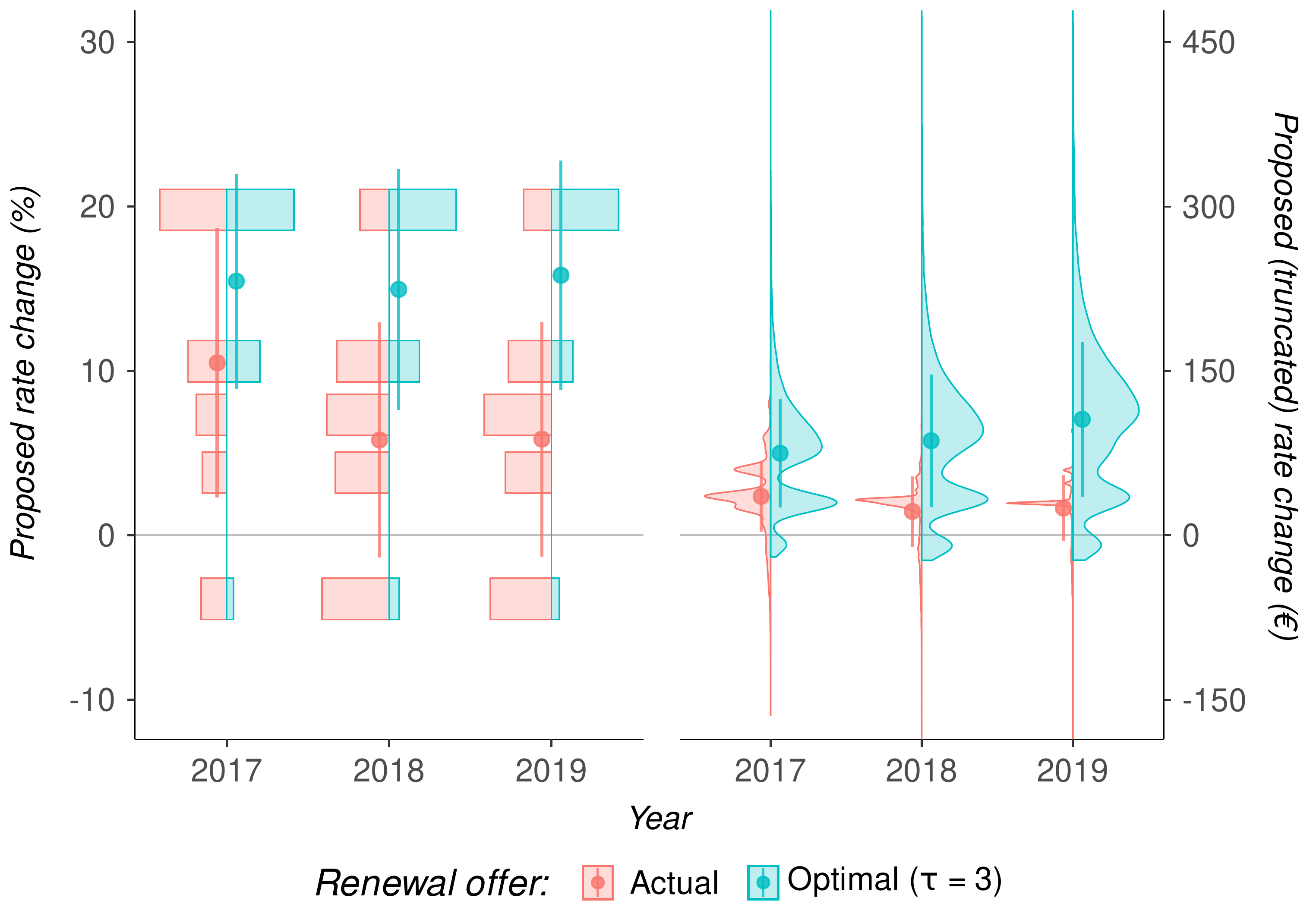}&
            \includegraphics[width=0.45\textwidth]{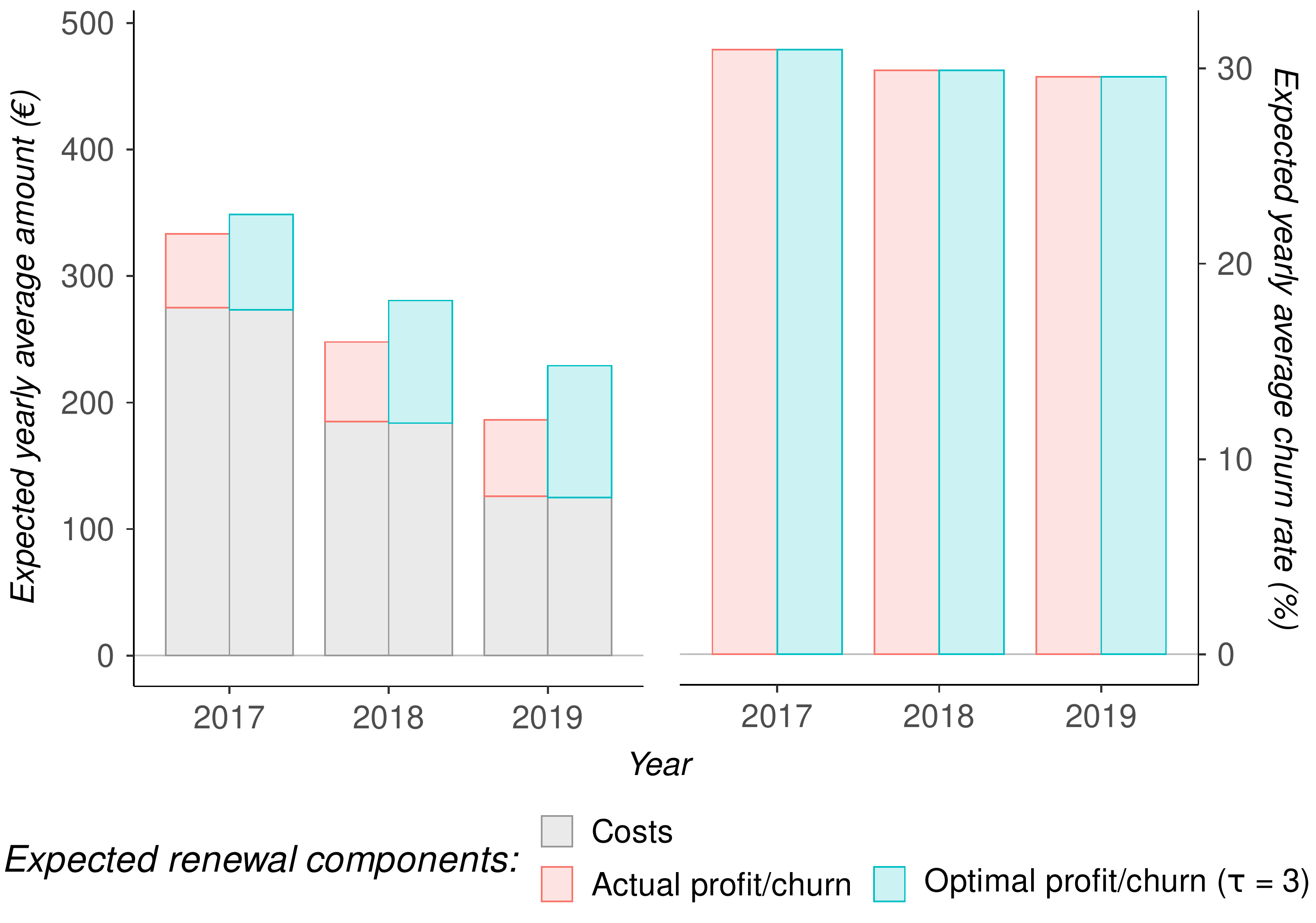}
        \end{tabular}\vspace{-6pt}
        \caption{Proposed rate changes and expected yearly premia for discrete global response model without multiple imputation}
        \label{Figure_11a}
    \end{subfigure}
    
    \vspace{3pt}
    
    \begin{subfigure}{\textwidth}
        \centering
        \begin{tabular}{c c}
            \centering
            \includegraphics[width=0.45\textwidth]{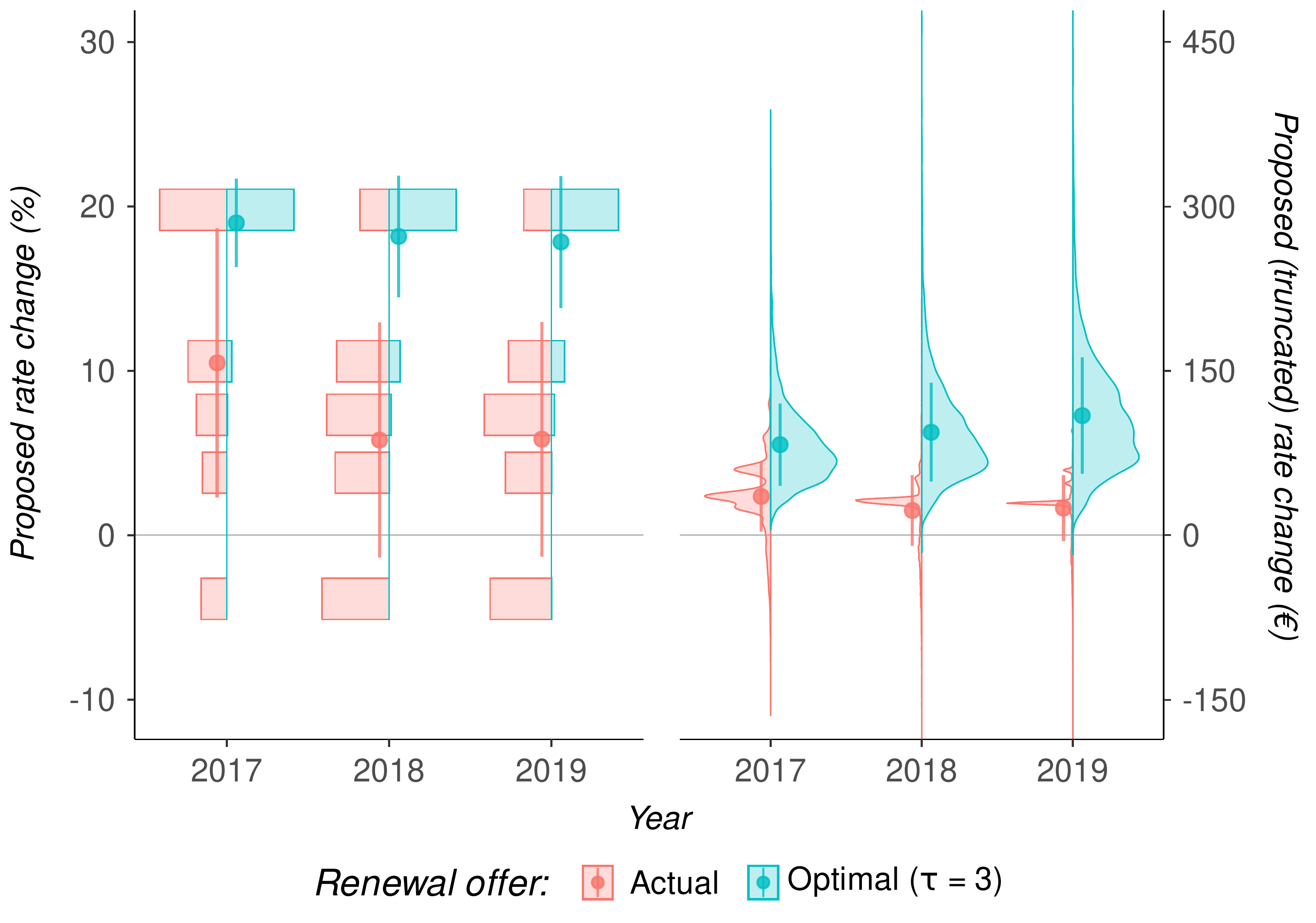}&
            \includegraphics[width=0.45\textwidth]{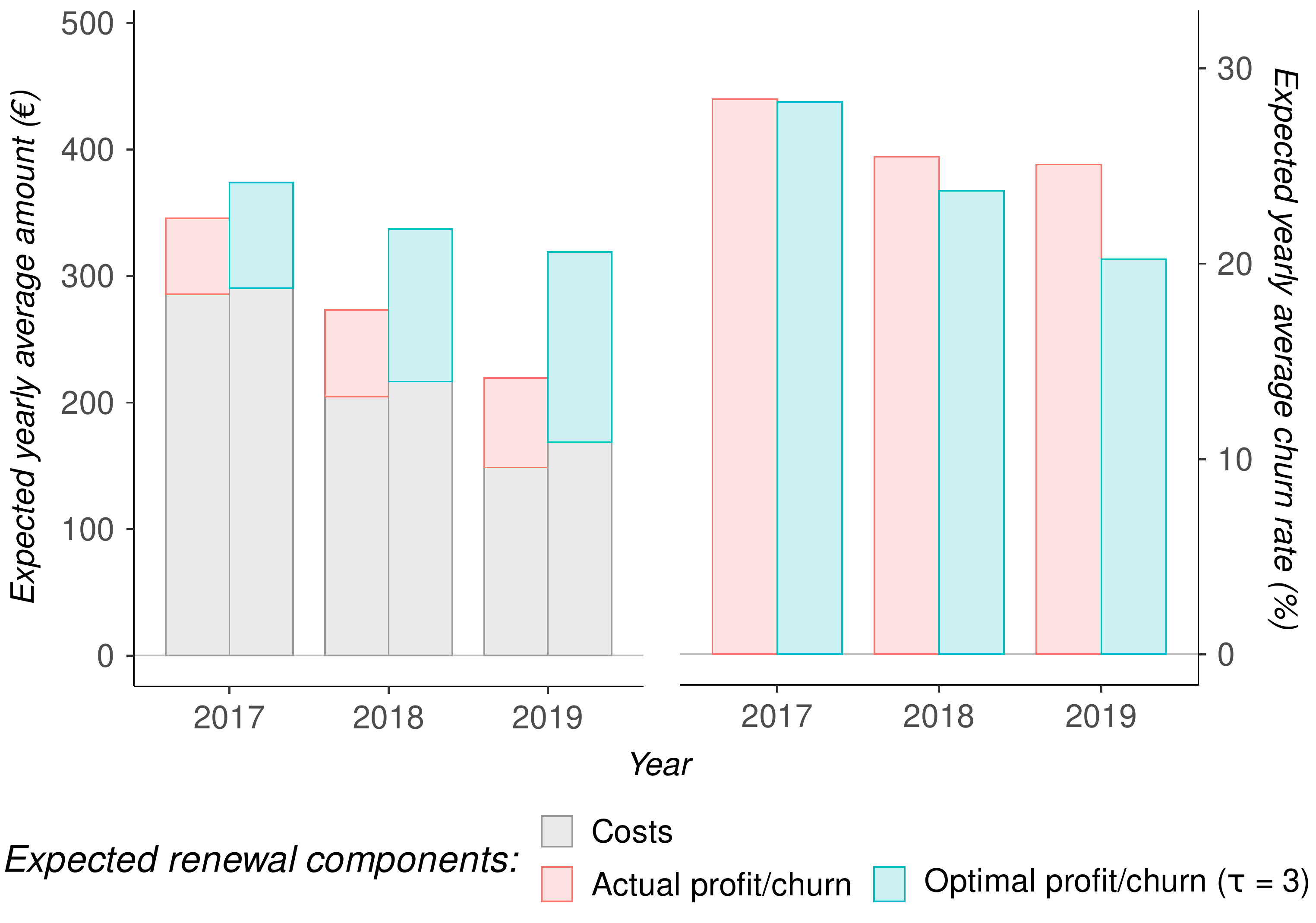}
        \end{tabular}\vspace{-6pt}
        \caption{Proposed rate changes and expected yearly premia for discrete global response model with XGBoost}
        \label{Figure_11b}
    \end{subfigure}
    
    \vspace{3pt}
    
    \begin{subfigure}{\textwidth}
        \centering
        \begin{tabular}{c c}
            \centering
            \includegraphics[width=0.45\textwidth]{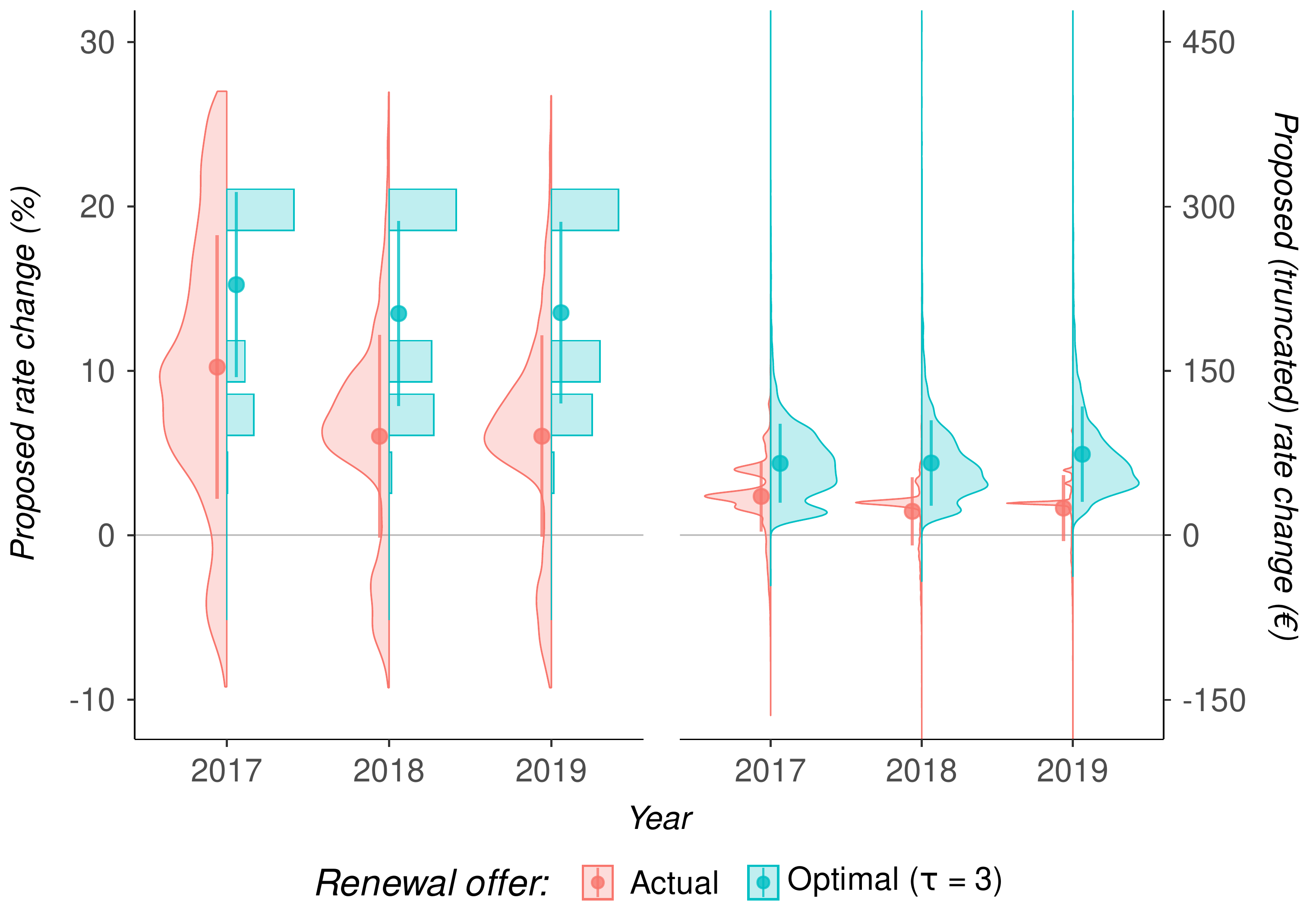}&
            \includegraphics[width=0.45\textwidth]{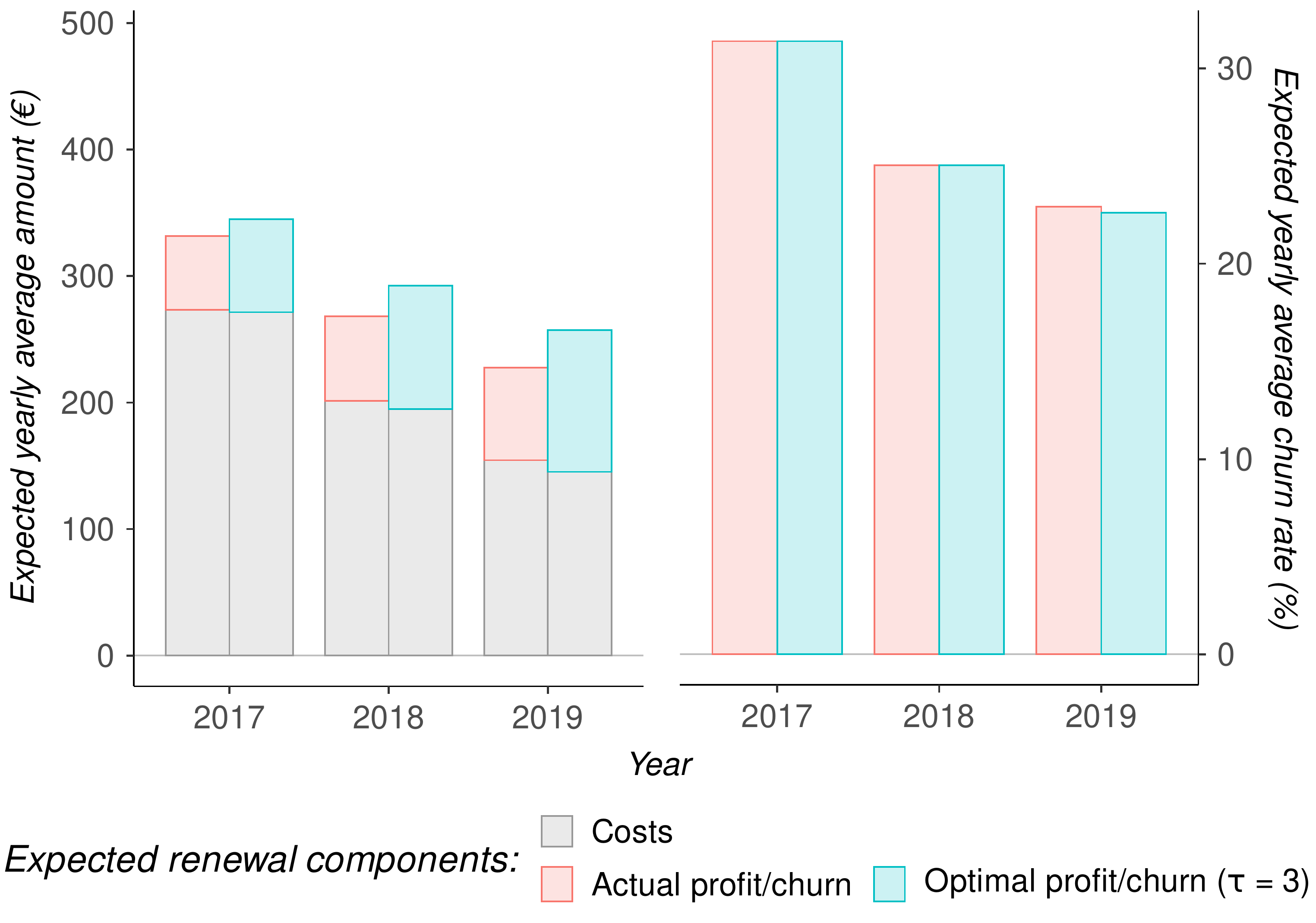}
        \end{tabular}\vspace{-6pt}
        \caption{Proposed rate changes and expected yearly premia for restricted continuous dose-response function}
        \label{Figure_11c}
    \end{subfigure}\vspace{-3pt}
    \caption{Proposed rate changes (left) and expected yearly average premia and churn rates (right) for $\tau = 3$ consecutive renewals with discrete rate changes without multiple imputation (panel (a)) and with XGBoost (panel (b)), and continuous rate changes restricted to the five categorical rate change medians (panel (c)) in automobile insurance.}
	\label{Figure_11}
\end{figure}

While the efficient frontiers clearly demonstrate the portfolio's vast profit potential, they do not explicitly account for temporal feedback. A low initial renewal offer may, for instance, increase a policy's future competitiveness and decrease its future customer churn at the expense of lower future profits, whereas a high initial renewal offer may increase its future profits at the expense of lower future competitiveness and higher future customer churn. A renewal offer may thus affect not only customers' current response but all of their future responses as well. To account for this temporal feedback, we therefore cast the previous optimization problem into a multi-period setting and numerically optimize
\begin{equation*}
    \max_{\{t_{i,j}\}_{i = 1, j = 1}^{N, \tau} \in \mathcal{T}^{N \tau}} \left\{ \sum_{i = 1}^{N} \sum_{j = 1}^{\tau} \left( \prod_{h = 1}^{j} \left[1 - \hat{Y_i}(t_{i,1}, \dots, t_{i,h}) \right] \right) \left( P_{i,j} - E_{i,j} \right) \right\}
\end{equation*}
subject to
\begin{equation*}
    \frac{1}{N} \sum_{i = 1}^{N} \hat{Y_i}(t_{i,1}, \dots, t_{i,j}) \leq \alpha_{j} \quad \textrm{for} \quad j = 1, \dots, \tau
\end{equation*}
for every policy $i$ with at least $\tau$ renewals in terms of the potential rate changes $t_{i,1}, \dots, t_{i,\tau}$. In other words, we select the renewal premia $P_{i,j} = P^{\textrm{Old}}_{i,1} \prod_{k = 1}^{j} (1 + t_{i,k})$ that maximize expected earnings after $\tau$ years and after yearly expenses $E_{i,j}$, and limit the overall churn rate to at most $\alpha_{j}$ in year $j$. Note that $t_{i,j}$ is again restricted to the medians of the five rate change categories in the discrete framework, while $t_{i,j}$ can take on any value in the range of observed treatment doses in the continuous framework. Using Linear and Non-Linear Programming techniques, we solve this optimization problem for a forecasting horizon of $\tau = 3$ years with $10{,}073$ customers and adopt the average churn rate expected for the actual renewal offers for the admissible overall churn rate each year, or $\alpha_{j} = \frac{1}{N} \sum_{i = 1}^{N} \hat{Y_i}(T_{i, 1}, \dots, T_{i, j})$ for $j = 1, \dots, \tau$. The rate changes and expected profits resulting from this routine are presented in \Cref{Figure_10a,Figure_B.4} and \Cref{Figure_10b}, respectively. These results are also shown for the discrete approach without multiple imputation and with XGBoost, and the continuous approach when restricting the potential rate changes to the five categorical medians in \Cref{Figure_11,Figure_B.5}.

While the two treatment approaches primarily appear to charge as much premium as possible, they do display some more subtle and intuitive results. From \Cref{Figure_10}, for instance, we observe that the proposed rate changes are slightly lower in the first period(s) and lead to substantially less profit compared to the later period(s), especially in the continuous framework. Lower rate changes are actually selected in the first period(s) to limit their indirect effect on the future demand and profits, since they affect a policy's competitiveness in the market in later periods. Moreover, while the actual renewal offers are typically much lower, the absolute optimal offers seem to suggest that an upward shift in the actual offers will substantially increase the insurer's profits. The majority of policies therefore appear underpriced in the portfolio, either to limit customer churn or simply because customers are more risky and thus more costly to insure than anticipated by the insurer. The continuous approach additionally leads to substantially higher expected profits with much lower overall expected churn rates than the discrete approach, since it can again distinguish between the rate changes within each rate change interval and is better able to describe the let sleeping dogs lie effect. Similarly, we observe in \Cref{Figure_11} that the expected profits again increase considerably in the discrete approach with XGBoost, more than in the restricted continuous approach and with somewhat lower expected churn rates. Without multiple imputation the expected profits turn out to decrease only slightly. Finally, we observe a slight tendency in \Cref{Figure_B.4,Figure_B.5} to propose lower rate changes for customers with a lower level of risk and whose policies are relatively competitive in the market. The optimization problem thus seems to encourage less profitable risks to churn, while remaining to some extent competitive in the market. As such, the optimal renewal offers appear to target customers' profitability by focusing on customers who are less sensitive to premium changes and simultaneously not too costly to insure.

The indirect demand effect thus seems to have a large impact on an insurer's profitability. Even though the discrete approach proves very successful at inferring customers' counterfactual responses, it seems to be rather crude and less flexible in estimating this indirect effect, especially when relying on the traditional logistic regression for the global response model. The continuous approach, on the other hand, allows for more flexibility in the let sleeping dogs lie effect and the customer price sensitivities, and it appears substantially more profitable for (multi-period) renewal optimization. Nonetheless, both approaches lead to a vast profit potential and identify a policy's competitiveness and a customer's level of risk as essential drivers of a customer's price sensitivity. It therefore seems vital for an insurer to account for these components and to incorporate offers of its competitors in the market in the design of its renewal premia. 
\section{Conclusion} \label{Section5}

In this paper, we have presented and applied a causal inference framework for measuring customer price sensitivities and deducing (multi-period) optimal premium renewal offers. While renewal premia are typically based on a cost-based pricing strategy in non-life insurance, we have employed a demand-based method in this paper to account for their indirect effect on a customer's willingness to stay. As such, we have extended the discrete treatment framework of \citet{guelman2014} by Extreme Gradient Boosting and multiple imputation, and have introduced a continuous treatment framework based on Extreme Gradient Boosting to the insurance literature. Finally, we have included the competitiveness of each policy in these frameworks to account for the highly competitive nature of insurance markets.

In our application of the two causal inference approaches, we considered a Dutch automobile insurance portfolio and six of the largest competitors in the market. Contrary to \citet{guelman2014}, this portfolio displayed a large inflection point in customer churn at small rate changes, supporting the common practice among insurers of not changing renewal premia (by too much) and letting sleeping dogs lie, to avoid raising price awareness in customers for only a small rate change. Using this portfolio, we conducted (generalized) propensity score matching for the quintiles of the observed renewal offers to infer customers' counterfactual responses and estimated a global response model to describe their potential churning behavior. We compared XGBoost to random forests, neural networks, or GAMs for the (generalized) propensity score model and found that complex, non-linear techniques, more so than a GAM, are in fact necessary to adequately capture the treatment assignment mechanism, with XGBoost being the most capable method. Furthermore, while the discrete approach achieved remarkably better balance in each of the five rate change intervals, it converged to the continuous approach with noticeably more variance as the number of intervals increased, since the latter can essentially be seen as a limiting case of the former. The continuous approach managed to exploit more information by avoiding any discretization or aggregation and allowed for far more flexibility in capturing the let sleeping dogs lie effect and the price sensitivities. In case of a global response model based on the traditional logistic regression, similar to \citet{guelman2014}, the discrete approach led to an efficient frontier with noticeably more expected total profit and portfolio retention than realized. However, a global response model with XGBoost was able to describe the let sleeping dogs lie effect much better and improved the efficient frontier considerably, whereas multiple imputation only led to a small improvement. The continuous approach, on the other hand, seemed to compensate for its lower balance with its ability to distinguish between the rate changes in each rate change interval and it improved the efficient frontier substantially more. Nonetheless, both frameworks identified a customer's level of risk and in particular a policy's competitiveness as fundamental drivers of the price sensitivities. A multi-period renewal optimization additionally revealed a small temporal feedback effect of previous rate changes on a policy's future competitiveness and thus demand, and indicated a vast profit potential, especially in the continuous approach. The two treatment frameworks presented in this paper, in particular the approach with continuous treatments, therefore appear promising for (multi-period) renewal pricing strategies of non-life insurers in competitive markets.

While this paper has primarily focused on logistic GLMs and to some extent on XGBoost for the discrete global response model, we can also adopt (other) techniques from the field of machine learning for this. These more complex methods may, for instance, allow for more flexible price sensitivities and better identified optimal renewal offers, similar to XGBoost. Alternatively, customers' risk factors can be introduced in the matching procedure to further increase the balance in each rate change interval. An even more interesting avenue for future research may be to adopt an ensemble of XGBoost, (causal) random forests, (deep) neural networks, and support vector machines in the two treatment frameworks next to the individual implementation of these methods. This combined method may, in turn, be able to improve our description of the indirect demand effect in automobile insurance. 

\section*{Acknowledgments}

The author gratefully acknowledges financial support from VIVAT. Any errors made or views expressed in this paper are the responsibility of the author alone. In addition, the author would like to thank Michel Vellekoop, Noud van Giersbergen, Peter Boswijk, and two anonymous referees for their valuable comments on this paper. 

\begingroup
    \setlength{\bibsep}{
    \if1\double
        7.5pt
    \else
        0.0pt
    \fi}
    \def\bibfont{\small}
    \bibliography{References_Paper} 
\endgroup

\appendix
{\renewcommand{\thesection}{Appendix \Alph{section}}
\renewcommand{\theequation}{\Alph{section}.\arabic{equation}}
\counterwithin*{equation}{section}
\renewcommand{\thefigure}{\Alph{section}.\arabic{figure}}
\counterwithin*{figure}{section}
\renewcommand{\thetable}{\Alph{section}.\arabic{table}}
\counterwithin*{table}{section}
\renewcommand{\thealgocf}{\Alph{section}.\arabic{algocf}}
\counterwithin*{algocf}{section}
\section{Extreme Gradient Boosting for decision trees} \label{Appendix_A}

Extreme Gradient Boosting, or XGBoost, has originally been developed by \citet{chen2016} as a generalized form of Gradient Boosting. It is typically applied iteratively to decision trees and relies on the ensemble method of boosting. The general principle behind boosting is that a combination of many weak learners yields a more powerful predictor than a single strong learner on its own \citep{schapire1990,freund1997}. These weak learners $h_{\ell}(\cdot)$ are estimated by minimizing a given loss function $\mathcal{L}(\cdot)$ over the combined predictor $f_{L}(\bm{X}_i) = \sum_{\ell = 0}^{L} h_{\ell}(\bm{X}_i)$ with risk factors $\bm{X}_i$. While the first learner $h_0(\cdot)$ attempts to optimally predict the responses $Y_i$ by minimizing $\mathcal{L}(Y_i, h_0(\bm{X}_i))$ directly, every subsequent learner tries to improve the previous learners in areas where they performed poorly by minimizing $\mathcal{L}(Y_i, f_{\ell - 1}(\bm{X}_i) + h_{\ell}(\bm{X}_i))$ over $h_{\ell}(\cdot)$. These weak learners usually perform only marginally better than random guessing and are formed through decision trees.

In brief, a decision tree partitions the predictor space using successive binary splits and predicts a constant for each distinct region. As such, they are able to identify highly complex interactions between predictors through these successive splits. Formally, decision trees $h(\cdot)$ are given by
\begin{equation*}
    h(\bm{X}_i) = \sum_{j = 1}^{J} \hat{\omega}_{R_{j}} \mathbbm{1}\left[ \bm{X}_i \in R_j \right],
\end{equation*}
where the predictor space $R$ has been partitioned into $J$ non-overlapping regions $R_1, \dots, R_J$ and $\hat{\omega}_{R_j}$ denotes the constant prediction for region $R_j$ \citep{chen2016,henckaerts2020}. Since it may be infeasible in practice to consider all possible partitions of the predictor space, algorithms have been developed that take a more greedy approach by considering only a single binary split at a time. Among these approaches are, for instance, the popular Classification And Regression Tree (CART) algorithm of \citet{breiman1984} and the C4.5 algorithm of \citet{quinlan1993}. These decision trees have been successfully incorporated in several ensemble methods as well, with Adaptive Boosting, or AdaBoost, of \citet{freund1997} being perhaps one of the most popular algorithms.

Although the concept of boosting is intuitively appealing, its loss function is originally restricted to either an exponential error or a negative Binomial log-likelihood. In addition, boosting can be prone to overfitting when considering a large amount of weak learners. \citet{friedman2001} has therefore introduced Gradient Boosting to allow for an arbitrary differentiable loss function and to include a shrinkage parameter in the boosting algorithm. \citet{friedman2002} has extended this approach further by injecting randomness in the learning process, leading to Stochastic Gradient Boosting. The resulting algorithm generalizes better to unseen data and has been applied in numerous studies, both within and outside of non-life insurance (see, e.g., \citet{lemmens2006}; \citet{burez2009}; \citet{guelman2012}; \citet{sung2012}; \citet{guelman2014}; \citet{zhang2017}; \citet{spedicato2018}; \citet{henckaerts2020}).

(Stochastic) Gradient Boosting essentially follows a two-step regularization procedure, shown for decision trees in \Cref{Algorithm_A.1}. In the first step, or lines $4$\,--\,$7$ of the algorithm, a decision tree of depth $d$ with at least $\nu$ instances in each node is fitted to the current pseudo-residuals $\rho_{\ell, i}$, or the negative gradient of the loss function evaluated at the current model fit, to identify the areas where the current predictor performs poorly. Given the optimal regions $R_{\ell, j}$ of this decision tree, its constants $\omega_{\ell, j}$ are estimated in lines $8$\,--\,$10$ by minimizing the loss function given the current predictor. Moreover, we draw a random sample of size $\delta N$ with $0 < \delta \leq 1$ from the data without replacement in line $3$ before calculating the pseudo-residuals to introduce randomness in the learning process. The regularization parameter $\eta$ is additionally used in line $11$ to shrink the update of the new predictor and to reduce overfitting. In practice, it is often tuned by a form of cross-validation, together with $d$, $\nu$, and $\delta$.

\begin{table}[t!]
\vspace{-3pt}
\centerline{\scalebox{0.80}{
\centering
\begin{minipage}{1.25\textwidth}
\centering
\begin{algorithm}[H]
    \caption{(Stochastic) Gradient Boosting for decision trees} \label{Algorithm_A.1}
    \SetAlgoLined
    Initialize $f_0(\bm{X}) = \arg \min_{\omega} \sum_{i = 1}^{N} \mathcal{L}(Y_i, \omega)$\\
    \For{$\ell = 1, \dots, L$}{
        Draw a random sample of size $\delta N$ without replacement from the data\\
        \For{$i = 1, \dots, \delta N$}{
            Calculate the pseudo-residuals $\rho_{\ell, i} = - \left[ \frac{\partial \mathcal{L}(Y_i, f(\bm{X}_i))}{\partial f(\bm{X}_i)} \right]_{f = f_{\ell - 1}}$\\
        }
        Fit a decision tree of depth $d$ with at least $\nu$ instances in each node to these pseudo-residuals and obtain its regions $R_{\ell, j}$ for $j = 1, \dots, J_{\ell}$\\
        \For{$j = 1, \dots, J_{\ell}$}{
            Estimate the constants $\hat{\omega}_{\ell, j} = \arg \min_{\omega} \sum\limits_{i: \bm{X}_i \in R_{\ell, j}} \mathcal{L}(Y_i, f_{\ell - 1}(\bm{X}_i) + \omega)$
        }
        Update $f_{\ell}(\bm{X}) = f_{\ell - 1}(\bm{X}) + \eta \sum_{j = 1}^{J_{\ell}} \hat{\omega}_{\ell, j} \mathbbm{1}\left[\bm{X} \in R_{\ell, j}\right]$\\
    }
    Output $\hat{f}(\bm{X}) = f_{L}(\bm{X})$
\end{algorithm}
\end{minipage}
}}\vspace{-4pt}
\end{table}

XGBoost now poses a more recent and much faster extension of this boosting methodology. It has originally been developed for decision trees and employs a second-order Taylor approximation of the loss function to incorporate its curvature $g_{\ell, i}$ in the minimization procedure (see \Cref{Algorithm_A.2}). Compared to Stochastic Gradient Boosting, it includes a penalty $\gamma$ on the complexity of the decision trees and allows for random selection of $\kappa K$ predictor variables with $0 < \kappa \leq 1$. It additionally allows for sparsity constraints on the coefficients by penalizing their $L_1$ and squared $L_2$ norm by $a$ and $\lambda$, respectively. The resulting algorithm became one of the most popular techniques used in the online machine learning competitions of Kaggle shortly after its introduction. Due to its increasing popularity and flexibility, XGBoost has also been applied in various studies in non-life insurance using, for instance, the package \texttt{xgboost} in \texttt{R} (see, e.g., \citet{spedicato2018}; \citet{pesantez-narvaez2019}; \citet{ferrario2019}; \citet{duval2019}; \citet{huang2019}).

While many loss functions $\mathcal{L}(\cdot)$ can be used for these algorithms, it depends on the application at hand which one is most suited. For customer churn prediction, for instance, the negative Bernoulli log-likelihood
\begin{equation} \label{Equation_A.1}
    \mathcal{L}(Y_i, f(\bm{X}_i)) = - \frac{1}{N} \left( Y_i \log \left( f(\bm{X}_i) \right) - (1 - Y_i) \log \left( 1 - f(\bm{X}_i) \right) \right)
\end{equation}
seems appropriate when the responses are binary, whereas the negative generalized Bernoulli, or multinoulli, log-likelihood
\begin{equation} \label{Equation_A.2}
    \mathcal{L}(T_i, f(\bm{X}_i)) = - \frac{1}{N} \sum_{c = 1}^{C} \mathbbm{1}\left[ T_i = t_{c} \right] \log\left( f_c(\bm{X}_i) \right)
\end{equation}
or the negative truncated Gaussian log-likelihood with standard Gaussian cumulative distribution function $\Phi(\cdot)$ and unbiased sample variance $\hat{\sigma}^2 = \frac{1}{N - 1} \sum_{j = 1}^{N} \left(T_j - f(\bm{X}_j) \right) ^ 2$
\begin{equation} \label{Equation_A.3}
    \mathcal{L}(T_i, f(\bm{X}_i)) = -\frac{1}{N} \left( - \frac{\ln(2\pi\hat{\sigma}) }{2} - \frac{(T_i - f(\bm{X}_i)) ^ 2}{2 \hat{\sigma}^2} - \ln\left( \Phi\left( \frac{\overline{T} - f(\bm{X}_i)}{\hat{\sigma}} \right) - \Phi\left( \frac{\underline{T} - f(\bm{X}_i)}{\hat{\sigma}} \right) \right) \right)\hspace{-2.4pt}
\end{equation}
appear more suited for (generalized) propensity score modeling when we have $C > 2$ categorical treatments or a continuum of treatments $[\mspace{2mu} \underline{T}, \overline{T} \mspace{2mu}]$, respectively. However, in case of a (generalized) propensity score model we are actually more interested in increasing the balance among the risk factors rather than predicting the true propensity score as accurately as possible. \citet{mccaffrey2004} and \citet{guelman2014} therefore argue that for the binary treatment case we should apply an early stopping rule to these propensity score models, to maximize the covariate balance. More specifically, they propose to adapt the number of decision trees to minimize the Average Standardized Absolute Mean (ASAM) difference in the risk factors of the model in cross-validation. We can extend this concept straightforwardly to the case of $C > 2$ treatment categories and $K$ risk factors by considering
\begin{equation} \label{Equation_A.4}
    \frac{1}{C} \sum_{c = 1}^{C} \left[ \frac{1}{K} \sum_{k = 1}^{K} \frac{\left| \sum_{i = 1}^{N} \frac{\mathbbm{1}\left[ T_i = t_{c} \right]}{\pi(t_{c}, \bm{X}_i)} X_{i, k} \Big/ \sum_{i = 1}^{N} \frac{\mathbbm{1}\left[ T_i = t_{c} \right]}{\pi(t_{c}, \bm{X}_i)} - \frac{1}{N} \sum_{i = 1}^{N} X_{i, k} \right|}{\sqrt{\frac{1}{N - 1} \sum_{i = 1}^{N} \left( X_{i, k} - \frac{1}{N} \sum_{i = 1}^{N} X_{i, k} \right) ^ 2}} \right]
\end{equation}
as the early stopping criterion, similar to \citet{mccaffrey2013}. In case of a continuum of treatments, we can also consider $t_{c}$ as simply a treatment dose interval. This, in turn, allows us to adopt XGBoost for causal inference in a flexible and intuitive manner.

\begin{table}[t!]
\vspace{-3pt}
\centerline{\scalebox{0.80}{
\centering
\begin{minipage}{1.25\textwidth}
\centering
\begin{algorithm}[H]
    \caption{Extreme Gradient Boosting (XGBoost) for decision trees} \label{Algorithm_A.2}
    \SetAlgoLined
    Initialize $f_0(\bm{X}) = \arg \min_{\omega} \sum_{i = 1}^{N} \mathcal{L}(Y_i, \omega)$\\
    \For{$\ell = 1, \dots, L$}{
        Draw a random sample of size $\delta N$ and $\kappa K$ risk factors without replacement from the data\\
        \For{$i = 1, \dots, \delta N$}{
            Calculate the pseudo-residuals $\rho_{\ell, i} = - \left[ \frac{\partial \mathcal{L}(Y_i, f(\bm{X}_i))}{\partial f(\bm{X}_i)} \right]_{f = f_{\ell - 1}}$ and their gradients $g_{\ell, i} = - \left[ \frac{\partial^2 \mathcal{L}(Y_i, f(\bm{X}_i))}{\partial f(\bm{X}_i)^2} \right]_{f = f_{\ell - 1}}$\\
        }
        Fit a regularized decision tree of depth $d$ with at least $\nu$ instances in each node to these pseudo-residuals and their gradients, and obtain its regions $R_{\ell, j}$ for $j = 1, \dots, J_{\ell}$\\
        Estimate the constants $\hat{\bm{\omega}}_{\ell} = \arg \min_{\bm{\omega}} \left\{ \sum_{j = 1}^{J_{\ell}} \left( \sum\limits_{i: \bm{X}_i \in R_{\ell, j}} \left[ \rho_{\ell, i} \omega_{j} + \frac{1}{2} g_{\ell, i} \omega_{j}^2 \right] + a | \omega_{j} | + \lambda \omega_{j}^2 \right) + \gamma J_{\ell} \right\}$ \\
        Update $f_{\ell}(\bm{X}) = f_{\ell - 1}(\bm{X}) + \eta \sum_{j = 1}^{J_{\ell}} \hat{\omega}_{\ell, j} \mathbbm{1}\left[\bm{X} \in R_{\ell, j}\right]$\\
    }
    Output $\hat{f}(\bm{X}) = f_{L}(\bm{X})$
\end{algorithm}
\end{minipage}
}}\vspace{-4pt}
\end{table} 
\newpage
\section{Supplementary material} \label{Appendix_B}

\begin{table}[h!]
    \caption{Sequential parameter grids and default values for random forest and neural network in discrete (continuous) treatment optimization with $10$-fold cross-validation, using the \texttt{ranger} and \texttt{nnet} package in \texttt{R}.}
    \label{Table_B.1}\vspace{-6pt}
	\centerline{\scalebox{0.80}{\begin{tabularx}{1.25\textwidth}{l X r X r X r}
	    \toprule \addlinespace[1ex] \vspace{1pt}
	    \textbf{Random forest parameter} && \multicolumn{1}{c}{\textbf{Search grid}} && \multicolumn{1}{c}{\textbf{Default}} && \multicolumn{1}{c}{\textbf{Optimal (G)PS}} \\ \hline \addlinespace[0.4ex]
		\textit{Tree-specific parameters} && && & \\ 
		\hspace{10pt}\textit{-\texttt{max.depth}} && $\{0, 1, 2, 4, \dots, 10, 25, 50\}$ &&  && $25$ ($1$) \\
		\hspace{10pt}\textit{-\texttt{min.node.size}} && $\{0, 1, \dots, 5, 10, 25, 50\}$ &&  && $0$ ($0$) \\ \hline \addlinespace[0.4ex]
		\textit{Stochastic features} && && & \\ 
		\hspace{10pt}\textit{-\texttt{sample.fraction}} && $\{0.1, 0.2, \dots, 1\}$ && $1$ && $0.1$ ($0.4$) \\
		\hspace{10pt}\textit{-\texttt{mtry}} && $\{1, 2, \dots, 8\}$ && $1$ && $4$ ($1$) \\ \hline \addlinespace[0.4ex]
		\textit{Number of trees} && && & \\
		\hspace{10pt}\textit{-\texttt{num.tree}} && $\{1, 2, \dots, 20, 30, 40, 50, 75, 100, 150, \dots, 5{,}000\}$ && $500$ && $350$ ($550$) \\ \hline \addlinespace[0.4ex]
	    \textbf{Neural network parameter} &&  &&  & \\ \hline \addlinespace[0.4ex]
		\textit{Network-specific parameters} && && & \\ 
		\hspace{10pt}\textit{-\texttt{size}} && $\{1, 2, 4, \dots, 10, 25, 50\}$ &&  && $4$ ($2$) \\
		\hspace{10pt}\textit{-\texttt{decay}} && $\{0, 0.1, 1, 10, 100, 250, 300, \dots, 1{,}000\}$ &&  && $100$ ($550$) \\ \hline \addlinespace[0.4ex]
		\textit{Number of iterations} && && & \\
		\hspace{10pt}\textit{-\texttt{maxit}} && $\{1, 2, \dots, 20, 30, 40, 50, 75, 100, 150, \dots, 5{,}000\}$ && $100$ && $250$ ($100$) \\
		\bottomrule
	\end{tabularx}}}
\end{table}

\begin{figure}[h!]
    \centering
	\begin{tabular}{c c}
        \centering
        \includegraphics[width=0.475\textwidth]{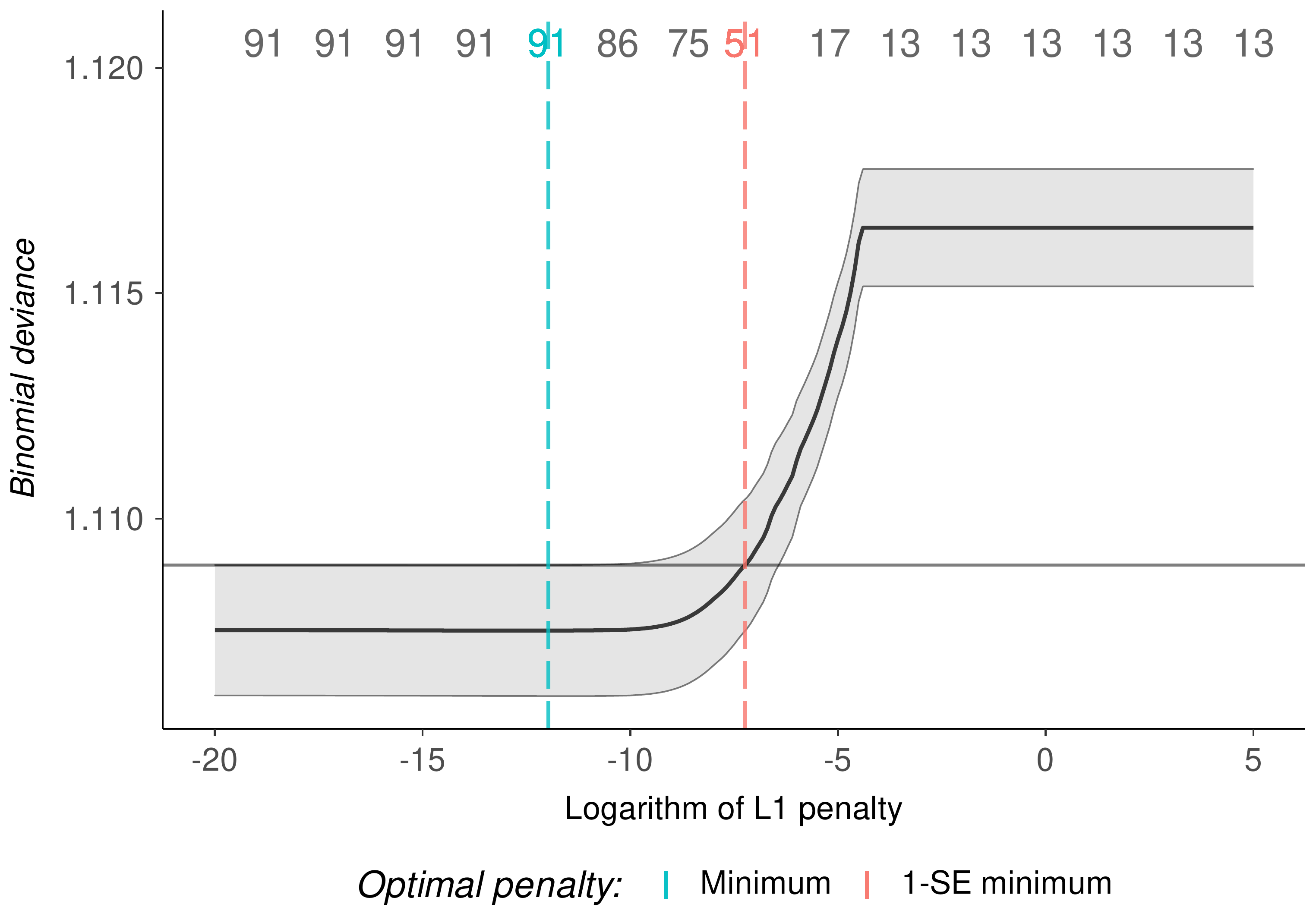}&
        \includegraphics[width=0.475\textwidth]{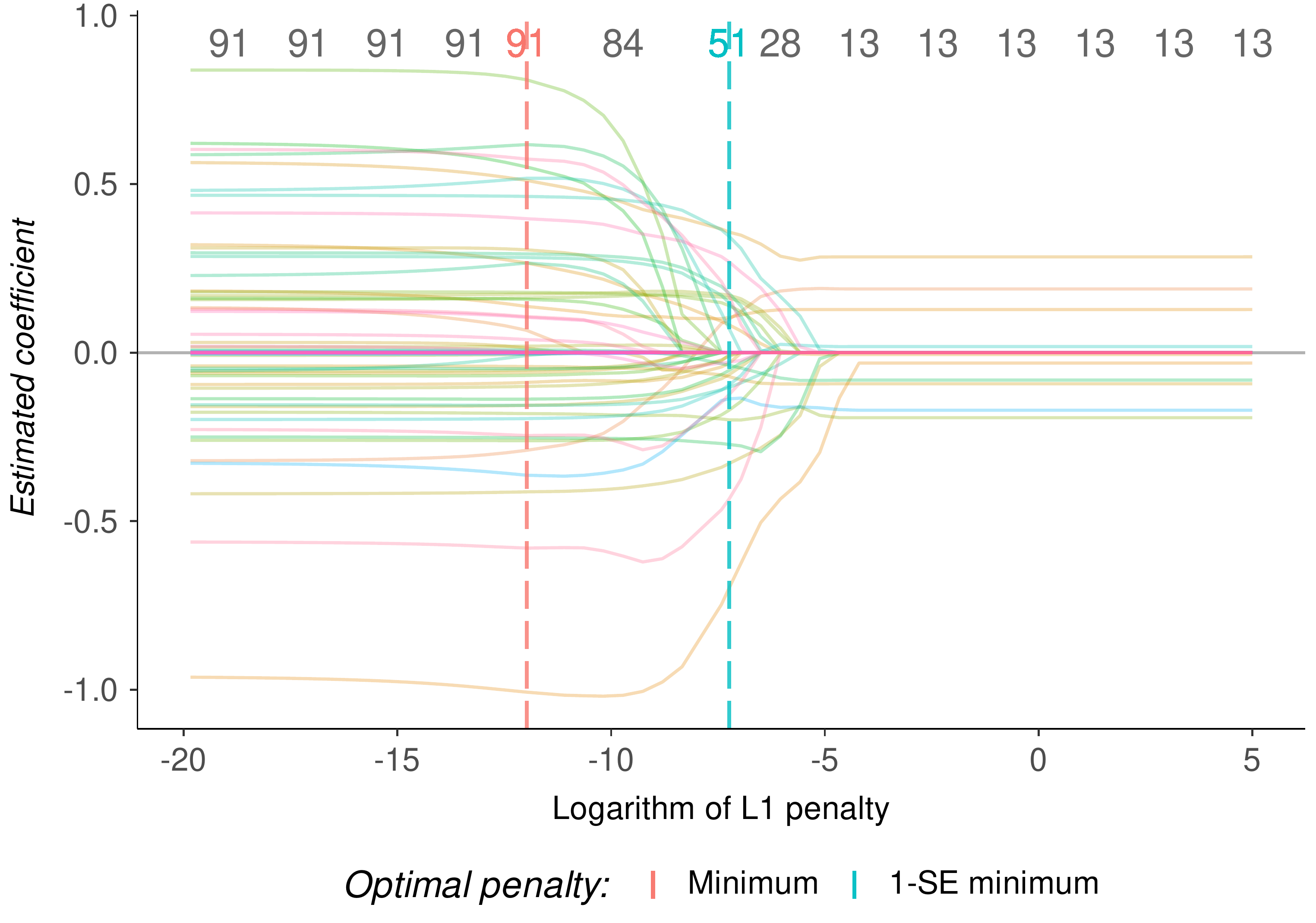}
    \end{tabular}\vspace{-9pt}
    \caption{LASSO cross-validation errors (left) and parameter estimates (right) with the number of non-zero coefficients denoted at the top for the discrete global response model in automobile insurance.}
	\label{Figure_B.1}\vspace{-8pt}
\end{figure}

\begin{table}[h!]
    \caption{Parameter estimates with standard errors in parenthesis for the discrete global response model after multiple imputation in automobile insurance.}
    \label{Table_B.2}\vspace{-6pt}
	\centerline{\scalebox{0.536}{\begin{tabularx}{1.872\textwidth}{l@{ }r r@{ }r@{}l c r@{ }r@{}l r@{ }r@{}l c r@{ }r@{}l r@{ }r@{}l r@{ }r@{}l r@{ }r@{}l }
	    \toprule \addlinespace[1ex] \vspace{1pt}
	    && && & & \multicolumn{6}{c}{\textbf{Interactions with \texttt{Competitiveness}}} & & \multicolumn{12}{c}{\textbf{Interactions with \texttt{Rate\_Change}}} \\ \cline{7-12} \cline{14-25} \addlinespace[1ex]
		\textbf{Risk factor}& & \multicolumn{3}{c}{\textbf{Coefficient}} & & \multicolumn{3}{c}{\textit{\texttt{Competitiveness}}} & \multicolumn{3}{c}{\textit{\texttt{Competitiveness}$^\mathit{2}$}} & & \multicolumn{3}{c}{$\mathit{[-9.28\%, 1.53\%]}$} & \multicolumn{3}{c}{$\mathit{(1.53\%, 6.06\%]}$} & \multicolumn{3}{c}{$\mathit{(8.58\%, 12.58\%]}$} & \multicolumn{3}{c}{$\mathit{(12.58\%, 27.01\%]}$} \\ \hline \addlinespace[0.4ex]
		\textit{\texttt{Constant}}& & -0.9305& (0.0240)&$^{***}$ &  & && & && & & && & && & && & && \\
		\textit{\texttt{Competitiveness}}& & 0.0190& (0.0628)& &  & && & && & & 0.0771& (0.0813)& & && & 0.1731& (0.0628)&$^{**}$ & && \\
		\textit{\texttt{Competitiveness}$^\mathit{2}$}& & && &  & && & && & & 0.3070& (0.1897)& & && & -0.0279& (0.2168)& & -0.2368& (0.1813)& \\
		\textit{\texttt{Rate\_Change}}& & && &  & && & && & & && & && & && & && \\
		\hspace{10pt}\textit{- }$\mathit{[-9.28\%, 1.53\%]}$& & -0.6224& (0.0403)&$^{***}$ &  & -0.0255& (0.0712)& & -0.0701& (0.1896)& & & && & && & && & && \\
		\hspace{10pt}\textit{- }$\mathit{(1.53\%, 6.06\%]}$& & 0.0493& (0.0217)&$^{*}$ &  & -0.0151& (0.0802)& & && & & && & && & && & && \\
		\hspace{10pt}\textit{- }$\mathit{(8.58\%, 12.58\%]}$& & 0.0329& (0.0361)& &  & -0.0134& (0.0726)& & && & & && & && & && & && \\
		\hspace{10pt}\textit{- }$\mathit{(12.58\%, 27.01\%]}$& & 0.1811& (0.0291)&$^{***}$ &  & -0.0176& (0.0726)& & && & & && & && & && & && \\
		\textit{\texttt{Premium\_New\_Base}}&$\mathit{\, (\times \, 10^{-2})}$ & -0.0009& (0.0034)& &  & 0.0118& (0.0073)& & && & & 0.1458& (0.0087)&$^{***}$ & && & -0.0045& (0.0046)& & -0.0058& (0.0045)& \\
		\textit{\texttt{Premium\_New\_Base}$^\mathit{2}$}&$\mathit{\, (\times \, 10^{-6})}$ & 0.0009& (0.0028)& &  & && & && & & -0.3857& (0.03777)&$^{***}$ & && & && & && \\
		\textit{\texttt{Undershooting\_1}}& & 0.0002& (0.0002)& &  & && & && & & && & 0.0003& (0.0003)& & && & 0.0005& (0.0002)&$^{*}$ \\
		\textit{\texttt{Undershooting\_1}$^\mathit{2}$}&$\mathit{\, (\times \, 10^{-6})}$ & -0.5628& (0.3447)& &  & && & && & & && & && & -0.0161& (0.4469)& & && \\
		\textit{\texttt{Undershooting\_2}}& & 0.0005& (0.0002)&$^{**}$ &  & -0.0004& (0.0002)&$^{*}$ & && & & 0.0010& (0.0002)&$^{***}$ & -0.0002& (0.0002)& & -0.0001& (0.0002)& & && \\
		\textit{\texttt{Undershooting\_2}$^\mathit{2}$}&$\mathit{\, (\times \, 10^{-5})}$ & && &  & && & && & & -0.1243& (0.0357)&$^{***}$ & -0.0075& (0.0107)& & && & -0.0223& (0.0257)& \\
		\textit{\texttt{Risk\_Level}}& & && &  & && & && & & && & && & && & && \\
		\hspace{10pt}\textit{- Low}& & -0.0062& (0.0207)& &  & 0.1200& (0.0653)& & && & & -0.0519& (0.0313)& & && & && & -0.0566& (0.0363)& \\
		\hspace{10pt}\textit{- Medium}& & -0.1150& (0.0387)&$^{**}$ &  & -0.1362& (0.0782)& & && & & -0.1407& (0.0574)&$^{*}$ & 0.2235& (0.0533)&$^{***}$ & 0.1568& (0.0509)&$^{**}$ & 0.2291& (0.0485)&$^{***}$ \\
		\hspace{10pt}\textit{- High}& & -0.0638& (0.1518)& &  & && & && & & && & && & && & 0.1677& (0.3131)& \\
		\textit{\texttt{Policy\_Type}}& & && &  & && & && & & && & && & && & && \\
		\hspace{10pt}\textit{- Employee}& & -0.0353& (0.0401)& &  & -0.0981& (0.1593)& & 0.0958& (0.4560)& & & -0.1032& (0.0826)& & && & & & & && \\
		\hspace{10pt}\textit{- Second car}& & -0.0867& (0.0469)& &  & && & && & & -0.2935& (0.0552)&$^{***}$ & -0.0431& (0.0747)& & 0.0875& (0.0549)& & -0.0200& (0.0586)& \\
		\bottomrule
		\multicolumn{25}{l}{\footnotesize Significance levels: $^{*}5\%$-level, $^{**}1\%$-level, $^{***}0.1\%$-level or less} \\
	\end{tabularx}}}
\end{table}\newpage

\vfill
\begin{figure}
    \centering
	\begin{tabular}{c c}
        \centering
        \includegraphics[width=0.475\textwidth]{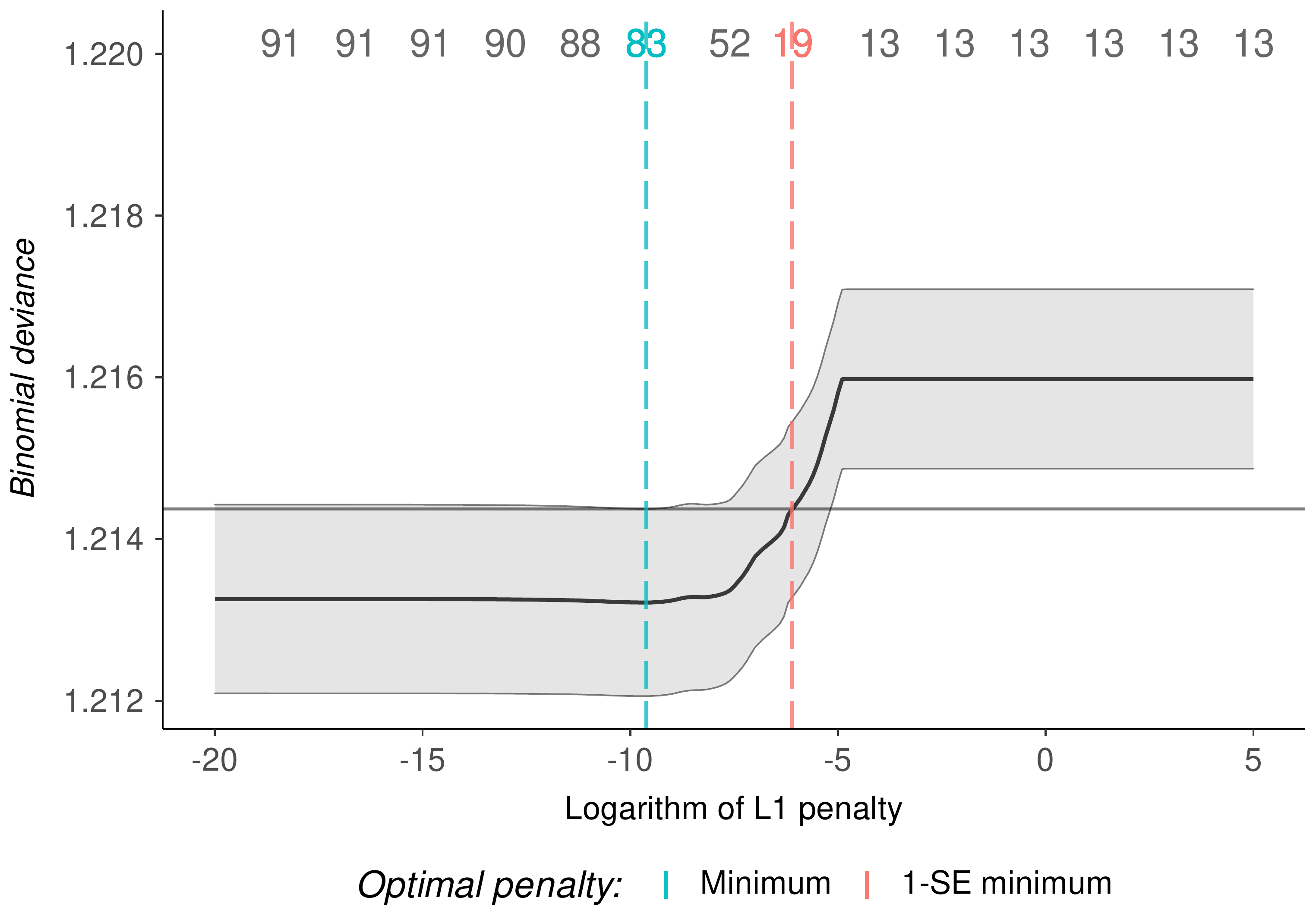}&
        \includegraphics[width=0.475\textwidth]{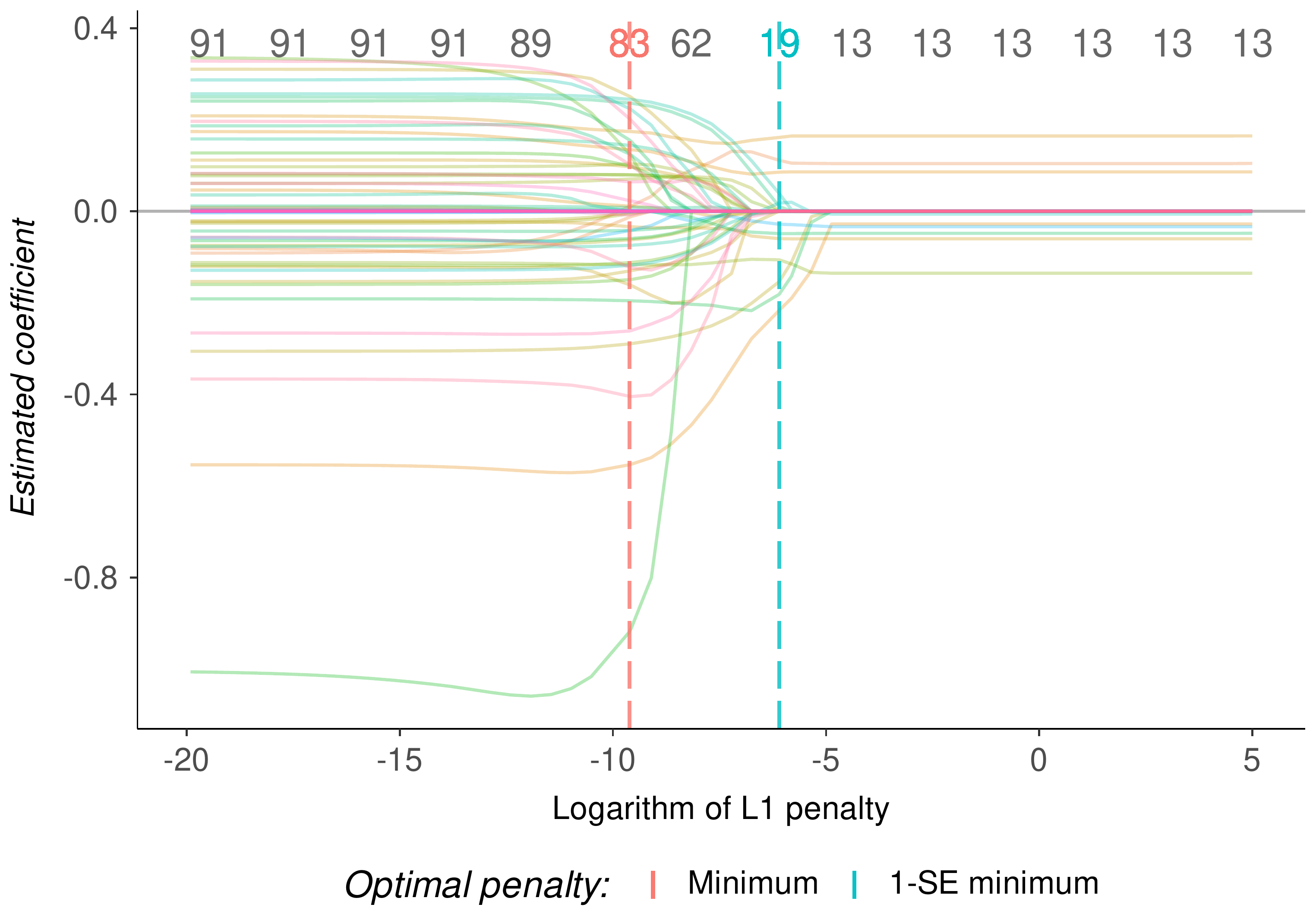}
    \end{tabular}\vspace{-6pt}
    \caption{LASSO cross-validation errors (left) and parameter estimates (right) with the number of non-zero coefficients denoted at the top for the discrete global response model without multiple imputation in automobile insurance.}
	\label{Figure_B.2}
\end{figure}

\vfill

\begin{table}
    \caption{Parameter estimates with standard errors in parenthesis for the discrete global response model without multiple imputation in automobile insurance.}
    \label{Table_B.3}\vspace{-6pt}
	\centerline{\scalebox{0.541}{\begin{tabularx}{1.857\textwidth}{l@{ }r r@{ }r@{}l c r@{ }r@{}l r@{ }r@{}l c r@{ }r@{}l r@{ }r@{}l r@{ }r@{}l r@{ }r@{}l }
	    \toprule \addlinespace[1ex] \vspace{1pt}
	    && && & & \multicolumn{6}{c}{\textbf{Interactions with \texttt{Competitiveness}}} & & \multicolumn{12}{c}{\textbf{Interactions with \texttt{Rate\_Change}}} \\ \cline{7-12} \cline{14-25} \addlinespace[1ex]
		\textbf{Risk factor}& & \multicolumn{3}{c}{\textbf{Coefficient}} & & \multicolumn{3}{c}{\textit{\texttt{Competitiveness}}} & \multicolumn{3}{c}{\textit{\texttt{Competitiveness}$^\mathit{2}$}} & & \multicolumn{3}{c}{$\mathit{[-9.28\%, 1.53\%]}$} & \multicolumn{3}{c}{$\mathit{(1.53\%, 6.06\%]}$} & \multicolumn{3}{c}{$\mathit{(8.58\%, 12.58\%]}$} & \multicolumn{3}{c}{$\mathit{(12.58\%, 27.01\%]}$} \\ \hline \addlinespace[0.4ex]
		\textit{\texttt{Constant}}& & -0.8776& (0.0116)&$^{***}$ &  & && & && & & && & && & && & && \\
		\textit{\texttt{Competitiveness}}& & 0.1528& (0.0239)&$^{***}$ &  & && & && & & && & && & && & && \\
		\textit{\texttt{Competitiveness}$^\mathit{2}$}& & && &  & && & && & & && & && & && & && \\
		\textit{\texttt{Rate\_Change}}& & && &  & && & && & & && & && & && & && \\
		\hspace{10pt}\textit{- }$\mathit{[-9.28\%, 1.53\%]}$& & -0.3062& (0.0193)&$^{***}$ &  & -0.0255& (0.0712)& & -0.0701& (0.1896)& & & && & && & && & && \\
		\hspace{10pt}\textit{- }$\mathit{(1.53\%, 6.06\%]}$& & 0.0713& (0.0120)&$^{***}$ &  & -0.0151& (0.0802)& & && & & && & && & && & && \\
		\hspace{10pt}\textit{- }$\mathit{(8.58\%, 12.58\%]}$& & -0.0008& (0.0117)& &  & -0.0134& (0.0726)& & && & & && & && & && & && \\
		\hspace{10pt}\textit{- }$\mathit{(12.58\%, 27.01\%]}$& & 0.1475& (0.0119)&$^{***}$ &  & -0.0176& (0.0726)& & && & & && & && & && & && \\
		\textit{\texttt{Premium\_New\_Base}}&$\mathit{\, (\times \, 10^{-2})}$ & -0.0009& (0.0034)& &  & && & && & & 0.1458& (0.0087)&$^{***}$ & && & && & && \\
		\textit{\texttt{Premium\_New\_Base}$^\mathit{2}$}&$\mathit{\, (\times \, 10^{-6})}$ & -0.0539& (0.0142)&$^{***}$ &  & && & && & & 0.6661& (0.0306)&$^{***}$ & && & && & && \\
		\textit{\texttt{Undershooting\_1}}& & 0.0002& (0.0001)&$^{*}$ &  & && & && & & && & && & && & && \\
		\textit{\texttt{Undershooting\_1}$^\mathit{2}$}&$\mathit{\, (\times \, 10^{-6})}$ & && &  & && & && & & && & && & && & && \\
		\textit{\texttt{Undershooting\_2}}& & 0.5075& (0.0495)&$^{***}$ &  & -0.6292& (0.1221)&$^{***}$ & && & & && & && & && & && \\
		\textit{\texttt{Undershooting\_2}$^\mathit{2}$}&$\mathit{\, (\times \, 10^{-5})}$ & && &  & && & && & & && & && & && & && \\
		\textit{\texttt{Risk\_Level}}& & && &  & && & && & & && & && & && & && \\
		\hspace{10pt}\textit{- Low}& & -0.0489& (0.0096)&$^{***}$ &  & && & && & & && & && & && & && \\
		\hspace{10pt}\textit{- Medium}& & -0.0338& (0.0209)& &  & && & && & & -0.2588& (0.0373)&$^{***}$ & 0.1644& (0.0349)&$^{***}$ & && & 0.1894& (0.0345)&$^{***}$ \\
		\hspace{10pt}\textit{- High}& & -0.0285& (0.0907)& &  & && & && & & && & && & && & 0.1677& (0.3131)& \\
		\textit{\texttt{Policy\_Type}}& & && &  & && & && & & && & && & && & && \\
		\hspace{10pt}\textit{- Employee}& & -0.0577& (0.0181)&$^{**}$ &  & -0.0981& (0.1593)& & 0.0958& (0.4560)& & & -0.1032& (0.0826)& & && & & & & && \\
		\hspace{10pt}\textit{- Second car}& & -0.0770& (0.0157)&$^{***}$ &  & && & && & & -0.2935& (0.0552)&$^{***}$ & -0.0431& (0.0747)& & 0.0875& (0.0549)& & -0.0200& (0.0586)& \\
		\bottomrule
		\multicolumn{25}{l}{\footnotesize Significance levels: $^{*}5\%$-level, $^{**}1\%$-level, $^{***}0.1\%$-level or less} \\
	\end{tabularx}}}
\end{table}\vfill\newpage

\begin{figure}
    \vspace{-24pt}
    \centering
    \begin{subfigure}{\textwidth}
        \centering
        \begin{tabular}{c c}
            \centering
            {\hspace{-45pt}\footnotesize\underline{\hspace{1.5mm}\textbf{Without multiple imputation$_{_{}}$}\hspace{1.5mm}}} & {\hspace{-85pt}\footnotesize\underline{\hspace{12.875mm}\textbf{With XGBoost$_{_{}}$}\hspace{12.875mm}}} \vspace{-13pt}\\
            \centering\hspace{-45pt}
            \includegraphics[width=0.675\textwidth]{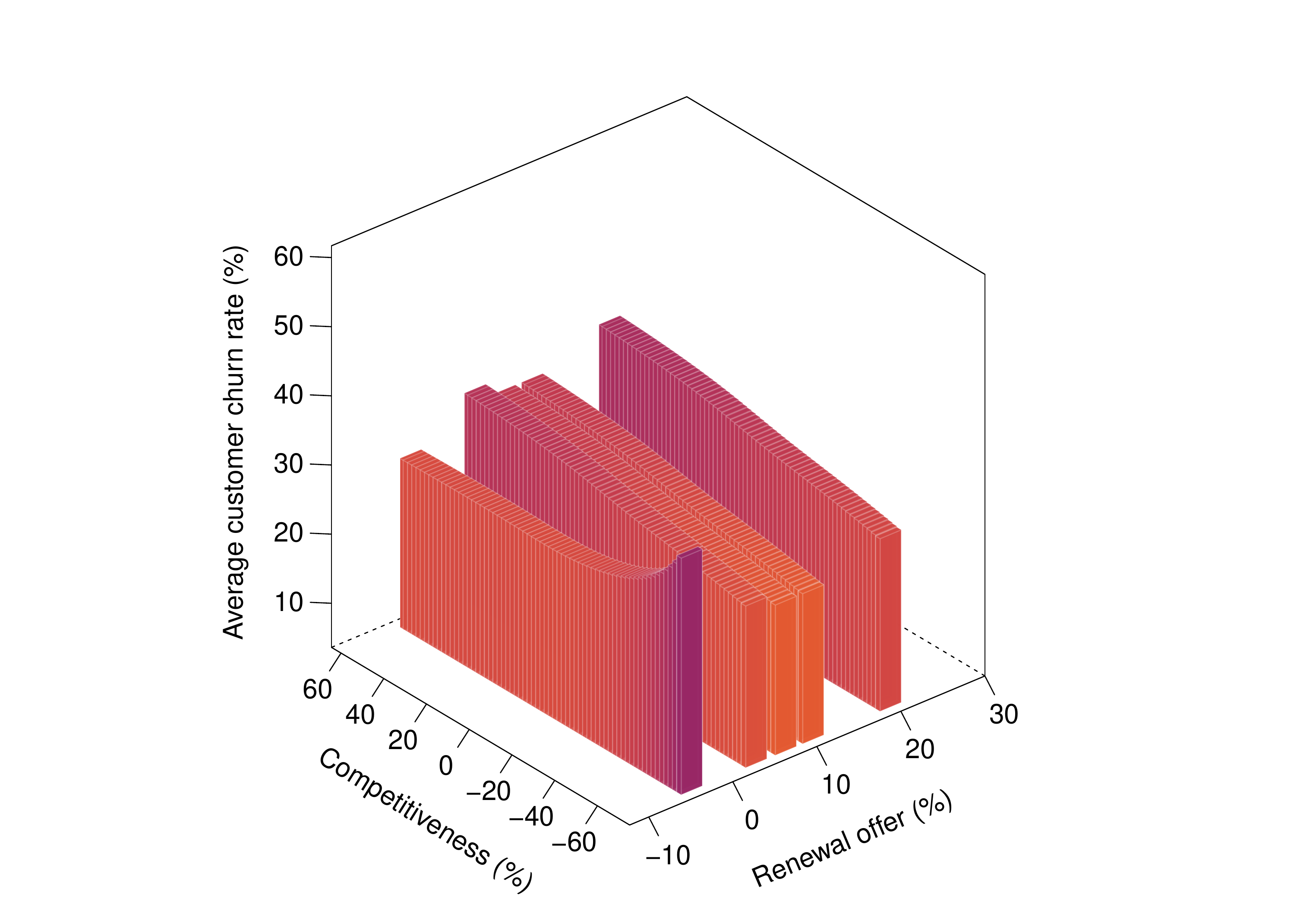}&\hspace{-85pt}
            \includegraphics[width=0.675\textwidth]{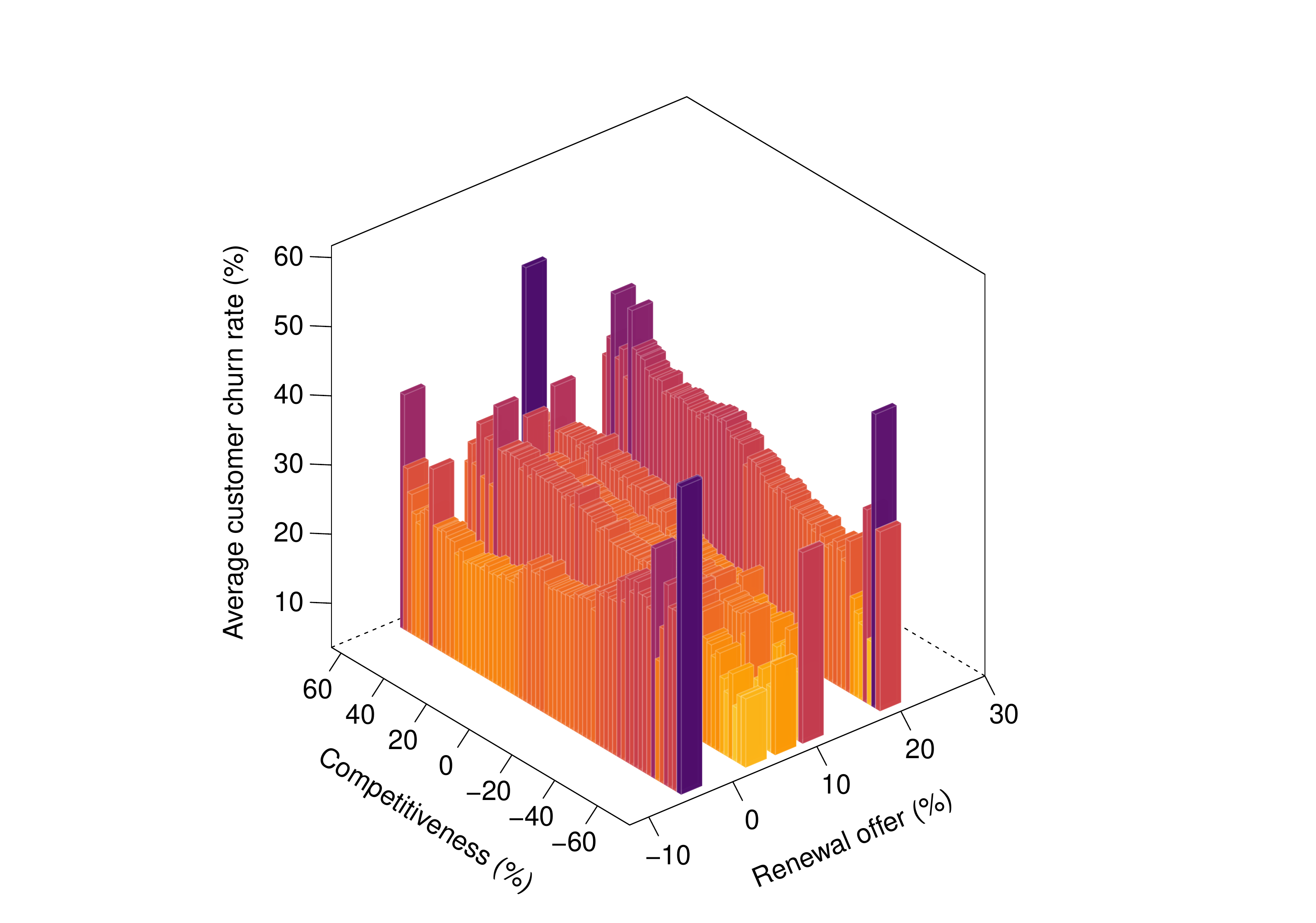}
        \end{tabular}\vspace{-8pt}
        \caption{Average customer churn for renewal offers and competitiveness}
	    \label{Figure_B.3a}
    \end{subfigure}

    \vspace{-9pt}

    \begin{subfigure}{\textwidth}
        \centering
        \begin{tabular}{c c}
            \centering\hspace{-45pt}
            \includegraphics[width=0.675\textwidth]{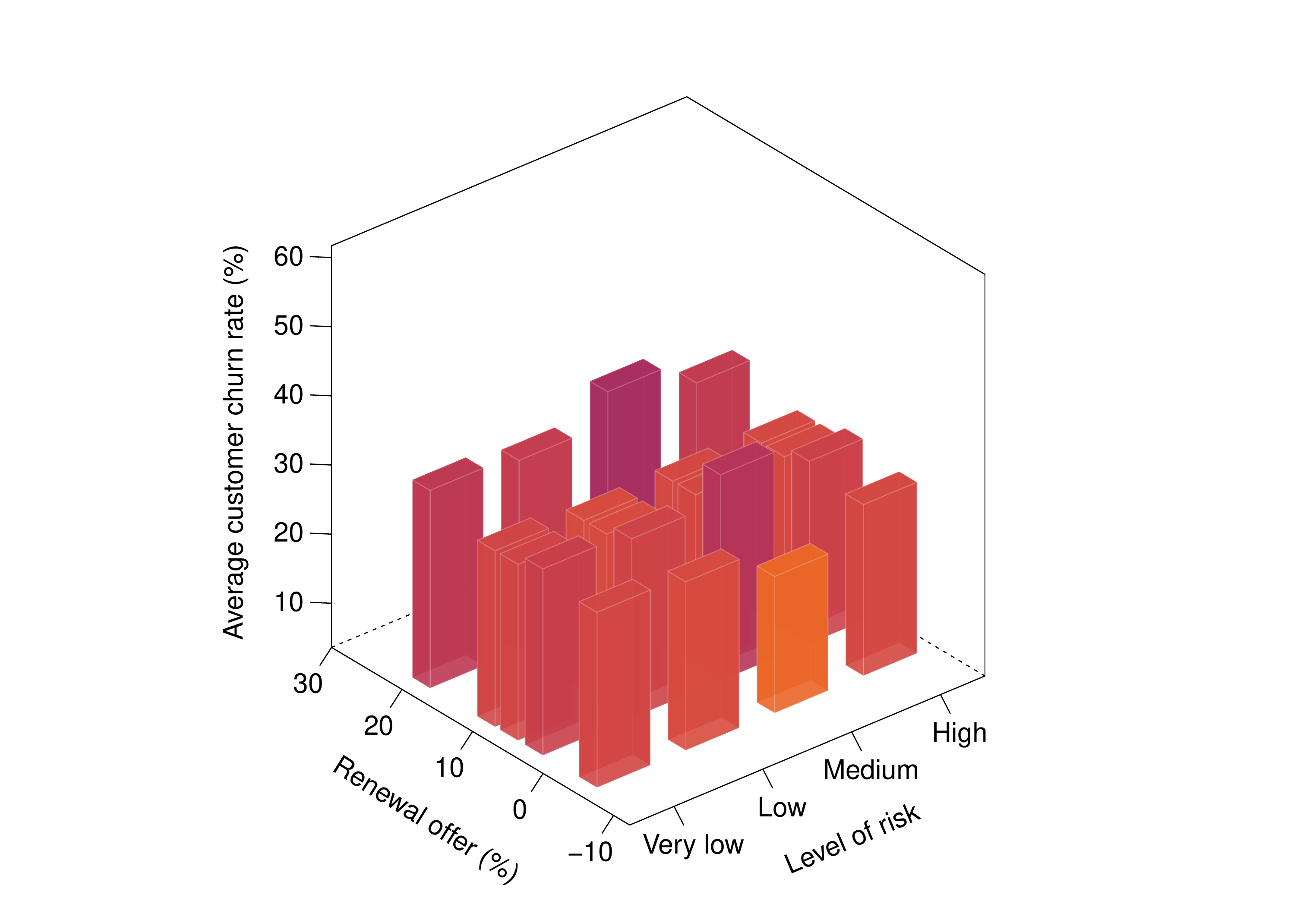}&\hspace{-85pt}
            \includegraphics[width=0.675\textwidth]{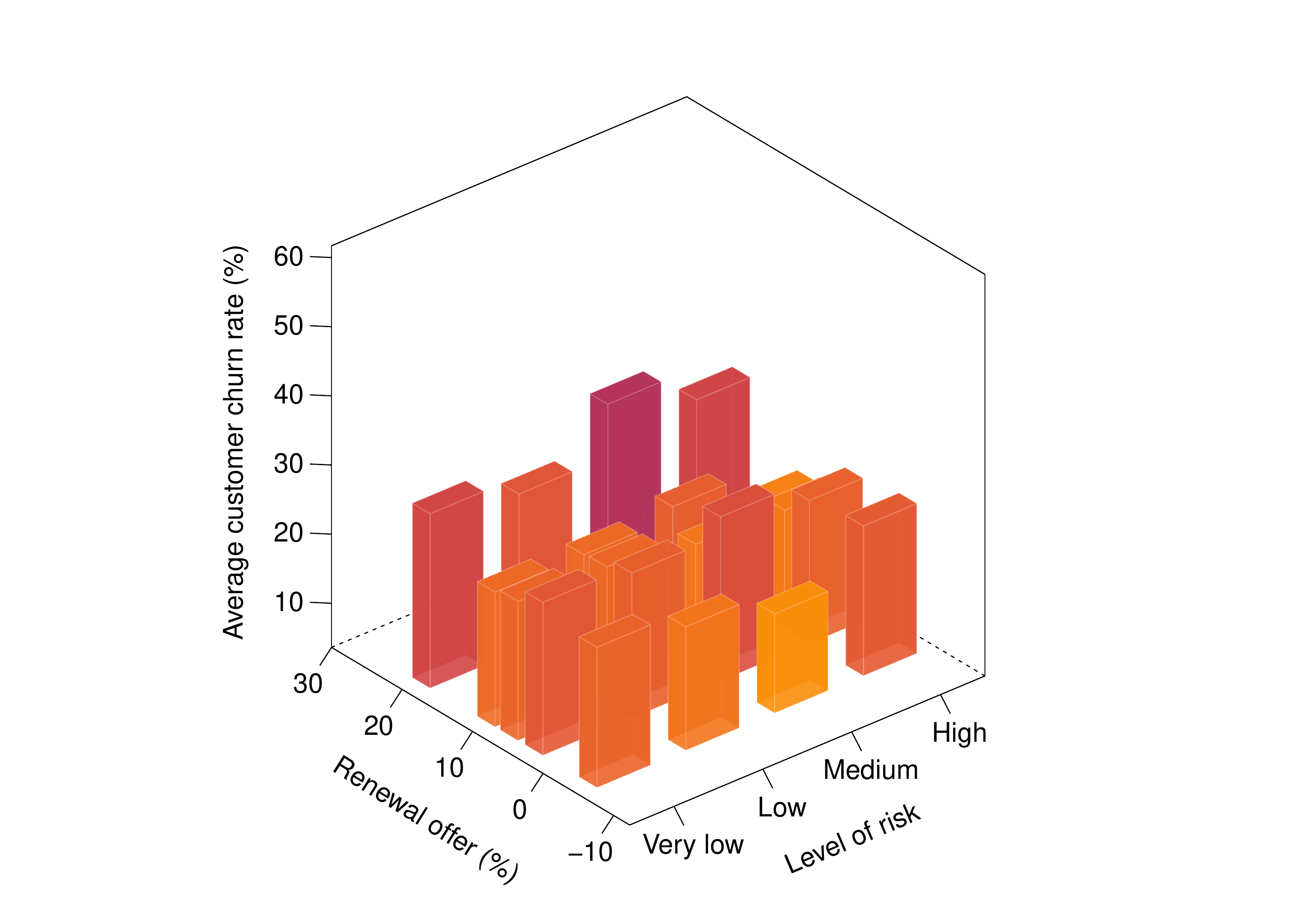}
        \end{tabular}\vspace{-8pt}
        \caption{Average customer churn for renewal offers and level of risk}
        \label{Figure_B.3b}
    \end{subfigure}
    
    \vspace{-9pt}
    
    \begin{subfigure}{\textwidth}
        \centering
        \begin{tabular}{c c}
            \centering\hspace{-45pt}
            \includegraphics[width=0.675\textwidth]{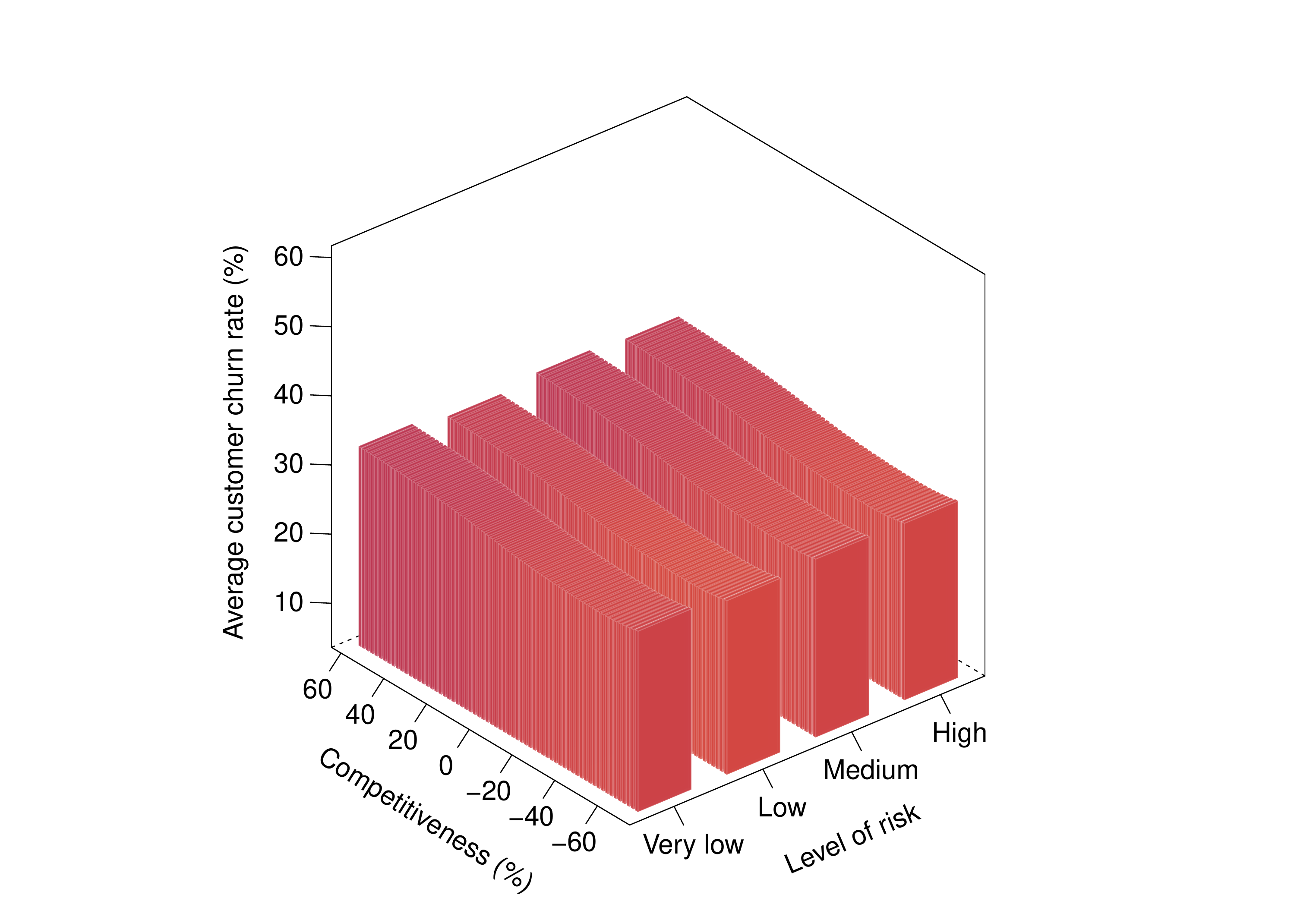}&\hspace{-85pt}
            \includegraphics[width=0.675\textwidth]{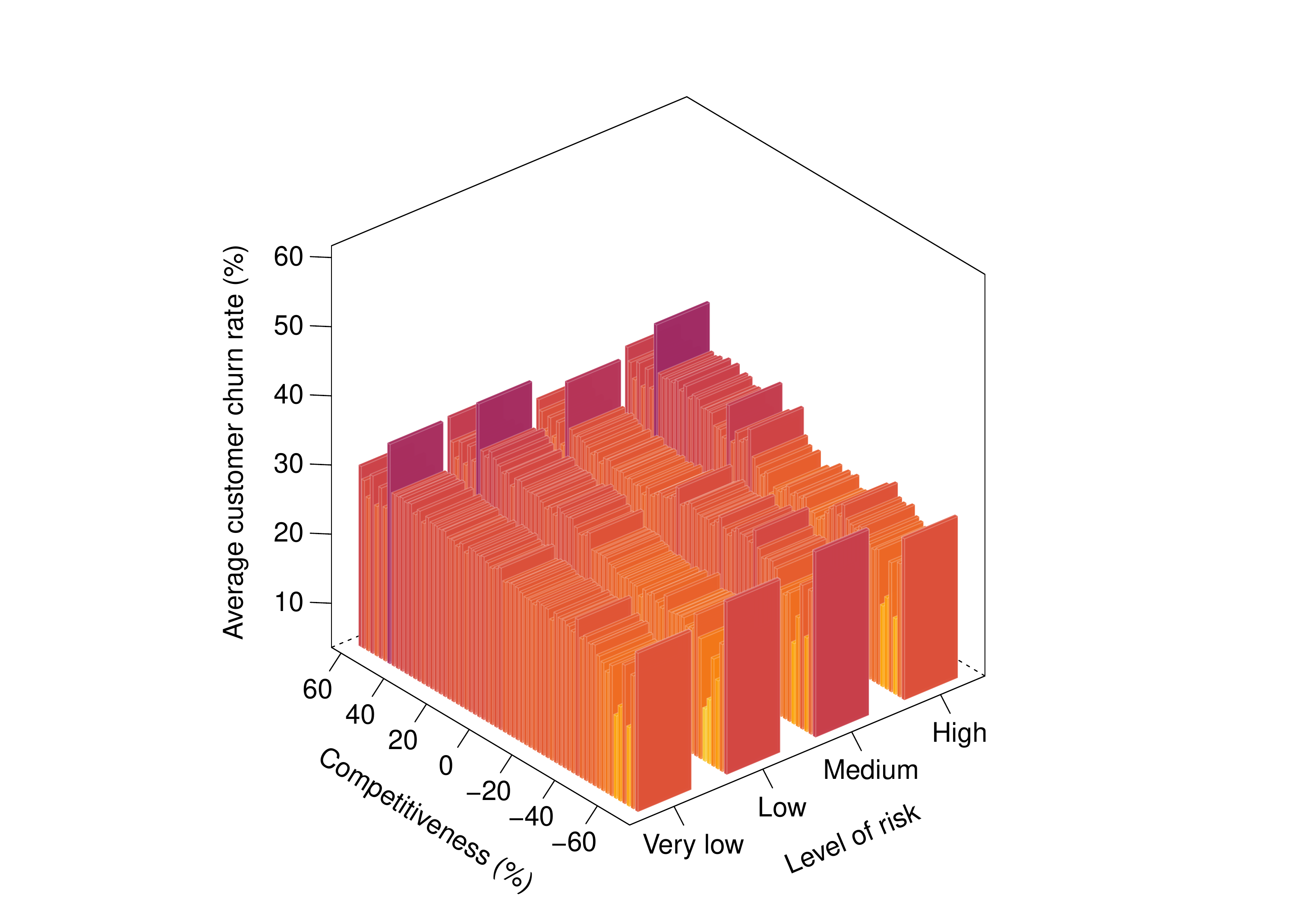}
        \end{tabular}\vspace{-8pt}
        \caption{Average customer churn for level of risk and competitiveness}
        \label{Figure_B.3c}
    \end{subfigure}\vspace{-3pt}
    \caption{Average customer churn estimate for each renewal offer and every competitiveness (panel (a)) and level of risk (panel (b)) as well as aggregated over all renewal offers (panel (c)) with discrete rate changes without multiple imputation (left) and with XGBoost (right) in automobile insurance.}
	\label{Figure_B.3}
\end{figure}\newpage

\begin{figure}
    \centering
    \begin{subfigure}{\textwidth}
        \centering
        \begin{tabular}{c c}
            \centering
            {\footnotesize\underline{\hspace{18.5mm}\textbf{Discrete$_{_{}}$}\hspace{18.5mm}}} & {\footnotesize\underline{\hspace{16.0mm}\textbf{Continuous$_{_{}}$}\hspace{16.0mm}}} \vspace{5pt}\\
            \centering
            \includegraphics[width=0.45\textwidth]{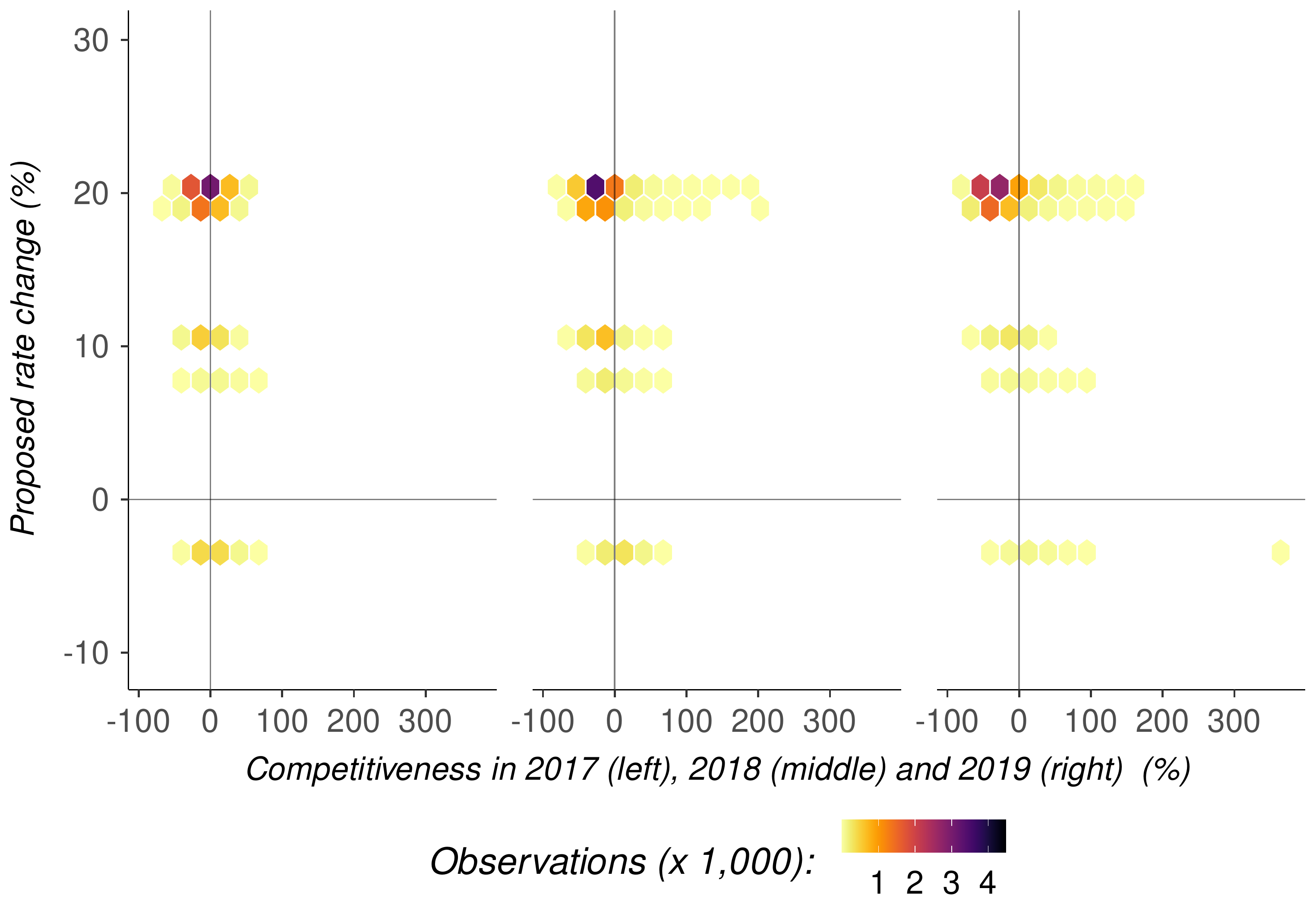}&
            \includegraphics[width=0.45\textwidth]{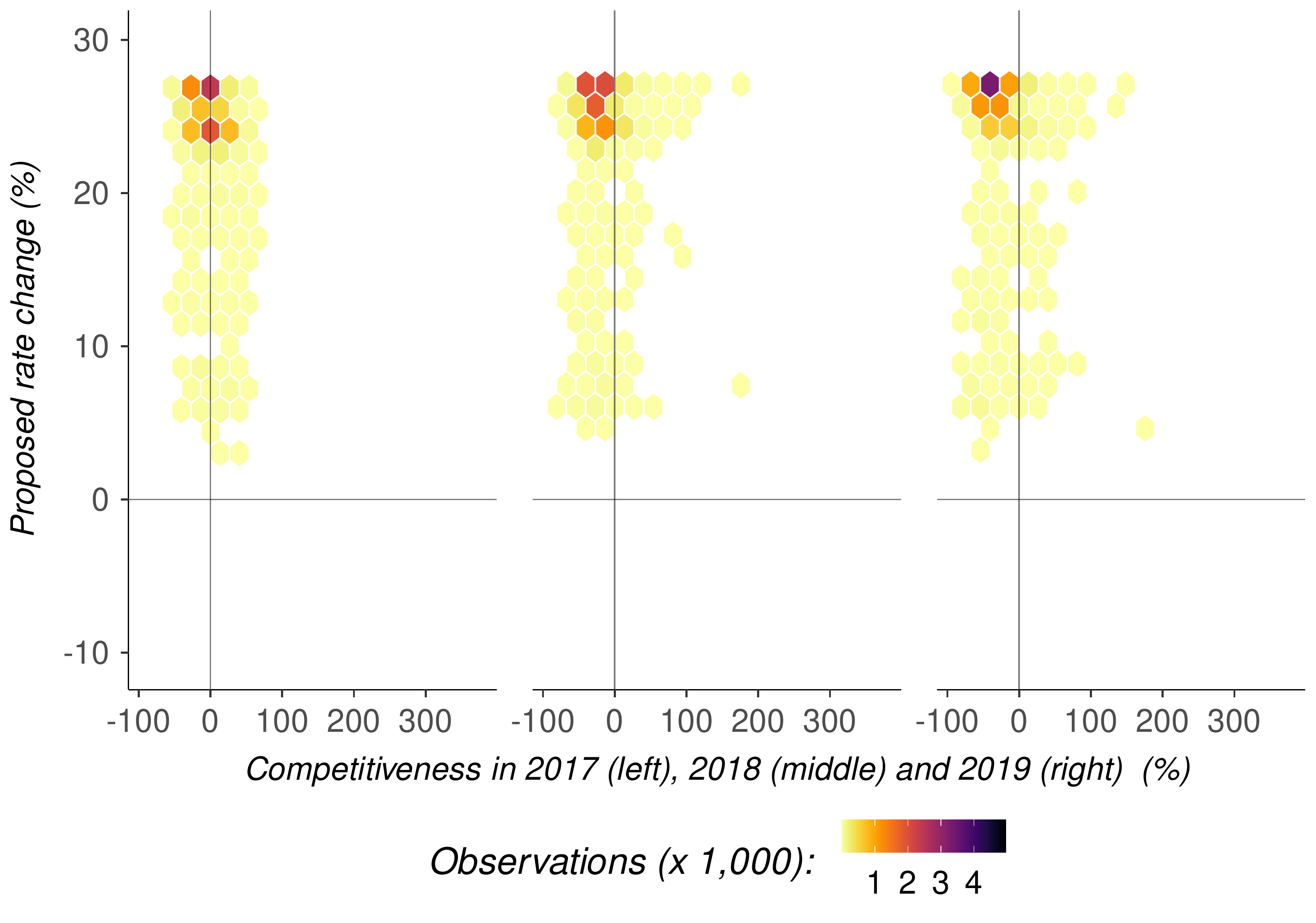}
        \end{tabular}\vspace{-6pt}
        \caption{Optimal rate changes over competitiveness}
        \label{Figure_B.4a}
    \end{subfigure}
    
    \vspace{3pt}
    
    \begin{subfigure}{\textwidth}
        \centering
        \begin{tabular}{c c}
            \centering
            \includegraphics[width=0.45\textwidth]{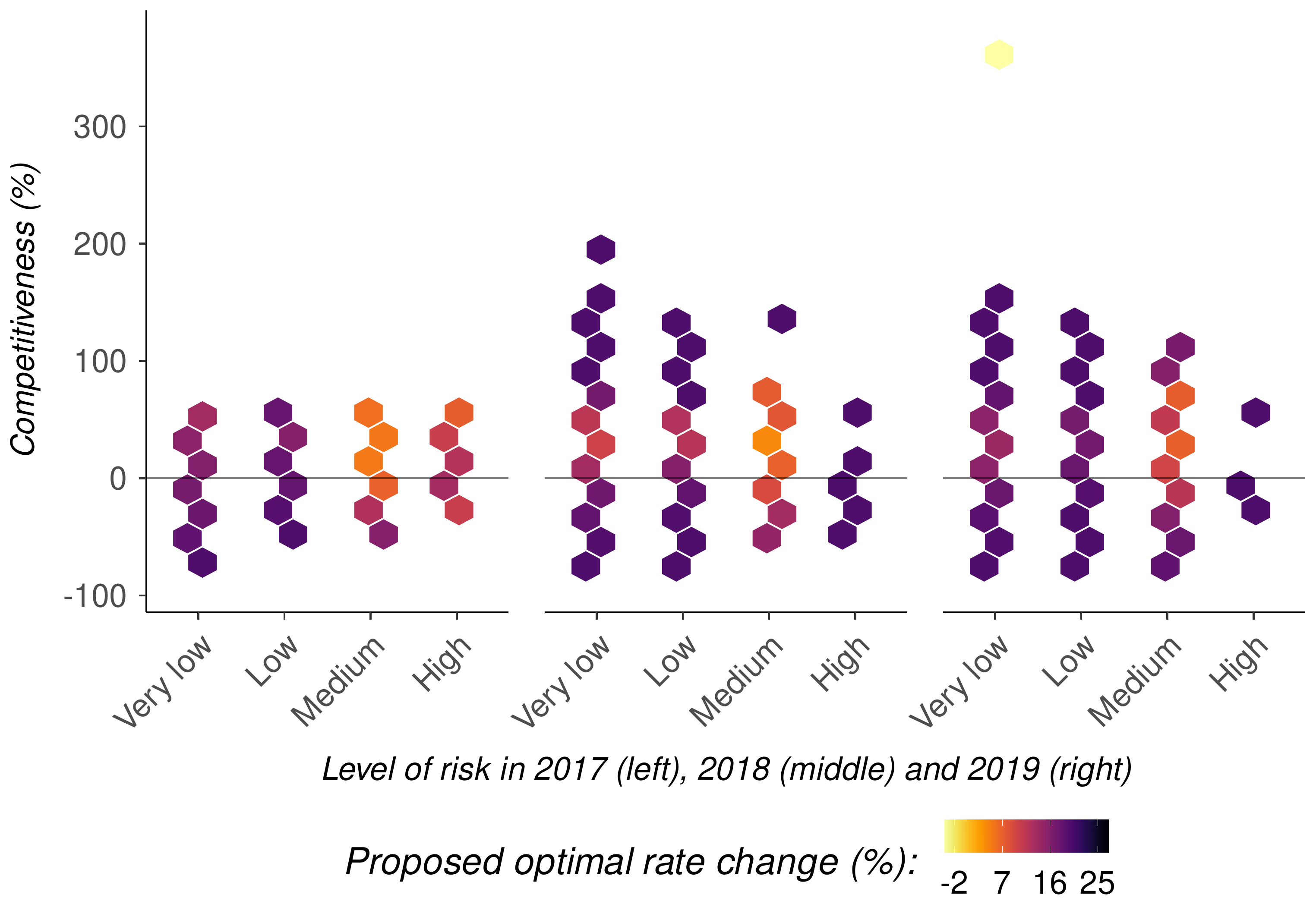}&
            \includegraphics[width=0.45\textwidth]{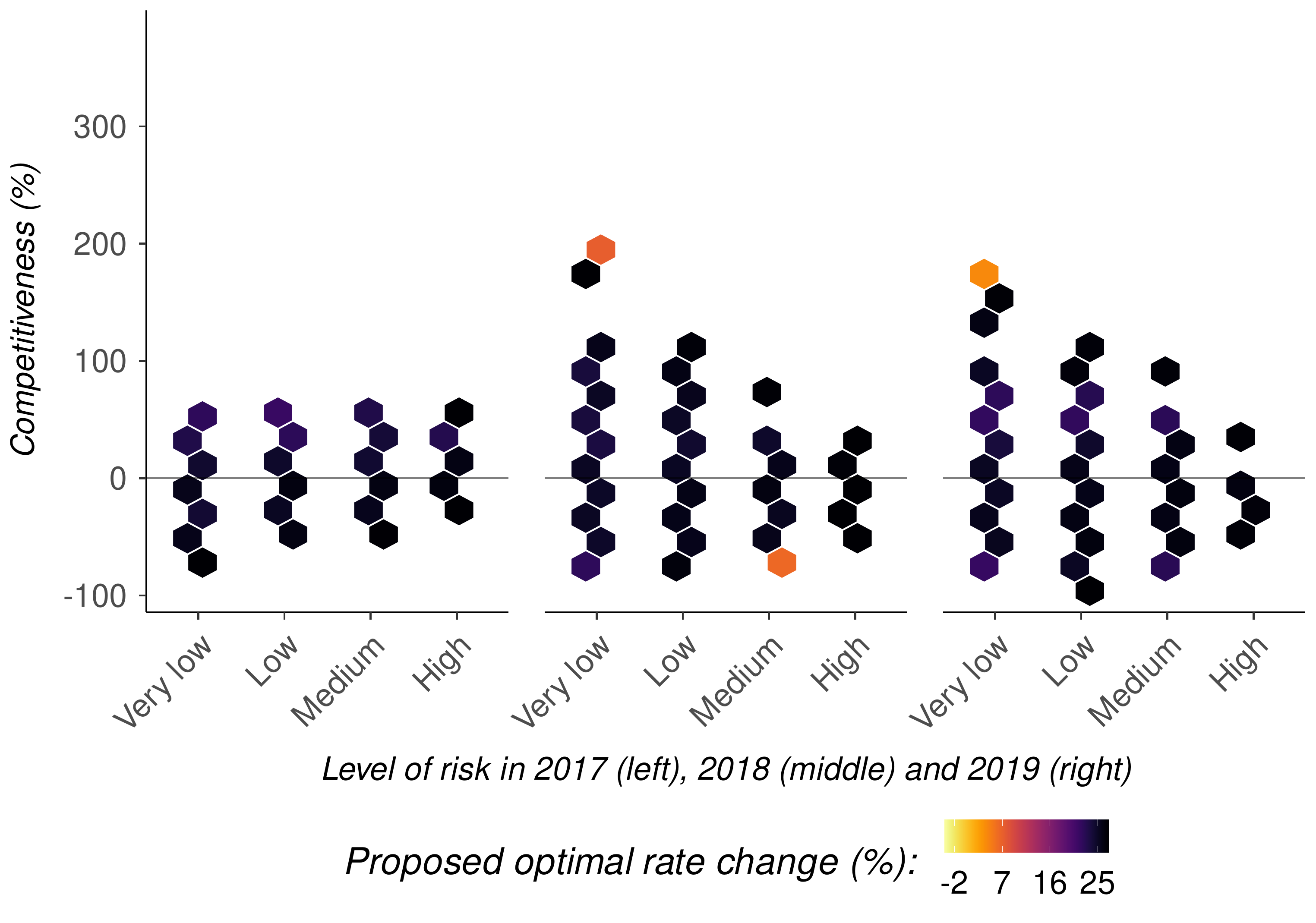}
        \end{tabular}\vspace{-6pt}
        \caption{Optimal rate changes over level of risk and competitiveness}
        \label{Figure_B.4b}
    \end{subfigure}\vspace{-3pt}
    \caption{Optimal rate changes over the competitiveness (panel (a)) jointly with the level of risk (panel (b)) for $\tau = 3$ consecutive renewals with discrete (left) and continuous (right) rate changes in automobile insurance.}
	\label{Figure_B.4}
\end{figure}\newpage

\begin{figure}
    \centering
    \begin{subfigure}{\textwidth}
        \centering
        \begin{tabular}{c c}
            \centering
            {\footnotesize\underline{\hspace{0.875mm}\textbf{Optimal over competitiveness$_{_{}}$}\hspace{0.875mm}}} & {\footnotesize\underline{\hspace{4.5mm}\textbf{Optimal over level of risk$_{_{}}$}\hspace{4.5mm}}} \vspace{5pt}\\
            \centering
            \includegraphics[width=0.45\textwidth]{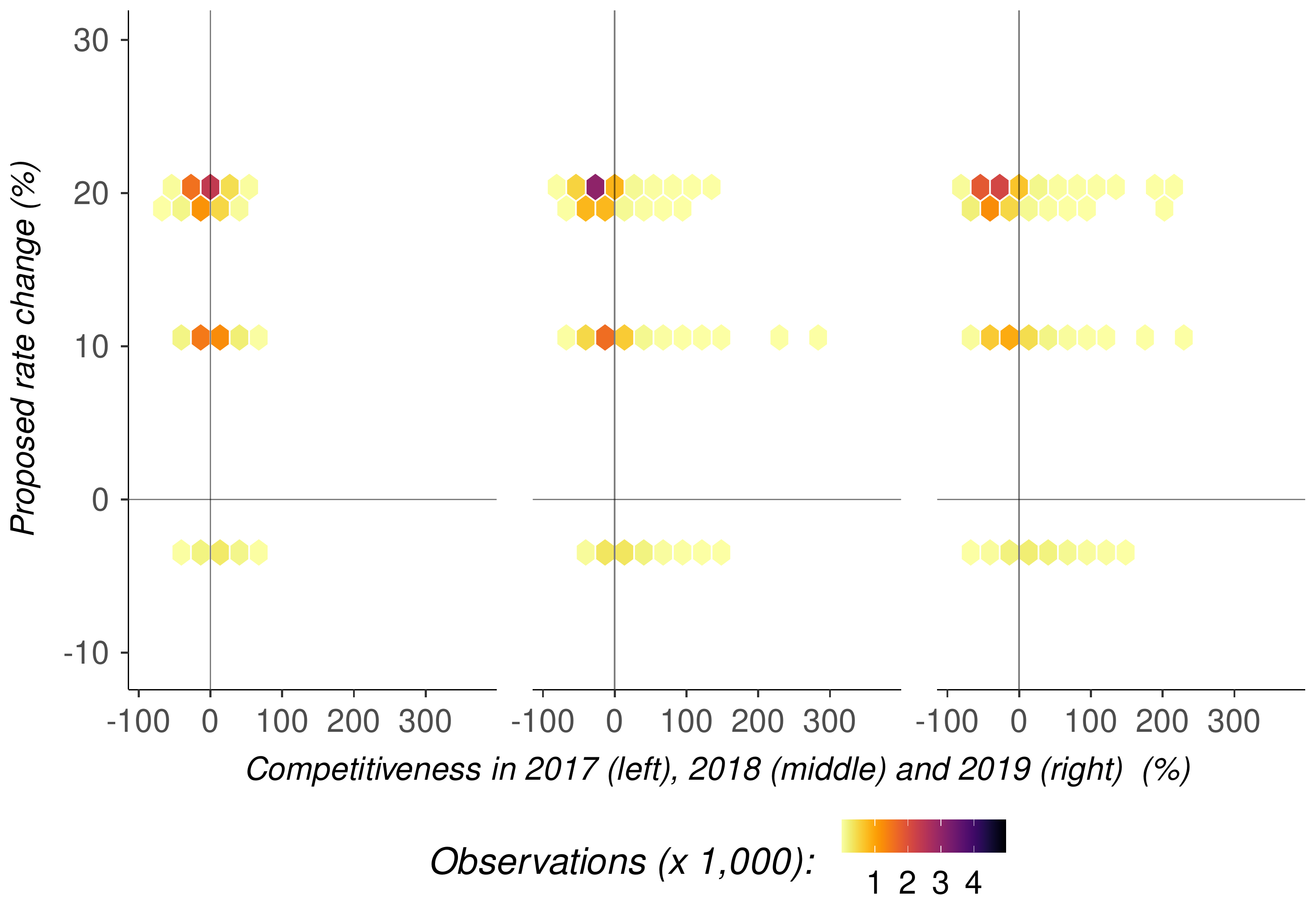}&
            \includegraphics[width=0.45\textwidth]{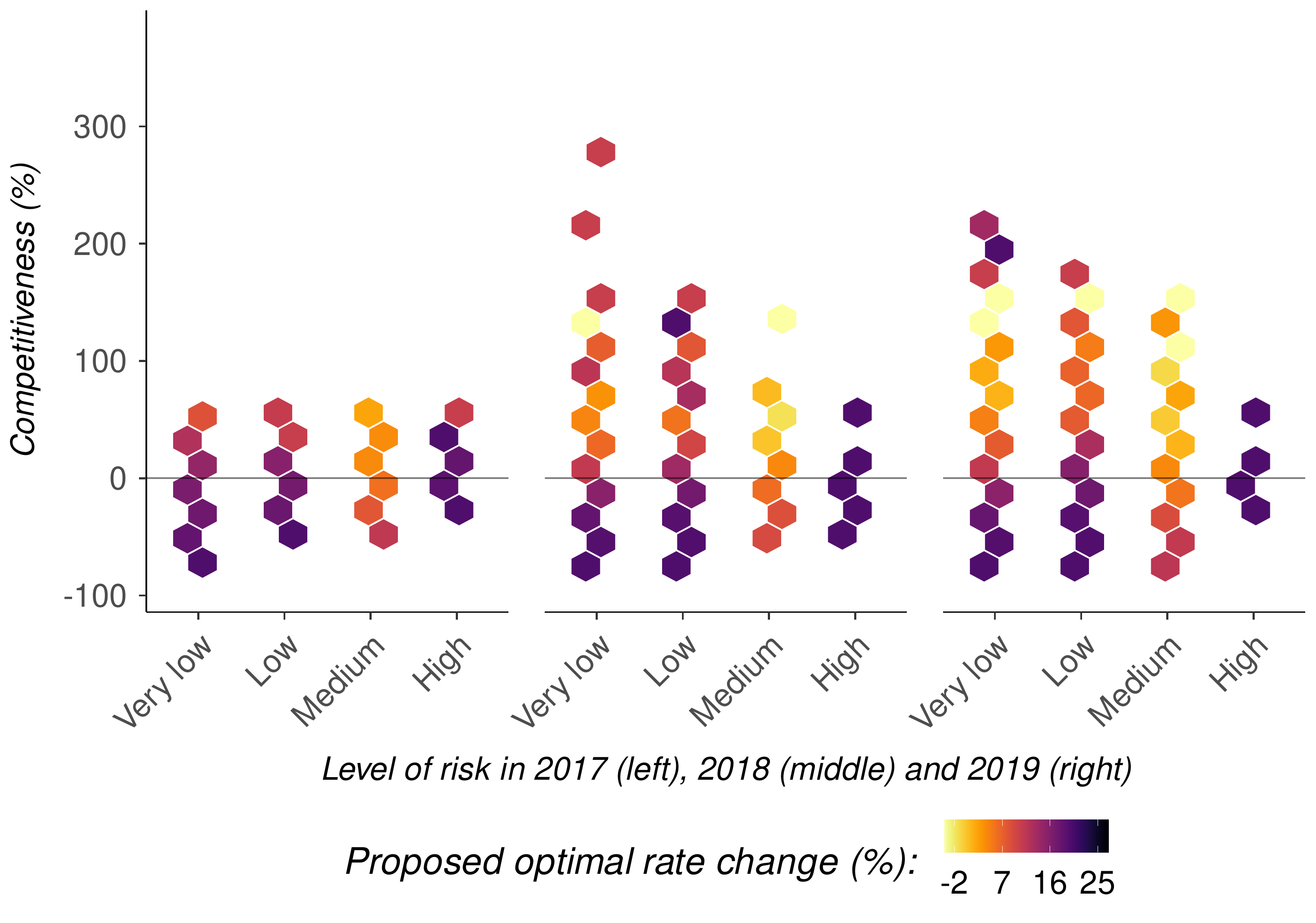}
        \end{tabular}\vspace{-6pt}
        \caption{Optimal rate changes for discrete global response model without multiple imputation}
        \label{Figure_B.5a}
    \end{subfigure}
    
    \vspace{3pt}
    
    \begin{subfigure}{\textwidth}
        \centering
        \begin{tabular}{c c}
            \centering
            \includegraphics[width=0.45\textwidth]{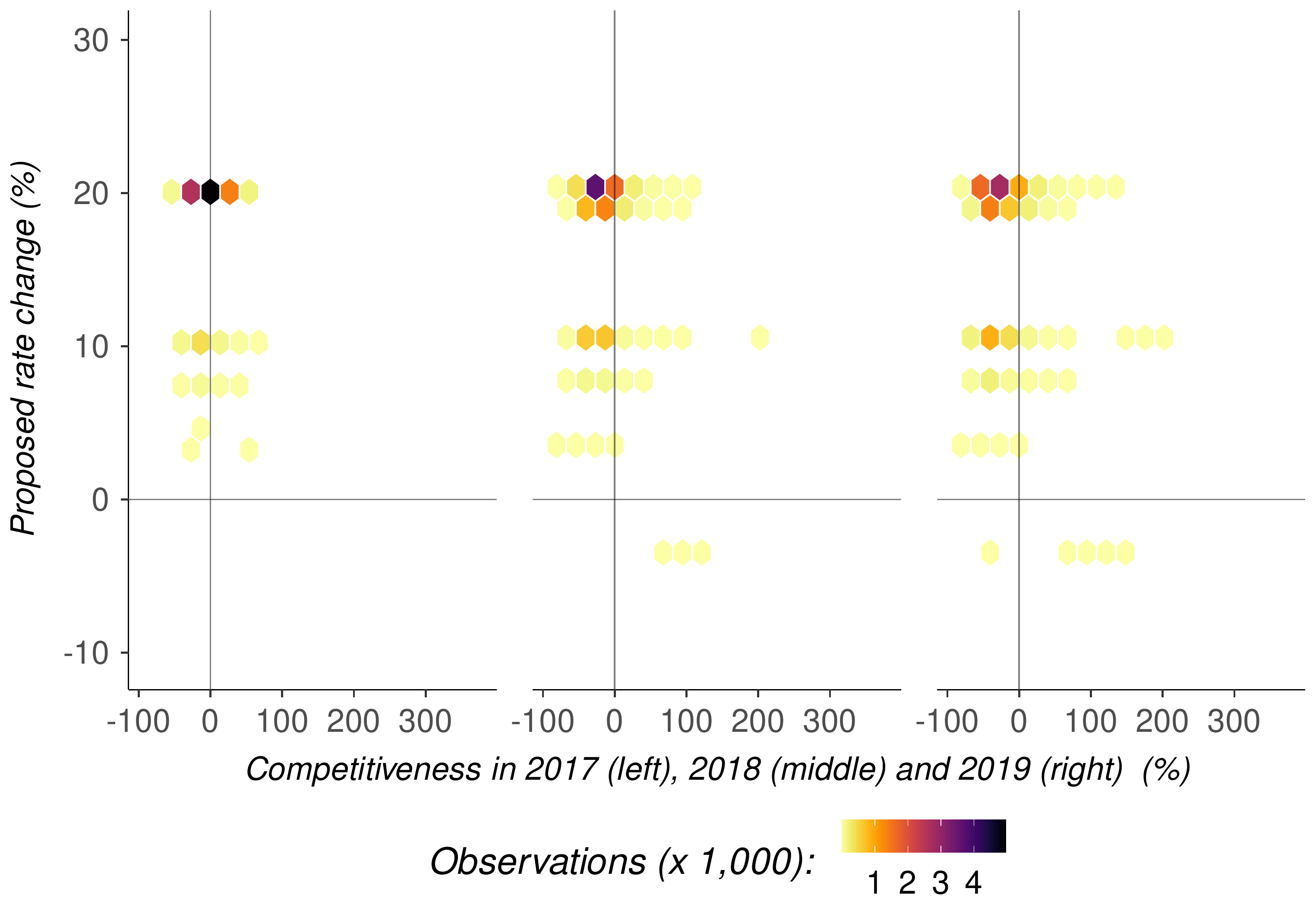}&
            \includegraphics[width=0.45\textwidth]{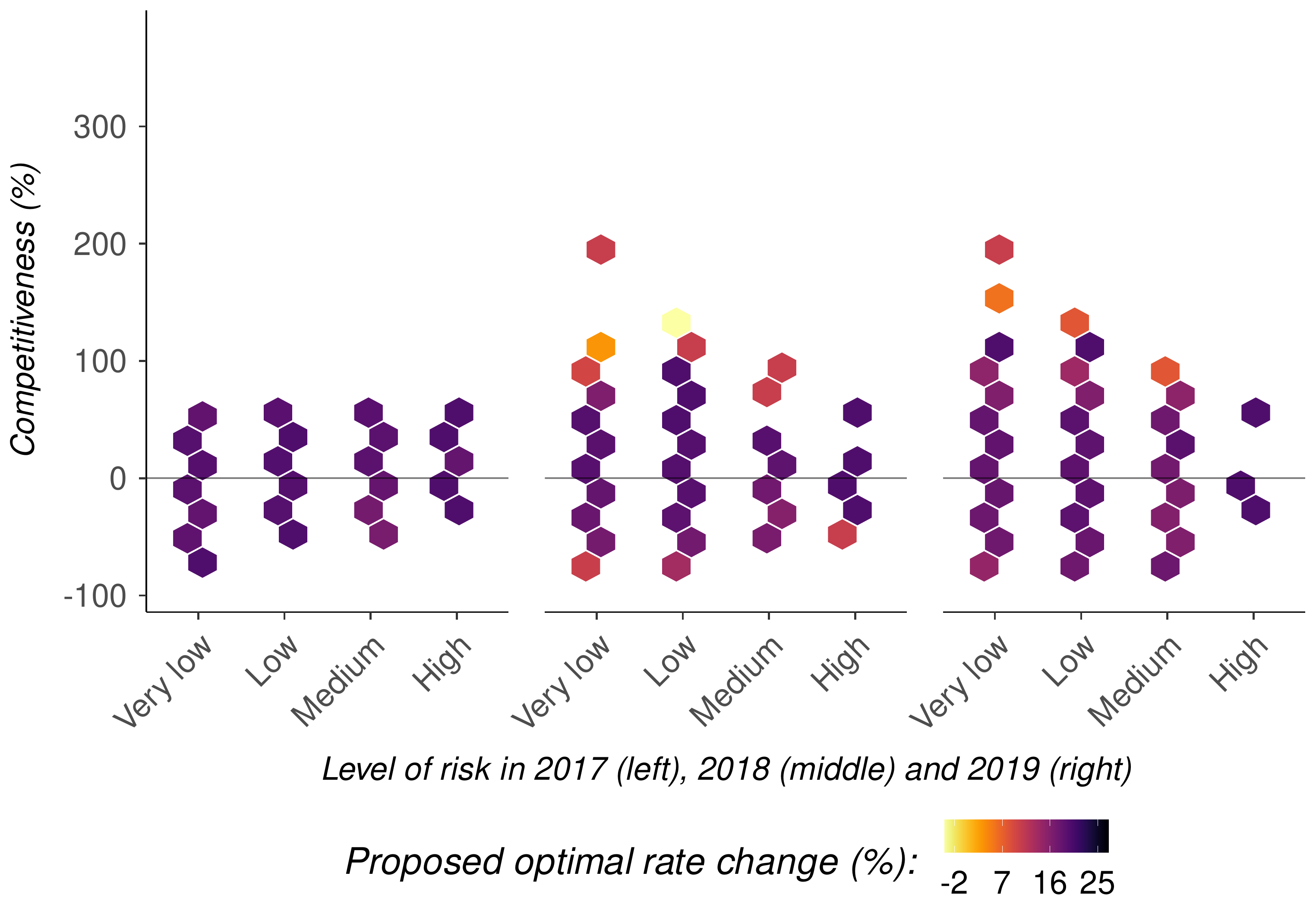}
        \end{tabular}\vspace{-6pt}
        \caption{Optimal rate changes for discrete global response model with XGBoost}
        \label{Figure_B.5b}
    \end{subfigure}
    
    \vspace{3pt}
    
    \begin{subfigure}{\textwidth}
        \centering
        \begin{tabular}{c c}
            \centering
            \includegraphics[width=0.45\textwidth]{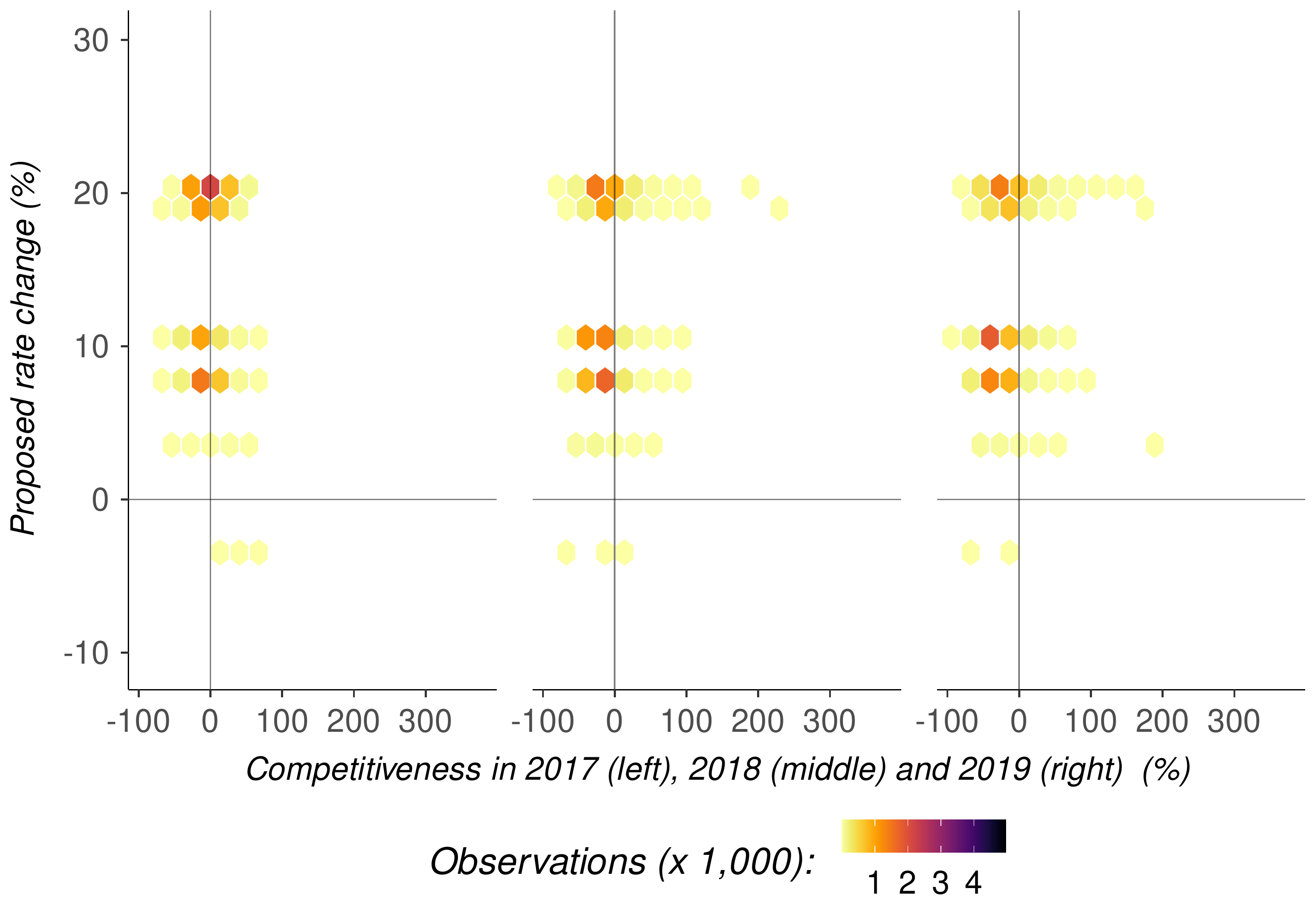}&
            \includegraphics[width=0.45\textwidth]{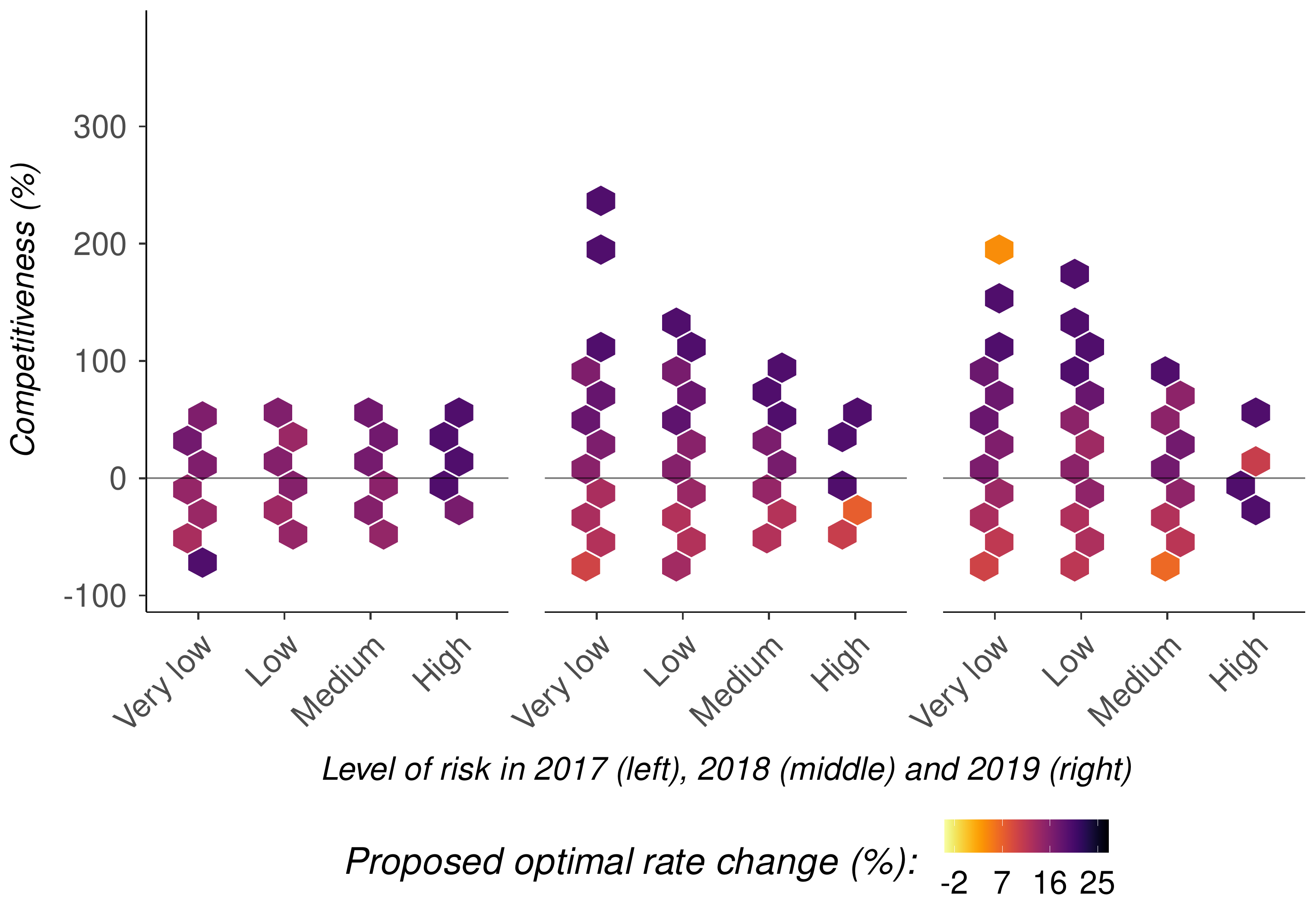}
        \end{tabular}\vspace{-6pt}
        \caption{Optimal rate changes for restricted continuous dose-response function}
        \label{Figure_B.5c}
    \end{subfigure}\vspace{-3pt}
    \caption{Optimal rate changes over the competitiveness (left) jointly with the level of risk (right) for $\tau = 3$ consecutive renewals with discrete rate changes without multiple imputation (panel (a)) and with XGBoost (panel (b)), and continuous rate changes restricted to the five categorical rate change medians (panel (c)) in automobile insurance.}
	\label{Figure_B.5}
\end{figure} 
}

\end{document}